\documentclass[prb, superscriptaddress, preprintnumbers,showpacs]{revtex4}
\usepackage{graphicx}
\usepackage{amsmath}
\usepackage{multirow}
\usepackage{bm}
\usepackage{color}
\usepackage{ulem}

\begin{document}
\normalem

\title{Dynamics in the quantum Hall effect and the phase diagram of graphene}
\date{\today}

\preprint{UWO-TH-08/10}

\author{E.V. Gorbar}
\email{gorbar@bitp.kiev.ua}
\affiliation{Bogolyubov Institute for Theoretical Physics, 03680, Kiev, Ukraine}

\author{V.P. Gusynin}
\email{vgusynin@bitp.kiev.ua}
\affiliation{Bogolyubov Institute for Theoretical Physics, 03680, Kiev, Ukraine}

\author{V.A. Miransky}
\email{vmiransk@uwo.ca}
\altaffiliation[On leave from ]{Bogolyubov Institute for Theoretical Physics, 03680, Kiev, Ukraine}
\affiliation{Department of Applied Mathematics, University of Western Ontario, London, Ontario N6A 5B7, Canada}

\author{I.A.~Shovkovy}
\email{i-shovkovy@wiu.edu}
\altaffiliation[On leave from ]{Bogolyubov Institute for Theoretical Physics, 03680, Kiev, Ukraine}
\affiliation{Physics Department, Western Illinois University, Macomb, Illinois 61455, USA}

\begin{abstract}
The dynamics responsible for lifting the degeneracy of
the Landau levels in the quantum Hall (QH) effect in graphene is
studied by utilizing a low-energy effective model with a contact
interaction. A detailed analysis of the solutions of the gap
equation for Dirac quasiparticles is performed at
both zero and nonzero temperatures. The characteristic feature
of the solutions is that the order parameters connected with the
QH ferromagnetism and magnetic catalysis scenarios necessarily
coexist. The solutions reproduce correctly the experimentally
observed novel QH plateaus in graphene in strong magnetic fields.
The phase diagram of this system in the plane of temperature and
electron chemical potential is analyzed. The phase transitions
corresponding to the transitions between different QH plateaus in
graphene are described.
\end{abstract}

\pacs{73.43.Cd, 71.70.Di, 81.05.Uw}

\maketitle

\section{Introduction}
\label{1}

In this paper, we analyze the dynamics in quantum Hall (QH) effect in graphene,
a single atomic layer of graphite.\cite{Geim2004Science} As was experimentally
discovered in Refs.~\onlinecite{Geim2005Nature,Kim2005Nature} and theoretically
predicted in Refs.~\onlinecite{Ando2002,Gusynin2005PRL,Peres2005}, an
anomalous quantization takes place in this case: the filling factors are
$\nu = \pm 4(n+ 1/2)$, where $n=0,1,2,\ldots$ is the Landau level index.
For each QH state, a four-fold (spin and sublattice-valley) degeneracy takes place.
These properties of the QH effect are intimately connected with relativistic-like features
in the graphene dynamics.\cite{Semenoff1984PRL,Haldane1988PRL,Khveshchenko2001PRL,
Gorbar2002PRB,SGB2004,Luk'yanchuk2004}

In recent experiments,\cite{Zhang2006,Jiang2007} it has been observed that in
a strong enough magnetic field, $B \gtrsim 20~\mbox{T}$, the new QH plateaus
with $\nu = 0, \pm 1$ and $\pm 4$ occur. This is attributed to the magnetic field
induced splitting of the $n =0$ and $n = 1$ Landau levels (LLs). It is
noticeable that while the degeneracy of the lowest LL (LLL), $n = 0$, is
completely lifted, only the spin degeneracy of the $n = 1$ LL is removed.

On theoretical side, there are now two leading scenarios for the description of
these plateaus. One of them is the QH ferromagnetism
(QHF).\cite{Nomura2006PRL,Goerbig2006,Alicea2006PRB,Sheng2007,LS} (The
dynamics of a Zeeman spin splitting enhancement considered in
Ref.~\onlinecite{Abanin2006PRL} is intimately connected with the QHF.)
The second one is the magnetic catalysis (MC) scenario in which Dirac masses
are spontaneously produced as a result of the excitonic
condensation.\cite{Gusynin2006catalysis,Herbut2006,Fuchs2006,Ezawa2006}
For a brief review of these two scenarios, see Ref.~\onlinecite{Yang2007}.

{The QHF scenario is connected with the theory of exchange-driven
spin splitting of Landau levels\cite{Fogler1995} and utilizes the dynamical
framework developed for bilayer QH systems.\cite{Arovas1999} 
The underlying physics relies on the fact that the spin and/or valley
degeneracy of the one-particle states is lifted by the repulsive
Coulomb interaction in a many-body system at half filling. The
argument is the same as that behind the Hund's rule in
atomic physics. The Coulomb energy of the system is lowered by
antisymmetrizing the coordinate part of the many-body wave function.
Because of the Fermi statistics of the charge carriers, the
corresponding lowest energy state must be symmetric in the
spin-valley degrees of freedom, i.e., it is spin and/or valley
polarized.}

{On the other
hand, the MC scenario is based on the phenomenon of an enhancement of the
density of states in infrared by a strong magnetic field, which catalyzes
electron-hole pairing (leading to excitonic condensates) in relativistic-like
systems. The essence of the MC phenomenon is the dimensional
reduction $D \to D - 2$ in the pairing dynamics on
the LLL with energy $E = 0$ (containing both electron and hole states).
In two dimensions, this reduction
implies a non-zero, proportional to $|eB|/2\pi\hbar c$, density of states 
in infrared. 
The latter is responsible for a Cooper-like electron-hole
pairing even at the weakest attractive
interaction between electrons and holes. This universal phenomenon was
revealed in Ref.~\onlinecite{Gusynin1995PRD} and was first considered in graphite
in Refs.~\onlinecite{Khveshchenko2001PRL,Gorbar2002PRB}.}

The difference between the QHF and MC scenarios is in utilizing
different order parameters in breaking an approximate $U(4)$ symmetry of the
Hamiltonian of graphene. This symmetry operates in the sublattice-valley and
spin spaces. While the QHF order parameters are described by densities of the
conserved charges connected with diagonal generators of the non-Abelian
subgroup $SU(4) \subset U(4)$, the order parameters in the MC scenario are
Dirac mass terms.

One may think that the QHF and MC order parameters should compete
with each other. However, as was recently {pointed out}
by three of the authors,\cite{Gorbar2007} the situation is quite
different: these two sets of the order parameters {\it
necessarily} coexist, which implies that they have the same
dynamical origin. The physics underlying their coexistence is
specific for relativistic-like dynamics that makes the QH dynamics
of the $U(4)$ breakdown in graphene to be quite different from
that in the bilayer QH systems\cite{Arovas1999} whose dynamics has
no relativistic-like features.

The main goal of this paper is a detailed study of the dynamics responsible
for lifting the degeneracy of the Landau levels in the quantum Hall effect in
graphene using the model of Ref.~\onlinecite{Gorbar2007}. To get the benchmark
results {that are} unobscured by the various types of possible
disorder,\cite{disorder0,disorder1,disorder2,LL-width} the analysis in this study
is done for graphene in the clean limit. By taking into account a considerable
improvement in samples quality seen in graphene suspended {
above a Si/SiO$_2$ gate electrode\cite{Bolotin} or above a graphite
substrate,\cite{Andrei200803} it is expected that the clean limit already
provides a reasonable qualitative description for some real devices
(the role of disorder in this dynamics will be briefly considered in Sec.~\ref{6}.)}

{The main tool in our analysis is a gap equation for the propagator of
Dirac quasiparticles.} {The highlights of the analysis are as follows:}

\begin{enumerate}

\item {The coexistence of the QHF and MC order parameters is
a robust phenomenon, which is mostly based on the kinematic and
symmetric properties of the QH dynamics in graphene.}

\item {The process of filling the LLs is described by varying the
electron chemical potential $\mu_0$. A set of the solutions of the
gap equation at a fixed $\mu_0$ is quite rich. The stable solution
is selected} {as the solution with the lowest free energy density.}
{The obtained results for the QH effect qualitatively agree with
the experimental data in Refs.~\onlinecite{Zhang2006,Jiang2007}.}

\item {The existence of two types of the Dirac masses in the QH
dynamics in graphene is established.} {Both of them play an
important role in the dynamics.}

\item {The phase diagram in the plane of temperature $T$ and electron
chemical potential $\mu_0$ is analyzed. The phase transitions corresponding
to the transitions between different QH plateaus are described.}

\end{enumerate}

The paper is organized as follows. In Sec.~\ref{2} we start by
describing the general features of the model itself as well as the
many-body approximation used in its analysis. After that, in
Sec.~\ref{3}, we derive the gap equation for Dirac quasiparticles
in graphene at zero and nonzero temperatures. The necessity of the
coexistence of the QHF and MC order parameters in the solutions of
the gap equation is shown. The analysis of the quasiparticle
dynamics at the LLL, which is relevant to the $\nu = 0, \pm 1$ QH
plateaus, is presented in Sec.~\ref{4}. There we first give a
detailed derivation of the analytic results of
Ref.~\onlinecite{Gorbar2007} at zero temperature. Then, we
consider the nonzero temperature case by utilizing numerical
calculations. In a similar fashion, in Sec.~\ref{5}, the
quasiparticle dynamics at the $n=1$ Landau level is analyzed. In
Sec.~\ref{6}, we summarize our findings in the form of the phase
diagram of graphene in the $T-\mu_0$ plane. The obtained phase
diagram is rich and it allows to describe all the recently
discovered novel plateaus (as well as the plateaus $\nu=\pm3$ and
$\nu=\pm5$ which have not been observed yet) in graphene in strong
magnetic fields.\cite{Zhang2006,Jiang2007} We also discuss the
correspondence between our results and the experimental data and
point out that the coexistence of the QHF and MC order parameters
could have important consequences for edge states, whose relevance
for the dynamics in graphene has been recently discussed in
Refs.~\onlinecite{Abanin2006PRL,Abanin2007PRL,Ong2007}. Detailed
derivations of some key results used in our analysis are presented
in four Appendices at the end of the paper.

\section{Model: General Description}
\label{2}

\subsection{Model: Hamiltonian and gap equation}
\label{2.1}

Our approach is based on the gap equation for the propagator of Dirac
quasiparticles. For the description of the dynamics in graphene, we will use the
model introduced recently in Ref.~\onlinecite{Gorbar2007}, which in turn is a modification
of the model in {Refs.~\onlinecite{Khveshchenko2001PRL,Gorbar2002PRB,Gusynin2006catalysis}.}
Let us start from the description of the latter. In this model, while quasiparticles are
confined to a 2-dimensional plane, the electromagnetic (Coulomb) interaction
between them is three-dimensional in nature. The low-energy quasiparticles
excitations in graphene are conveniently described in terms of a four-component
Dirac spinor $\Psi_{s}^T = \left( \psi_{KAs},\psi_{KBs},\psi_{K^\prime Bs},
\psi_{K^\prime As}\right)$ which combines the Bloch states with spin indices
$s=\pm$ on the two different sublattices ($A$, $B$) of the hexagonal
graphene lattice and with momenta near the two inequivalent valley points
($K$, $K^\prime$) of the two-dimensional Brillouin zone. The free quasiparticle
Hamiltonian can be recast in a relativistic-like form with the Fermi
velocity $v_F\approx 10^6~\mbox{m/s}$ playing the role of the speed of light:
\begin{equation}
H_0=v_F\int d^2{r}\,\overline{\Psi}\left(\gamma^1\pi_x+\gamma^2\pi_y\right)\Psi,
\label{free-hamiltonian}
\end{equation}
where $\mathbf{r} =(x,y)$ is the position vector in the plane of
graphene and $\overline{\Psi} =\Psi^\dagger\gamma^0$ is the Dirac
conjugated spinor. In Eq.~(\ref{free-hamiltonian}), $\gamma^\nu$
with $\nu=0,1,2$ are $4 \times 4$ gamma matrices belonging to a
reducible representation of the Dirac algebra, namely,
$\gamma^\nu=\tilde{\tau}^3\otimes (\tau^3,i\tau^2,-i\tau^1)$,
where the Pauli matrices $\tilde{\tau}^{i}$ and $\tau^{i}$, with
$i=1,2,3$, act in the subspaces of the valleys ($K$, $K^\prime$)
and sublattices ($A$, $B$), respectively.\cite{gamma} The matrices
satisfy the usual anticommutation relations
$\left\{\gamma^\mu,\gamma^\nu\right\}=2g^{\mu\nu}$, where
$g^{\mu\nu}=\mbox{diag}\,(1,-1,-1)$ and $\mu,\nu=0,1,2$. The
canonical momentum $\bm{\pi} \equiv (\pi_x, \pi_y)=
-i\hbar\bm{\nabla}+ {e\mathbf{A}}/c$ includes the vector potential
$\mathbf{A}$ corresponding to a magnetic field $B_{\perp}$, which
is the component of the external magnetic field $\mathbf{B}$
orthogonal to the $xy$-plane of graphene.

The Coulomb interaction term has the form
\begin{eqnarray}
H_{C} &=& \frac{1}{2}\int d^2{r} d^2{r}^\prime {\Psi}^{\dagger}(\mathbf{r})
\Psi(\mathbf{r})U_{C}(\mathbf{r}-\mathbf{r}^\prime)
{\Psi}^{\dagger}(\mathbf{r}^\prime) \Psi(\mathbf{r}^\prime),
\label{Coulomb}
\end{eqnarray}
where $U_{C}(\mathbf{r})$ is the Coulomb potential in a magnetic
field, calculated in the random phase approximation (RPA) in
Ref.~\onlinecite{Gorbar2002PRB}, see Eq.~(46) there. The
Hamiltonian $H = H_0+H_{C}$ possesses a global $U(4)$ symmetry
discussed in the next subsection. The electron chemical potential
$\mu_0$ is introduced by adding the term $-\mu_0
\Psi^{\dagger}\Psi$ to the Hamiltonian density. This term also
preserves the $U(4)$ symmetry. The Zeeman interaction is included
by adding the term $\mu_{B}B\Psi^{\dagger} \sigma^3 \Psi$, where
$B \equiv |\mathbf{B}|$ and $\mu_B=e\hbar/(2mc)$ is the Bohr
magneton. Here we took into account that the Lande factor for
graphene is $g_L\simeq2$ (our convention is $e > 0$). The spin
matrix $\sigma^3$ has eigenvalue $+1$ ($-1$) for the states with
the spin directed along (against) the magnetic field
$\mathbf{B}$.\cite{Zeeman} Such states will be called spin up
(down) states. Because of the Zeeman term, the $U(4)$ symmetry is
broken down to a symmetry $U(2)_{+} \times U(2)_{-}$, where the
subscript $\pm$ labels the spin of the states on which this
subgroup operates (see the next subsection).

\begin{figure}
\begin{center}
\includegraphics[width=.4\textwidth]{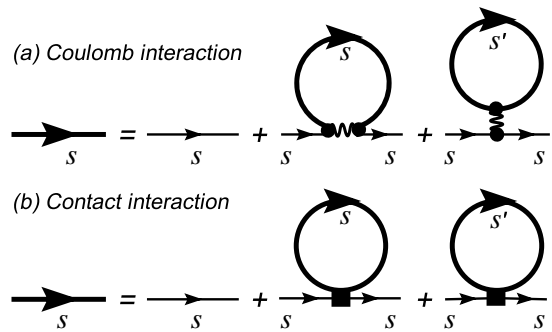}
\caption{The diagrammatic form of the {gap} equation in the
Hartree-Fock (mean field) approximation. The upper (lower) diagram
corresponds to the form of the gap equation with the long-range Coulomb
(contact) interaction. The indices denote quasiparticle spins.}
\label{fig.SD_eq}
\end{center}
\end{figure}

The dynamics will be treated in the Hartree-Fock (mean field) approximation,
which is conventional and appropriate in this case.\cite{Khveshchenko2001PRL,
Gorbar2002PRB,Nomura2006PRL,Goerbig2006,Gusynin2006catalysis} Then, at zero
temperature and in the clean limit (no impurities), the gap equation takes the
form:
\begin{equation}
G^{-1}(u,u^\prime) = S^{-1}(u,u^\prime)
+i\hbar\gamma^0G(u,u^\prime)\gamma^0 \delta(t-t^{\prime})
U_{C}(\mathbf{r}-\mathbf{r}^\prime)
- i\hbar\gamma^0 \mbox{tr} \left[ \gamma^0 G(u,u) \right]
\delta^{3}(u - u^{\prime})U_{C}^{(F)}(0),
\label{SD}
\end{equation}
where $u \equiv (t,\mathbf{r})$, $t$ is the time coordinate,
$U_{C}^{(F)}(0)$ is the Fourier transform of $U_{C}(\mathbf{r})$ at
${\bf k} = 0$, $G(u,u^\prime)=
\hbar^{-1}\langle 0|T\Psi(u)\bar{\Psi}(u^\prime)|0\rangle$ is
the {\em full} quasiparticle propagator, and
\begin{equation}
iS^{-1}(u,u^\prime)=\left[(i\hbar\partial_t+\mu_0 - \mu_BB\sigma^3)\gamma^0
-v_{F}(\bm{\pi}\cdot\bm{\gamma})\right]\delta^{3}(u- u^\prime)
\label{inversebare}
\end{equation}
is the inverse {\em bare} quasiparticle propagator.
Note that while the second term on the right hand side of Eq.~(\ref{SD})
describes the exchange interaction, the third one is the Hartree term describing
the direct interaction. The diagrammatic form of the gap equation is
shown in Fig.~\ref{fig.SD_eq}(a).

{As will be shown in Sec. \ref{3} below, in order to determine all
the order parameters,  
the analysis of the gap equation (\ref{SD}) has to be done
beyond the LLL approximation, which is a
formidable problem.} Because of this, we follow the approach of
Ref.~\onlinecite{Gorbar2007} and replace the Coulomb potential $U_{C}(\mathbf{r})$
in the gap equation by the contact interaction $G_{\rm int}\delta^{2}(\mathbf{r})$.
Thus, we arrive at
\begin{equation}
G^{-1}(u,u^\prime) = S^{-1}(u,u^\prime)
+ i\hbar G_{\rm int}\gamma^0G(u,u)\gamma^0 \delta^{3}(u - u^\prime) 
- i\hbar G_{\rm int}\gamma^0\, \mbox{tr}[\gamma^0G(u,u)]\delta^{3}(u- u^\prime),
\label{gap}
\end{equation}
where $G_{\rm int}$ is a dimensionful coupling constant. As we
will see later, in the analysis it would be more convenient to use
a dimensionless coupling constant $\lambda=G_{\rm int} \Lambda
/(4\pi^{3/2} \hbar^2 v_{F}^2)$ instead of $G_{\rm int}$ (note that
$\Lambda$ is the energy cutoff parameter which is required when a
contact interaction is used). The corresponding diagrammatic form
of the equation is given in Fig.~\ref{fig.SD_eq}(b). A similar
approximation is commonly used in Quantum Chromodynamics (QCD),
where the long range gluon interaction is replaced by the contact
Nambu--Jona-Lasinio one. This leads to a good description of many
features of the nonperturbative dynamics in low energy region of
QCD (for a review see, for example, Ref.~\onlinecite{book}).  By
taking into account the universality of the MC phenomenon and the
fact that the symmetry and kinematic structures of equations
(\ref{SD}) and (\ref{gap}) are the same, we expect that
approximate gap equation (\ref{gap}) should be at least
qualitatively reliable for the description of the LLL and the
first few LLs, say, with $n = \pm 1$.

As to the value of the cutoff $\Lambda$, note that, in a strong magnetic field,
the Landau scale
\begin{equation}
\epsilon_B \equiv \sqrt{2\hbar|eB_{\perp}|v_{F}^2/c} \simeq
424\sqrt{|B_{\perp}[\mbox{T}]|}~\mbox{K}
\label{Lscale}
\end{equation}
is the only relevant energy scale in the dynamics with the Coulomb interaction.
This suggests that the ultraviolet cutoff $\Lambda$ should be taken of order
$\epsilon_{B}$ in the approximation with the contact interaction.
The dimensionful coupling constant $G_{\rm int}$ then becomes
{$G_{\rm int} \sim 4\pi^{3/2}\hbar^2 v^{2}_F\lambda/\epsilon_{B}$.}

{Before concluding this section, the following remark concerning the
present approximation is in order. While there is Debye screening at nonzero
chemical potential, the situation is more complicated near the Dirac point with
$\mu_0 =0$. In that case, while for subcritical values of the Coulomb coupling
constant\cite{footnote1a} the polarization effects lead only to its screening without
changing the {\it form} of the Coulomb interactions at large distances,\cite{Katsnelson2006}
they lead to a drastic change of the form of the interactions for a supercritical
coupling.\cite{Shitov2007} In the present work, the dynamics with a subcritical
coupling is utilized, when no dynamical gaps are generated without a magnetic
field ({this is in agreement with the experiments}\cite{Geim2005Nature,Kim2005Nature}).
In our approximation, utilizing smeared contact interactions with an ultraviolet cutoff
$\Lambda \sim \epsilon_B$, the contribution of large energies $\omega > \epsilon_B$
is suppressed much stronger than for the subcritical Coulomb like interactions.
However, because the dominant contribution in the gap equation comes
from energies $\omega \ll \epsilon_B$, we expect that the present approximation
is qualitatively reliable even near the Dirac point.}

\subsection{Model: Symmetry and order parameters}
\label{2.2}

The Hamiltonian $H = H_0 + H_C$, with $H_0$ and $H_C$ given in
Eqs.~(\ref{free-hamiltonian}) and (\ref{Coulomb}), respectively,
possesses the $U(4)$ symmetry with the following 16 generators
(see for example Refs.~\onlinecite{Gorbar2002PRB,Gusynin2006catalysis}):
\begin{equation}
\frac{\sigma^\alpha}{2}\otimes I_4,\quad
\frac{\sigma^\alpha}{2i}\otimes\gamma^3,\quad
\frac{\sigma^\alpha}{2}\otimes\gamma^5,\quad\mbox{and}\quad
\frac{\sigma^\alpha}{2}\otimes\gamma^3\gamma^5,
\end{equation}
where $I_4$ is the $4 \times 4$ Dirac unit matrix and $\sigma^\alpha$, with
$\alpha=0, 1, 2, 3$, are four Pauli matrices connected with the spin degrees
of freedom ($\sigma^0$ is the $2 \times 2$ unit matrix). In the representation
used in the present paper (for the definition of the $\gamma^{\nu}$ matrices,
see the previous subsection), the Dirac matrices $\gamma^3$ and $\gamma^5
\equiv i\gamma^0 \gamma^1 \gamma^2 \gamma^3 $ are
\begin{equation}
\gamma^3= i\left(\begin{array}{cc} 0& I \\ I& 0\end{array}\right),\quad
\gamma^5= i\left(\begin{array}{cc} 0&I\\-I&0\end{array}\right), \label{35}
\end{equation}
where $I$ is the $2\times 2$ unit matrix. Note that while the Dirac matrices
$\gamma^0$ and $\pmb \gamma = (\gamma^{1}, \gamma^{2})$ anticommute with
$\gamma^3$ and $\gamma^5$, they commute with the diagonal matrix
$\gamma^3\gamma^5 = - \gamma^5\gamma^3$,
\begin{equation}
\gamma^3\gamma^5 = \left(\begin{array}{cc} I& 0 \\ 0& -I\end{array}\right).
\label{gamma35}
\end{equation}
The matrix $\gamma^3\gamma^5$ is called a pseudospin operator.

The total Hamiltonian
\begin{equation}
H_{\rm tot} \equiv
H + \int d^2{r}\,[\mu_{B}B\Psi^{\dagger}\sigma^3\Psi -
\mu_{0} \Psi^{\dagger}\Psi]
\label{tot}
\end{equation}
possesses a lower symmetry. Because of the Zeeman term
$\mu_{B}B\Psi^{\dagger}\sigma_3\Psi$, the $U(4)$ symmetry is broken down to
the ``flavor'' symmetry $U(2)_{+} \times U(2)_{-}$, where the subscript $\pm$
corresponds to spin up and spin down states, respectively. The generators of the
$U(2)_{s}$, with $s=\pm$, are $I_4 \otimes P_{s}$, $-i\gamma^3 \otimes P_{s}$,
$\gamma^5 \otimes P_{s}$, and $\gamma^3\gamma^5 \otimes P_{s}$, where
$P_{\pm}=(1\pm \sigma^3)/2$ are the projectors on spin up and down states.

Our goal is to find all solutions of Eq.~(\ref{gap}) both with intact and spontaneously
broken $SU(2)_{s}$ symmetry, where $SU(2)_{s}$ is the largest non-Abelian subgroup
of the $U(2)_{s}$. The Dirac mass term $\tilde{\Delta}_{s}\bar{\Psi}P_{s}\Psi \equiv
\tilde{\Delta}_{s}\Psi^{\dagger}\gamma^{0}P_{s}\Psi$, where $\tilde{\Delta}_{s}$ is a
Dirac gap (mass),\cite{footnote1} is assigned to the triplet representation of the
$SU(2)_{s}$, and the generation of such a mass would lead to a spontaneous {
breakdown} of the flavor $SU(2)_{s}$ symmetry down to the $\tilde{U}(1)_{s}$
with the generator $\gamma^3\gamma^5 \otimes
P_{s}$.\cite{Gusynin2006catalysis,Khveshchenko2001PRL,Gorbar2002PRB}
There is also a Dirac mass term of the form $\Delta_{s}\bar{\Psi}\gamma^3\gamma^{5}P_{s}\Psi$
that is a singlet with respect to $SU(2)_{s}$, and therefore its generation would
not break this symmetry. On the other hand, while the triplet mass term is even
under time reversal $\cal{T}$, the singlet mass term is $\cal{T}$-odd (for a
recent review of the transformation properties of different mass terms in
graphene, see Ref.~\onlinecite{Gusynin2007review}). Note that the possibility of
a singlet Dirac mass like $\Delta$ was first discussed in relation to graphite
about 20 years ago.\cite{Haldane1988PRL}

The masses $\Delta_{s}$ and $\tilde{\Delta}_{s}$ are related to the MC order
parameters $\langle {\bar{\Psi}\gamma^3\gamma^{5}P_{s}\Psi} \rangle$ and
$\langle {\bar{\Psi}P_{s}\Psi} \rangle$. In terms of the Bloch components of
the spinors, the corresponding operators take the following forms:
\begin{eqnarray}
\label{singlet_mass}
\Delta_{s}: &\quad&
{\bar{\Psi}\gamma^3 \gamma^5 P_{s} \Psi} =
  \psi_{K  As}^\dagger\psi_{K As}
- \psi_{K^{\prime} As}^\dagger \psi_{K^{\prime}As}
- \psi_{K Bs}^\dagger \psi_{K Bs}
+ \psi_{K^{\prime}Bs}^\dagger \psi_{K^{\prime} Bs},\\
\label{triplet_mass}
\tilde{\Delta}_{s}:&\quad&
{\bar{\Psi} P_{s} \Psi} =
  \psi_{K  As}^\dagger\psi_{K As}
+ \psi_{K^{\prime} As}^\dagger \psi_{K^{\prime}As}
- \psi_{K Bs}^\dagger \psi_{K Bs}
- \psi_{K^{\prime}Bs}^\dagger \psi_{K^{\prime} Bs}.
\end{eqnarray}
The expressions on the right hand side further clarify the
physical meaning of the Dirac mass parameters as the Lagrange
multipliers that control various density imbalances of electrons
at the two valleys and the two sublattices. {In particular, the
order parameter (\ref{triplet_mass}), connected with the triplet
Dirac mass, describes the charge density imbalance between
the two sublattices, i.e., a charge density
wave.\cite{Khveshchenko2001PRL,Gusynin2006catalysis}}

As revealed in Ref.~\onlinecite{Gorbar2007}, and will be discussed in detail in the
next section, these MC order parameters necessarily coexist with QHF ones
in the solutions of the gap equation (\ref{gap}). More precisely, for a fixed spin,
the full inverse quasiparticle propagator takes the following general form [compare
with Eq.~(\ref{inversebare})]:
\begin{eqnarray}
iG^{-1}_{s}(u,u^\prime)&=& \left[(i\hbar\partial_t+\mu_{s} +
\tilde{\mu}_{s}\gamma^3\gamma^5)\gamma^0 - v_{F}(\bm{\pi}\cdot\bm{\gamma})
-\tilde{\Delta}_{s} + \Delta_{s}\gamma^3\gamma^5\right]\delta^{3}(u-u^\prime),
\label{full-inverse}
\end{eqnarray}
where the parameters $\mu_{s}$, $\tilde{\mu}_{s}$, $\Delta_{s}$,
and $\tilde{\Delta}_{s}$ are determined from gap equation
(\ref{gap}). Note that the {full} electron chemical
potentials $\mu_{\pm}$ include the Zeeman energy $\mp Z$ with
\begin{equation}
Z \simeq \mu_BB = 0.67B[\mbox{T}]~\mbox{K}.
\label{Zeeman}
\end{equation}
The chemical potential $\tilde{\mu}_{s}$ is related to the density of the
conserved pseudospin charge $\Psi^{\dagger}\gamma^3\gamma^{5}P_{s}\Psi$,
which is assigned to the triplet representation of the $SU(2)_{s}$. Therefore, unlike
the masses $\Delta_{s}$ and $\tilde{\Delta}_{s}$, the chemical potentials $\mu_3
\equiv (\mu_{+} - \mu_{-})/2$ and $\tilde{\mu}_{s}$ are related to the conventional
QHF order parameters: the spin density $\langle {\Psi^{\dagger}\sigma^3 \Psi} \rangle$ and
the pseudospin density $\langle {\Psi^{\dagger}\gamma^3\gamma^5P_{s}\Psi} \rangle $,
respectively. In terms of the Bloch components, the corresponding operators
take the following forms:
\begin{eqnarray}
\label{singlet_mu}
\mu_3:&\quad&
 {\Psi^{\dagger}\sigma^3 \Psi} =\frac{1}{2}
 \sum_{\kappa=K  ,K^{\prime}} \sum_{a =A ,B}
\left(  \psi_{\kappa a +}^\dagger\psi_{\kappa a +}
- \psi_{\kappa a-}^\dagger\psi_{\kappa a-}\right),\\
\label{triplet_mu}
\tilde{\mu}_{s}:&\quad&
{\Psi^{\dagger}\gamma^3\gamma^5P_{s}\Psi} =
  \psi_{K  As}^\dagger\psi_{K As}
- \psi_{K^{\prime} As}^\dagger \psi_{K^{\prime}As}
+ \psi_{K Bs}^\dagger \psi_{K Bs}
- \psi_{K^{\prime}Bs}^\dagger \psi_{K^{\prime} Bs}.
\end{eqnarray}
By comparing the last expression with Eq.~(\ref{triplet_mass}), we see that while
the triplet MC order parameter related to $\tilde{\Delta}_{s}$ describes the charge
density imbalance between the two graphene sublattices,
the pseudospin density
(related to $\tilde{\mu}_{s}$)  describes the charge density imbalance between
the two valley points in the Brillouin zone. 
On the other hand, as seen
from Eq. (\ref{singlet_mu}), $\mu_3$ is related to the conventional ferromagnetic order
parameter $\langle {\Psi^{\dagger}\sigma^3 \Psi} \rangle$ .

The following remark is in order. Because of the relation
$\gamma^5 = i\gamma^0 \gamma^1 \gamma^2 \gamma^3$, the operator in
Eq. (\ref{triplet_mu}) can be rewritten as
$i{\bar \Psi}\gamma^1 \gamma^2 P_{s}\Psi$. The latter has the same
form as the anomalous magnetic moment operator in Quantum Electrodynamics
(QED). However, unlike QED, in graphene, it describes not the polarization
of the spin degrees of freedom but that of the pseudospin ones,
related to the valleys and sublattices. Because of that, this operator
can be called the anomalous magnetic pseudomoment operator.

Let us describe the breakdown of the $U(4)$ symmetry down to the $U(2)_{+}
\times U(2)_{-}$ flavor symmetry, responsible for a spin gap, in more detail.
Because of the Zeeman term, this breakdown is not spontaneous but explicit.
The point however is that, as was shown in Ref.~\onlinecite{Abanin2006PRL},
a magnetic field leads to a strong enhancement of the spin gap in graphene.
Such an enhancement is reflected in a large chemical potential
$\mu_3 = (\mu_{+} -\mu_{-})/2\gg Z$ and the corresponding QHF order
parameter $\langle {\Psi^{\dagger}\sigma^3 \Psi} \rangle$. But as was pointed out
in Ref.~\onlinecite{Gorbar2007} and will be shown below in Sec.~\ref{4}, it is not all.
There is also a large contribution to the spin gap connected with
the flavor singlet Dirac mass $\Delta_3 \equiv (\Delta_{+} -\Delta_{-})/2$ and the
corresponding MC order parameter $\langle {\bar{\Psi}\gamma^{3}\gamma^{5}\sigma^3
\Psi} \rangle $. This feature leads to important consequences for the dynamics of edge
states in graphene (see Secs.~\ref{4} and \ref{6}).

As will be shown in Subsec.~\ref{4.1}, the spin gap may remain large even in
the limit
when the Zeeman energy $Z = \mu_{B}B$ goes to zero. In this limit, a genuine
spontaneous breakdown of the $U(4)$ takes place. In the realistic case with a
nonzero but small $Z$, one can say that a quasi-spontaneous breakdown of the
$U(4)$ is realized.

The $U(2)_{+} \times U(2)_{-}$ is an exact symmetry of the total Hamiltonian
$H_{\rm tot}$ (\ref{tot}) of the continuum effective theory. However, as was
pointed out in Ref.~\onlinecite{Alicea2006PRB} 
{(see also Refs. \onlinecite{LS}, \onlinecite{Herbut2006}, and 
\onlinecite{Aleiner2007})}, 
it is not exact for the Hamiltonian on
the graphene lattice. In fact there are small on-site repulsion
interaction terms which break the
$U(2)_{+} \times U(2)_{-}$ symmetry down to a $U(1)_{+} \times Z_{2+}
\times U(1)_{-} \times Z_{2-}$ subgroup, where the elements of the
discrete group $Z_{2s}$ are
$\gamma^5 \otimes P_{s} + I_{4}\otimes P_{-s}$ and the unit matrix.
Unlike {a
spontaneous breakdown of continuous symmetries, a spontaneous breakdown of
the discrete symmetry $Z_{2\pm}$, with the order parameters
$\langle {\bar{\Psi}P_{\pm}\Psi} \rangle $ and
$\langle\Psi^{\dagger}\gamma^3\gamma^{5}P_{\pm}\Psi\rangle$,
is not forbidden by the Mermin-Wagner theorem at finite
temperatures in a planar system.\cite{MW} This observation is of relevance
for the description of the ground state responsible for the $\nu = \pm 1$
plateaus (see Subsec.~\ref{4.2}).}

{Thus, there are six order parameters describing the breakdown of
the $U(4)$ symmetry: the two singlet order parameters connected with $\mu_3$ and
$\Delta_3$ and the four triplet ones connected with $\tilde{\mu}_{\pm}$ and
$\tilde{\Delta}_{\pm}$.}

By extracting the location of the poles in full propagator $G(u,u^\prime)$, which is
given in terms of the sum over separate LL contributions in Eq.~(\ref{interaction-A1})
in Appendix~\ref{A}, it is straightforward to derive the dispersion relations
for the quasiparticles in graphene. The dispersion relations for LLs with
$n \geq 1$ are
\begin{eqnarray}
\hspace{-2mm}\omega^{(\sigma)}_{ns} &=&-\mu_{s} + \sigma\tilde{\mu}_{s}
\pm\sqrt{n \epsilon_{B}^2 + (\tilde{\Delta}_{s}+\sigma\Delta_{s})^2}\,,
\label{higherLLs}
\end{eqnarray}
where $\sigma=\pm 1$ and the two signs in front of the square root
correspond to the energy levels above and below the Dirac point.
In the case of the LLL, which is special, the corresponding
dispersion relations read
\begin{equation}
\omega^{(\sigma)}_{s}= -\mu_{s} + \sigma
\left(\tilde{\mu}_{s}\,\mbox{sign}(eB_{\perp})+\,\tilde{\Delta}_{s}\right)
+ \Delta_{s}\,\mbox{sign}(eB_{\perp}).
\label{LLLenergylevels}
\end{equation}
As shown in Subsec.~\ref{A2} in Appendix~\ref{A}, the parameter
$\sigma$ in Eqs.~(\ref{higherLLs}) and (\ref{LLLenergylevels}) is
connected with the eigenvalues of the diagonal pseudospin matrix
$\gamma_3\gamma_5$ in Eq.~(\ref{gamma35}). For the LLs with $n
\geq 1$, the value $\sigma = \pm 1$ in (\ref{higherLLs})
corresponds to the eigenvalues $\mp 1$ of $\gamma^3\gamma^5$. On
the other hand, for LLL, the value $\sigma = \pm 1$ in
(\ref{LLLenergylevels}) corresponds to
$\mbox{sign}(eB_\perp)\times \,(\mp 1)$, with $\mp 1$ being the
eigenvalues of $\gamma^3\gamma^5$.

One can see from Eqs.~(\ref{higherLLs}) and (\ref{LLLenergylevels}) that
at a fixed spin, the terms with $\sigma$ are responsible for splitting of LLs.
We will return to this issue in Sec.~\ref{4}.

\section{Gap Equation: Explicit Form at $T=0$ and $T\neq 0$ and
coexistence of QHF and MC order parameters}
\label{3}

In this section, we will present the explicit equations for the Dirac masses
and the chemical potentials at zero and finite temperature. In particular, it
will be shown that the QHF and MC order parameters necessarily coexist.

The equations for the Dirac masses $\Delta_{s}$ and $\tilde{\Delta}_{s}$ and the
chemical potentials $\mu_{s}$ and $\tilde{\mu}_{s}$ follow from the matrix form
of the gap equation in Eq.~(\ref{gap}) and expression (\ref{full-inverse}).
Their derivation, while straightforward, is rather tedious. It is considered
in Appendix~\ref{A} in detail. At zero temperature, the equations are
\begin{eqnarray}
\tilde{\Delta}_{s} &=& \frac{A}{2}
\Bigg\{
-\left[\mbox{sign}(\mu_{s}-\tilde{\mu}_{s})\theta(|\mu_{s}-\tilde{\mu}_{s}|-E_{0s}^+)
- \mbox{sign}(\mu_{s}+\tilde{\mu}_{s})\theta(|\mu_{s}+\tilde{\mu}_{s}|-E_{0s}^-)\right]
\mbox{sign}(eB_{\perp}) \nonumber \\
&&+\sum_{n=0}^{\infty}\left[\frac{(\tilde{\Delta}_{s}+
\Delta_{s})\theta(E_{ns}^+-|\mu_{s}-\tilde{\mu}_{s}|)}{E_{ns}^+}
+\frac{(\tilde{\Delta}_{s}-\Delta_{s})\theta(E_{ns}^--|\mu_{s}+\tilde{\mu}_{s}|)}
{E_{ns}^-}\right]
[1+\theta(n-1)]\Bigg\},
\label{E1-a}
\end{eqnarray}
\begin{eqnarray}
\Delta_{s} &=& \frac{A}{2}
\Bigg\{
-\left[\mbox{sign}(\mu_{s}-\tilde{\mu}_{s})\theta(|\mu_{s}-\tilde{\mu}_{s}|-E_{0s}^+) +
\mbox{sign}(\mu_{s}+\tilde{\mu}_{s})\theta(|\mu_{s}+\tilde{\mu}_{s}|-E_{0s}^-)\right]
\mbox{sign}(eB_{\perp}) \nonumber \\
&&+ \sum_{n=0}^{\infty}
\left[\frac{(\tilde{\Delta}_{s}+\Delta_{s})\theta(E_{ns}^+-|\mu_{s}-\tilde{\mu}_{s}|)}
{E_{ns}^+}-\frac{(\tilde{\Delta}_{s}-\Delta_{s})\theta(E_{ns}^-
-|\mu_{s}+\tilde{\mu}_{s}|)}{E_{ns}^-}\right][1+\theta(n-1)]
\Bigg\},
\label{E2-a}
\end{eqnarray}
\begin{eqnarray}
\tilde{\mu}_{s} &=& \frac{A}{2}
\Bigg\{
\left[\frac{(\tilde{\Delta}_{s}+\Delta_{s})\theta(E_{0s}^+
-|\mu_{s}-\tilde{\mu}_{s}|)}{E_{0s}^+} +
\frac{(\tilde{\Delta}_{s}-\Delta_{s})\theta(E_{0s}^--
|\mu_{s}+\tilde{\mu}_{s}|)}{E_{0s}^-}\right]
\mbox{sign}(eB_{\perp})\nonumber \\
&&+\sum_{n=0}^{\infty}
\left[-\mbox{sign}(\mu_{s}-\tilde{\mu}_{s})\theta(|\mu_{s}-\tilde{\mu}_{s}|-E_{ns}^+) +
\mbox{sign}(\mu_{s}+\tilde{\mu}_{s}) \theta(|\mu_{s}+\tilde{\mu}_{s}|-E_{ns}^-)\right]
 [1+\theta(n-1)]
\Bigg\},
\label{E3-a}
\end{eqnarray}
\begin{eqnarray}
\mu_{s} &=& \bar{\mu}_{s} + X + \frac{A}{2}
\Bigg\{
-\left[\frac{(\tilde{\Delta}_{s}+\Delta_{s})\theta(E_{0s}^+-|\mu_{s}-\tilde{\mu}_{s}|)}{E_{0s}^+} -
\frac{(\tilde{\Delta}_{s}-\Delta_{s})\theta(E_{0s}^-
-|\mu_{s}+\tilde{\mu}_{s}|)}{E_{0s}^-}\right]\mbox{sign}(eB_{\perp})\nonumber \\
&&+\sum_{n=0}^{\infty}
\left[\mbox{sign}(\mu_{s}-\tilde{\mu}_{s})\theta(|\mu_{s}-\tilde{\mu}_{s}|-E_{ns}^+) +
\mbox{sign}(\mu_{s}+\tilde{\mu}_{s})\theta(|\mu_{s}+\tilde{\mu}_{s}|-E_{ns}^-)\right]  [1+\theta(n-1)]
\Bigg\}, \label{E4-a}
\end{eqnarray}
where the step function is defined by $\theta(x)= 1$ for $x\geq 0$ and $\theta(x)= 0$
for $x<0$. Regarding the other notation, $\bar{\mu}_{\pm} \equiv \mu_0 \mp Z$ is the
bare electron chemical potential which includes the Zeeman energy $Z = \mu_{B}B$,
and $E_{ns}^{\pm}=\sqrt{n \epsilon_{B}^2+ (\tilde{\Delta}_{s} \pm \Delta_{s})^2}$
are quasiparticle energies. In {these} equations, we introduced a new
energy scale, $A$, that plays an important role throughout the analysis.
It is determined by the value of the magnetic field and the coupling
constant strength,
\begin{equation}
A\equiv \frac{G_{\rm int}|eB_{\perp}|}{8\pi\hbar c} =
\frac{\sqrt{\pi}\lambda\epsilon_{B}^2}{4\Lambda}.
\label{Apar}
\end{equation}
The second term on the right hand side in Eq.~(\ref{E4-a}) is defined as
follows:
\begin{equation}
X = \sum_{s=\pm}\,X_{s},
\label{X}
\end{equation}
where
\begin{eqnarray}
X_{s} &=& -2A
\Bigg\{
 -\left[\frac{(\tilde{\Delta}_{s}+\Delta_{s})\theta(E_{0s}^+
-|\mu_{s}-\tilde{\mu}_{s}|)}{E_{0s}^+} -
\frac{(\tilde{\Delta}_{s}-\Delta_{s})\theta(E_{0s}^-
-|\mu_{s}+\tilde{\mu}_{s}|)}{E_{0s}^-}\right] \mbox{sign}(eB_{\perp})
\nonumber \\
&&+\sum_{n=0}^{\infty}
\left[\mbox{sign}(\mu_{s}-\tilde{\mu}_{s})\theta(|\mu_{s}-\tilde{\mu}_{s}|-E_{ns}^+) +
\mbox{sign}(\mu_{s}+\tilde{\mu}_{s})\theta(|\mu_{s}+\tilde{\mu}_{s}|-E_{ns}^-)\right]
 [1+\theta(n-1)]
\Bigg\}.
\label{Xs}
\end{eqnarray}
The following comment is in order here. Because of the Hartree term
in the gap equation (\ref{gap}), the equations for the spin up and spin down parameters
do not decouple: they are mixed via the $X$ term in Eq.~(\ref{E4-a}). Fortunately,
it is the only place affected by the Hartree term. As shown in Appendix~\ref{B},
this fact strongly simplifies the analysis of the system of equations (\ref{E1-a})--(\ref{E4-a}).
This point also clearly reflects the essential difference between the roles played
by the exchange and Hartree interactions in the quasiparticle dynamics
of graphene. While the former dominates in producing the QHF and MC order
parameters, the latter participates only in the renormalization of the electron
chemical potential, which is relevant for the filling of LLs.

Since the step functions in the above set of equations depend on $\mu_{s} \pm \tilde{\mu}_{s}$
and $\tilde{\Delta}_{s} \pm \Delta_{s}$, it is more convenient to rewrite the gap
equations for the following set of parameters
\begin{eqnarray}
\Delta_{s}^{(\pm)} = \Delta_{s}\pm\tilde{\Delta}_{s}\, , \qquad
\mu_{s}^{(\pm)} = \mu_{s}\pm\tilde{\mu}_{s}.
\label{newparam}
\end{eqnarray}
In the numerical analysis, we always consider a nonzero temperature. This is
implemented by utilizing the Matsubara formalism. Using the identities
\begin{eqnarray}
T\sum_{n=-\infty}^{\infty}
\frac{1}{[(2n+1)\pi T+i\mu]^2+E^2} &=&
\frac{1}{2E}\,\frac{\sinh(E/T)}{\cosh(E/T)+\cosh(\mu/T)},
\\
T\sum_{n=-\infty}^{\infty} \frac{-i(2n+1)\pi T+\mu}{[(2n+1)\pi T+i\mu]^2+E^2}
&=& -\frac{1}{2}\frac{\sinh (\mu/T)}{\cosh(E/T)+ \cosh(\mu/T)},
\end{eqnarray}
it is straightforward to write the equations at nonzero temperature. One can check
that the prescription for modifying Eqs.~(\ref{E1-a})--(\ref{E4-a}) at $T \ne 0$ is to
replace
\begin{eqnarray}
\mbox{sign} (\mu_{s}^{(\pm)}) \theta(|\mu_{s}^{(\pm)}| -
E_{ns}^{\mp}) & \to & \frac{\sinh \frac{\mu_{s}^{(\pm)}}{T}}{\cosh
\frac{E_{ns}^{\mp}}{T} + \cosh \frac{\mu_{s}^{(\pm)}}{T}},
\label{finiteT1}
\\
\theta(E_{ns}^{\pm} - |\mu_{s}^{(\mp)}|) & \to & \frac{\sinh
\frac{E_{ns}^{\pm}}{T}}{\cosh \frac{E_{ns}^{\pm}}{T} + \cosh
\frac{\mu_{s}^{(\mp)}}{T}}.
\label{finiteT2}
\end{eqnarray}
This leads to the following set of equations:
\begin{eqnarray}
\Delta_{s}^{(\pm)} &=& A f_1\left(\Delta_{s}^{(\pm)},\mu_{s}^{(\mp)}\right),
\label{Deltas}
\\
\mu_{s}^{(\pm)} &=& \bar{\mu}_{s} +
A f_2\left(\Delta_{s}^{(\mp)},\mu_{s}^{(\pm)}\right)
+2 A f_2\left(\Delta_{s}^{(\pm)},\mu_{s}^{(\mp)}\right)
+2 A f_2\left(\Delta_{-s}^{(\pm)},\mu_{-s}^{(\mp)}\right)
+2 A f_2\left(\Delta_{-s}^{(\mp)},\mu_{-s}^{(\pm)}\right),
\label{mus}
\end{eqnarray}
where $\Delta^{(\pm)}_{s}$ and $\mu_{s}^{(\pm)}$ are given in
Eq.~(\ref{newparam}), and
\begin{eqnarray} \label{f1}
f_1\left(\Delta^{(\pm)}_{s},\mu^{(\mp)}_{s}\right)&=&
\frac{\sinh\left(\frac{\Delta^{(\pm)}_{s}}{T}\right) -
s_{\perp}\sinh\left(\frac{\mu^{(\mp)}_{s}}{T}\right)}
{\cosh\left(\frac{\Delta^{(\pm)}_{s}}{T}\right)+
\cosh\left(\frac{\mu^{(\mp)}_{s}}{T} \right)}+\sum_{n=1}^{\infty}
\frac{2\Delta^{(\pm)}_{s}\sinh\left(\frac{E_{ns}^{\pm}}{T}\right)}
{E_{ns}^{\pm}\left[\cosh\left(\frac{E_{ns}^{\pm} }{T}\right)+
\cosh\left(\frac{\mu^{(\mp)}_{s}}{T} \right)\right]},\\
\label{f2} f_2\left(\Delta^{(\pm)}_{s},\mu^{(\mp)}_{s}\right)&=&
\frac{s_{\perp}\sinh\left(\frac{\Delta^{(\pm)}_{s}}{T}\right) -
\sinh\left(\frac{\mu^{(\mp)}_{s}}{T}\right)}
{\cosh\left(\frac{\Delta^{(\pm)}_{s}}{T}\right)+
\cosh\left(\frac{\mu^{(\mp)}_{s}}{T} \right)}
-\sum_{n=1}^{\infty}\frac{2\sinh\left(\frac{\mu^{(\mp)}_{s}}{T}\right)}
{\cosh\left(\frac{E_{ns}^{\pm}}{T}\right)+ \cosh\left(\frac{\mu^{(\mp)}_{s}}{T}
\right)},
\end{eqnarray}
with $s_{\perp} \equiv \mbox{sign}(eB_{\perp})$ and
$E_{ns}^{\pm}= \sqrt{n \epsilon_{B}^2+ \left(\Delta_{s}^{(\pm)}\right)^2}$.

Let us now show that the QHF and MC order parameters should always coexist
in this dynamics. Suppose that Eqs.~(\ref{Deltas}) and (\ref{mus}) have a
solution with some of the chemical potentials $\mu^{\mp}_{s}$ being nonzero
but the Dirac masses being zero, $\Delta^{(\pm)}_{s} = 0$. Then, the
left hand side of Eq.~(\ref{Deltas}) is equal to zero. On the other hand,
taking into account expression (\ref{f1}) for the function $f_1$, we find that for
$\Delta^{(\pm)}_{s} = 0$ the right hand side of this equation takes the form
\begin{equation}
f_1\left(0,\mu^{(\mp)}_{s}\right) =\frac{-s_{\perp}
\sinh\left(\frac{\mu^{(\mp)}_{s}}{T}\right)} {1 +
\cosh\left(\frac{\mu^{(\mp)}_{s}}{T} \right)} =-s_{\perp}
\tanh\left(\frac{\mu^{(\mp)}_{s}}{2T}\right),
\end{equation}
and {it} could be zero only if {\it all} chemical potentials 
$\mu^{(\mp)}_{s}$
disappear, in contradiction with our assumption. Therefore we conclude that the
QHF and MC order parameters in this dynamics necessarily coexist indeed.
This is perhaps one of the central observations in this study.

Which factors underlie this feature of the graphene dynamics in a
magnetic field? It is the relativistic nature of the free
Hamiltonian $H_0$ in Eq.~(\ref{free-hamiltonian}) and the special
features of the LLs associated with it. To see this, note that
while the triplet Dirac mass $\tilde{\Delta}_{s}$ multiplies the
unit Dirac matrix $I_4$, the triplet chemical potential
$\tilde{\mu}_{s}$ comes with the matrix $\gamma^3\gamma^5\gamma^0$
in the inverse propagator $G^{-1}_{s}$ in
Eq.~(\ref{full-inverse}). Let us trace how these two structures
are connected with each other. The point is that there are terms
with $i\gamma^1\gamma^2\mbox{sign}(eB_{\perp})$ matrix in the
expansion of the propagator $G_{s}$ over LLs [see
Eq.~(\ref{Dsn-new}) in Appendix~\ref{A}]. Taking into account the
definition $\gamma^5 = i\gamma^0\gamma^1\gamma^2\gamma^3$, we have
$i\gamma^1\gamma^2 = \gamma^3\gamma^5\gamma^0$. Then, through the
exchange term $\sim \gamma^0 G_{s} \gamma^0$ in gap equation
(\ref{gap}), the $\tilde{\Delta}_{s}$ term in the inverse
propagator $G^{-1}_{s}$ necessarily induces the term with the
chemical potential $\tilde{\mu}_{s}$. In the same way, the singlet
Dirac mass $\Delta_{s}$ in $G^{-1}_{s}$ is connected with the
singlet chemical potential $\mu_{s}$.

{These arguments are based on the kinematic structure of gap equation (\ref{gap}),
which is the same as that for equation (\ref{SD}) with the Coulomb interaction.
Taking into account the universality of the MC phenomenon, we conclude that
the coexistence of the QHF and MC order parameters is a robust feature of the
QH dynamics in graphene.}

{The necessity of the coexistence of the QHF and MC order
parameters can be} {clearly seen in the case of the
dynamics on the LLL.} {As follows from
Eq.~(\ref{interaction-A1}) in Appendix \ref{A}, the LLL propagator
contains only the combinations $-\mu_{s} + \Delta_{s}{\rm
sign}(eB_{\perp})$ and $\tilde{\mu}_{s}\mbox{sign}(eB_{\perp}) +
\tilde{\Delta}_{s}$. Therefore, in this case, the QHF and MC
parameters not only coexist but they are not independent,
which in turn reflects the fact that the sublattice and valley
degrees of freedom are not independent on the LLL.
In
particular, by using Eqs.~(\ref{singlet_mass}),
(\ref{triplet_mass}), (\ref{singlet_mu}), and
(\ref{triplet_mu}), one can easily check that, because of the
projector 
${\cal{P}_{-}} = 
[1 - i\gamma^{1}\gamma^{2}\mbox{sign}(eB_{\perp})]/2$ 
in the LLL propagator [see Eqs.~(\ref{Dsn-new}) and (\ref{Dn})], the
operators $\Psi^{\dagger}P_{s}\Psi$ and $\bar{\Psi}\gamma^3
\gamma^5 P_{s} \Psi$ ($\Psi^{\dagger}\gamma^3\gamma^5 P_{s}\Psi$
and $\bar{\Psi}P_{s} \Psi$), determining the order parameters
related to $\mu_{s}$ and $\Delta_{s}$ ($\tilde{\mu}_{s}$ and
$\tilde{\Delta}_{s}$), {coincide up to a sign factor}
$\mbox{sign}(eB_{\perp})$.\cite{footnoteLLL}} 
{This fact in particular implies that in order to determine
all the order parameters, it is necessary to analyze the gap 
equation beyond the LLL approximation.}

{The important point, however, is that this special
feature of the LLL takes place only 
on an infinite plane.} In real graphene
samples with boundaries the situation is different: the
QHF and MC parameters on the LLL become
independent.\cite{edge_states,edge_states_long} As is discussed in
Sec.~\ref{6}, this leads to important consequences for the
dynamics of edge states on the LLL.

\section{Dynamics on LLL: $\nu=0$, $\nu=\pm 1$, and
$\nu=\pm 2$ plateaus}
\label{4}

As was already discussed in Introduction, at magnetic fields $B \alt 10~\mbox{T}$,
the plateaus with the filling factors $\nu = \pm 4(n + 1/2)$ are observed in the QH
effect in graphene.\cite{Geim2005Nature,Kim2005Nature} At stronger magnetic
fields, new plateaus, with $\nu=0$ and $\nu=\pm 1$ occur: while
the former arises at $B \gtrsim 10~\mbox{T}$, the latter appear
at $B \gtrsim 20~\mbox{T}$.\cite{Zhang2006,Jiang2007} In this
section, we will describe the dynamics underlying these
new plateaus, and the plateaus $\nu=\pm 2$ corresponding to the
gap between the LLL and the $n = 1$ LL, by using the solutions
of the gap equation presented in the next subsection.
{We will consider positive $\nu$
and $\mu_0$ (the dynamics with negative $\nu$ and
$\mu_0$ is related by electron-hole symmetry and will not be discussed
separately). As will be shown below, there is} {a large number}
{of the solutions corresponding to the same $\mu_0$.
In order to find the {most stable of them},
we compare the free energy density $\Omega$ for
the solutions. The derivation of the expression
for $\Omega$ is presented in Appendix \ref{C}.}

\subsection{Overview of analytic solutions at LLL}
\label{4.0}

{The $\nu=0$, $\nu=\pm 1$ and $\nu=\pm 2$ plateaus are connected
with a process of doping of the LLL, which is described by varying
the electron chemical potential $\mu_0$. Therefore we start our analysis
by reviewing the solutions to the gap equations in the case when $\mu_0$
is much less than the Landau energy scale, i.e.,  $\mu_0\ll \epsilon_{B}$.
At zero temperature the corresponding gap equations are analyzed analytically
in Appendix~\ref{B}. It is concluded there that only the following three
stable solutions are realized:}
\vspace{3mm}

(i) The solution with {\em singlet} Dirac masses for both spin up and spin
down quasiparticles,
\begin{equation}
\begin{split}
& \tilde{\Delta}_{+}=\tilde{\mu}_{+}=0,\qquad \mu_{+}=\bar{\mu}_{+}-A,\qquad
\Delta_{+}=s_{\perp}M,
\\
& \tilde{\Delta}_{-}=\tilde{\mu}_{-}=0,\qquad \mu_{-}=\bar{\mu}_{-}+ A,\qquad
\Delta_{-}=-s_{\perp}M.
\end{split}
\label{i}
\end{equation}
[By definition $M\equiv A/(1-\lambda)$ and
$\lambda\equiv 4A\Lambda/(\sqrt{\pi}\epsilon_{B}^2)$,
see Eq.~(\ref{gap-solution-1}) and its derivation in Appendix~\ref{B}.]
This solution is energetically
most favorable for $0 \le \mu_0 < 2A+Z$.\cite{footnote2}
{It is one of several solutions with nonvanishing singlet Dirac
masses and
we call it the $S1$ solution (here $S$ stands for {\em singlet}). Because
of the opposite signs of both the masses $\Delta_{+}$ and $\Delta_{-}$ and
the chemical
potentials $\mu_{+}$ and $\mu_{-}$, the explicit breakdown of the $U(4)$
symmetry down to $U(2)_{+} \times U(2)_{-}$ by the Zeeman term is
strongly enhanced by the dynamics. Since all triplet order parameters
vanish, the flavor $U(2)_{+} \times U(2)_{-}$ symmetry is intact in the
state described by this solution. As discussed in Subsec.~\ref{4.1} below,
the $S1$ solution corresponds to the $\nu = 0$ plateau.}

(ii) The {\em hybrid} solution with a {\em triplet} Dirac mass for spin up and
a {\em singlet} Dirac mass for spin down quasiparticles,
\begin{equation}
\begin{split}
& \tilde{\Delta}_{+} = M,\qquad \tilde{\mu}_{+}=As_{\perp},\qquad \mu_{+} =
\bar{\mu}_{+} - 4A,\qquad \Delta_{+}=0,
\\
& \tilde{\Delta}_{-}=0,\qquad \tilde{\mu}_{-}=0,\qquad \mu_{-}=\bar{\mu}_{-}-3A,\qquad
\Delta_{-}=-s_{\perp}M.
\end{split}
\label{ii}
\end{equation}
It is most favorable for $2A+Z \le \mu_0 < 6A+Z$. We call it the
$H1$ solution (here $H$ stands for {\em hybrid}, meaning that the
solution is a mixture of the singlet and triplet parameters). In
this case, while the $SU(2)_{+}\subset U(2)_{+}$ symmetry
connected with spin up is spontaneously broken down to $U(1)_{+}$
(whose generator is $\gamma^3\gamma^5 \otimes P_{+}$), the
$SU(2)_{-}\subset U(2)_{-}$ symmetry connected with spin down
remains intact. As will be shown in Subsec.~\ref{4.2}, the $H1$
solution corresponds to the $\nu = 1$ plateau.

(iii) The solution with equal {\em singlet} Dirac masses for both spin up and
spin down quasiparticles
\begin{equation}
\begin{split}
& \tilde{\Delta}_{+}=\tilde{\mu}_{+}=0,\qquad \mu_{+}=\bar{\mu}_{+}-7A,\qquad
\Delta_{+}=-s_{\perp}M,
\\
& \tilde{\Delta}_{-}=\tilde{\mu}_{-}=0,\qquad  \mu_{-}=\bar{\mu}_{-}- 7A,\qquad
\Delta_{-}=-s_{\perp}M.
\end{split}
\label{iii}
\end{equation}
It is most favorable for $\mu_0 > 6A+Z$. We call it the $S2$
solution. (Note that the dynamics in the $n=1$ LL will set an
upper limit for the range where the $S2$ solution is the ground
state, see Sec~\ref{5} below.)  In the state given by the $S2$
solution, the $U(4)$ symmetry is broken down to $U(2)_{+} \times
U(2)_{-}$ only by the Zeeman term. Indeed, the singlet masses and
the dynamical contributions to the chemical potentials are of the
same sign for both spin orientations and thus have no effect on
breaking any symmetry. As will be shown in Subsec.~\ref{4.3}, the
$S2$ solution corresponds to the $\nu = 2$ plateau connected with
the gap between the filled LLL and the empty $n=1$ LL.

\begin{figure}
\begin{center}
\includegraphics[width=.45\textwidth]{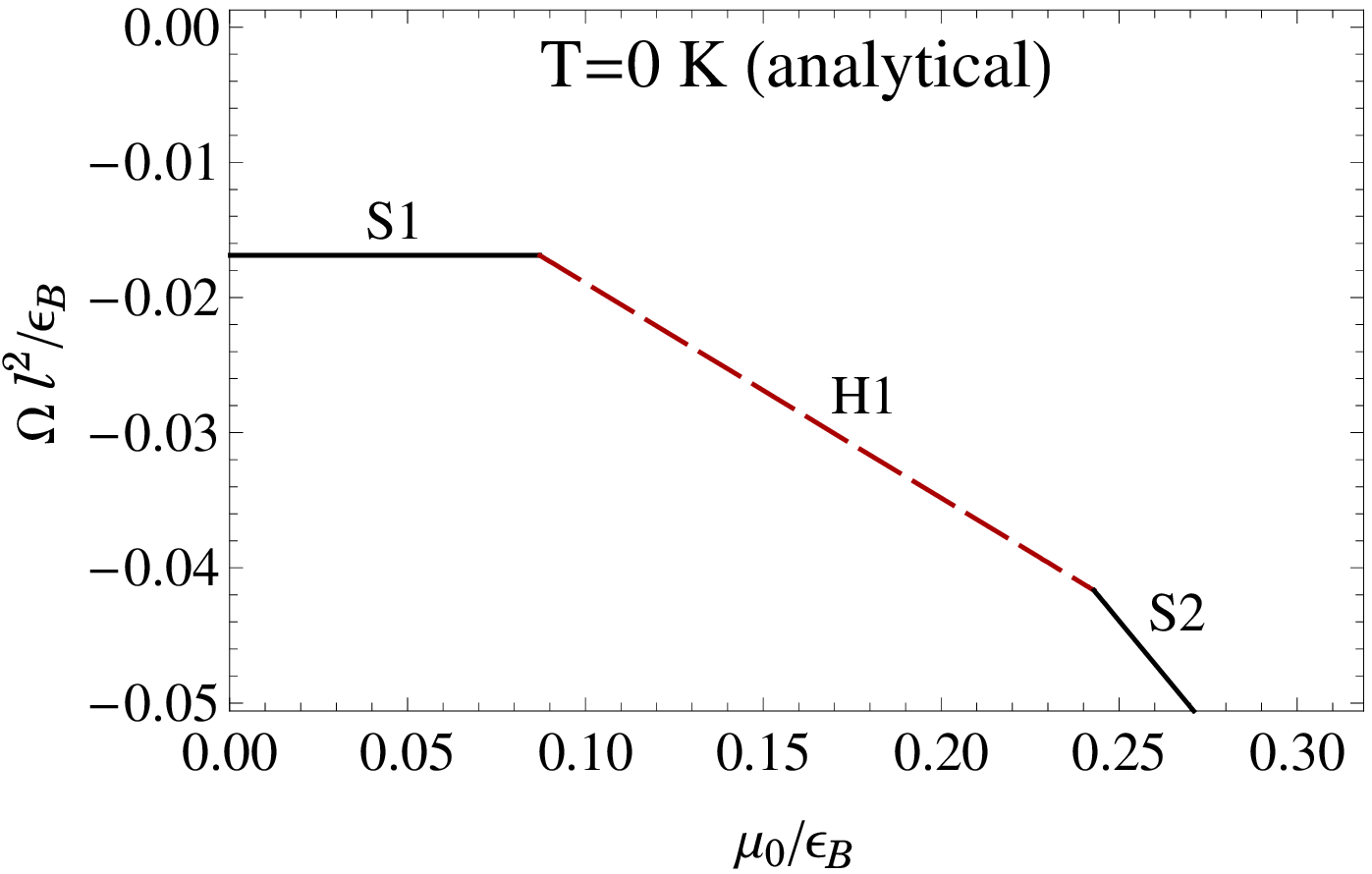}
\hspace{.04\textwidth}
\includegraphics[width=.45\textwidth]{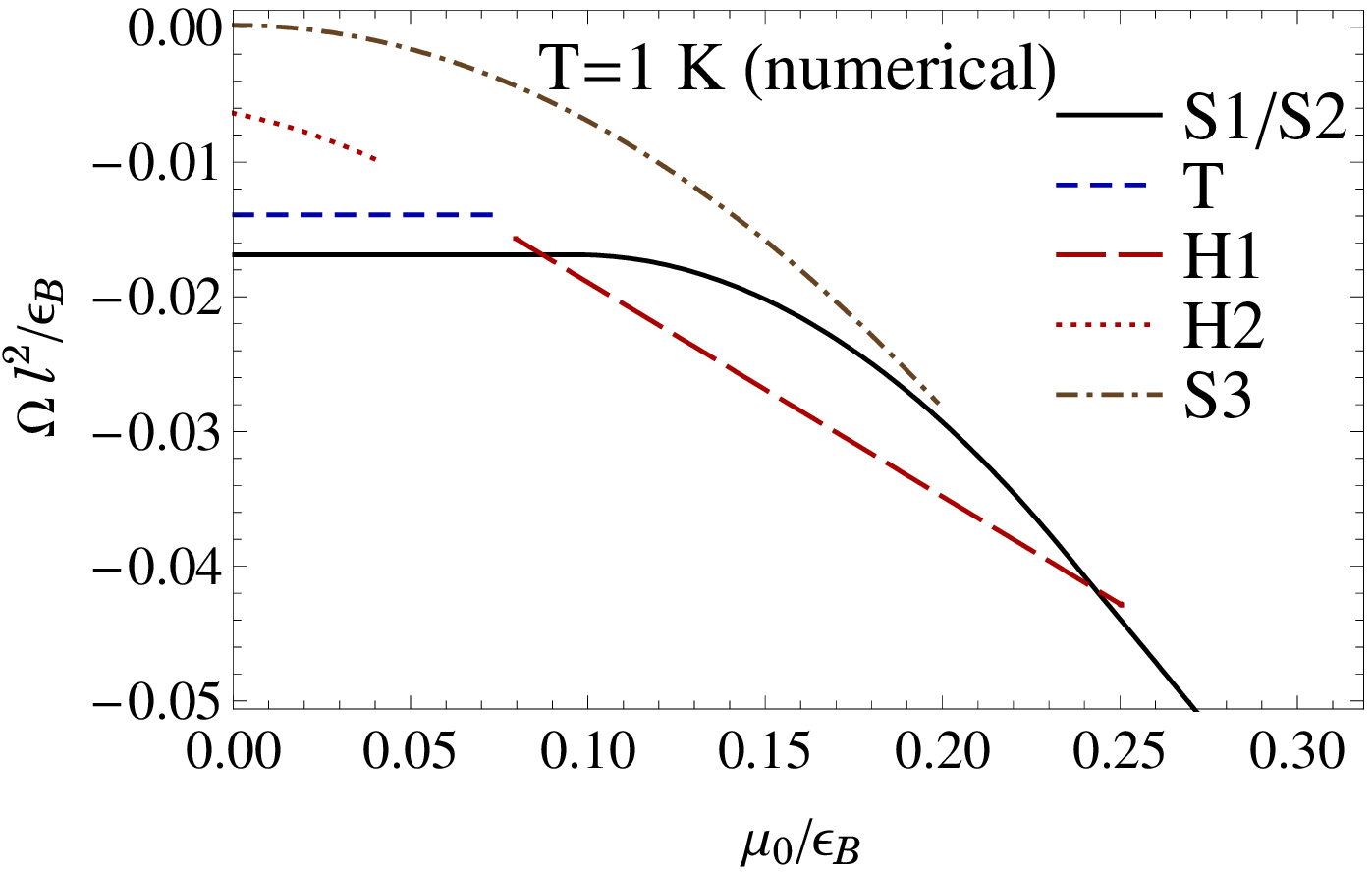}
\caption{Free energy density versus the electron chemical potential $\mu_0$
for several different solutions, found analytically (left panel) and numerically
(right panel) in a range of $\mu_0$ relevant to the dynamics in the lowest
Landau level. The numerical results are shown for a nonzero but small
temperature, $T=1$~K. The values of the electron chemical potential are
given in units of the Landau energy scale $\epsilon_B$, and the free energy
densities are given in units of $\epsilon_B/l^2$, where $l=\sqrt{\hbar c/|eB_\perp|}$
is the magnetic length.}
\label{fig.V.eff.T1}
\end{center}
\end{figure}

The free energy densities for the above three solutions are given
by the following expressions {(see Subsec.~\ref{B6} in
Appendix \ref{B}):}
\begin{eqnarray}
\label{O1}
\Omega &=& -\frac{|eB_{\perp}|}{2\pi \hbar c} \left( M + A + 2Z + h \right),
\quad\mbox{for}\quad 0<\mu_0 < 2A+Z,\\
\label{O2}
\Omega &=& -\frac{|eB_{\perp}|}{2\pi \hbar c} \left( M - A + Z + h + \mu_0 \right),
\quad\mbox{for}\quad 2A+Z < \mu_0 < 6A+Z,\\
\label{O3}
\Omega &=& -\frac{|eB_{\perp}|}{2\pi \hbar c} \left( M - 7A + h + 2\mu_0 \right),
\quad\mbox{for}\quad 6A+Z < \mu_0 ,
\end{eqnarray}
where the parameter $h$ is defined in Eq.~(\ref{h}). We note that
although the parameters of the solutions jump abruptly at the
transition points, $\mu_0=2A+Z$ and $\mu_0=6A+Z$, their free
energy densities match exactly. We conclude, therefore, that first
order phase transitions take place at these values of the electron
chemical potential $\mu_0$.

{The free energy densities in Eqs.~(\ref{O1})-(\ref{O3})
are shown as functions of the chemical potential $\mu_0$ in the
left panel in Fig.~\ref{fig.V.eff.T1}. In order to plot the
results, we took $M = 4.84 \times 10^{-2}\epsilon_{B}$ and $A =
3.90 \times 10^{-2}\epsilon_{B}$ which coincide with the values of
the corresponding dynamical parameters in the numerical analysis.
For comparison, the numerical results at nonzero but sufficiently
small temperature are shown in the right panel of
Fig.~\ref{fig.V.eff.T1}. As we see, the agreement is very good. It
is interesting to note that the singlet-type numerical solution,
given by the solid line, spans both the $S1$ and $S2$ solutions,
as well as the intermediate (metastable) branch connecting them.
In addition to the $S1$, $H1$, and $S2$ solutions, numerical
results for several other (metastable) solutions are shown. The
metastable solutions are discussed in Subsec.~\ref{4.4} below.}

\subsection{Numerical analysis at LLL}
\label{4.numerical}

In this subsection, we give the key details regarding our numerical analysis.

Throughout this paper the default choice of the magnetic field in the numerical
calculations is $B=35~\mbox{T}$. The corresponding Landau energy scale
is $\epsilon_{B}|_{B=35~{\rm T}}\approx 2510~\mbox{K}$. In order to do the
numerical calculations in the model at hand, we use a simple regularization
method that renders the formally defined divergent sum in Eq.~(\ref{f1})
finite. In particular, we redefine the corresponding function as follows:
\begin{equation}
f_1\left(\Delta^{(\pm)}_{s},\mu^{(\mp)}_{s}\right)=
\frac{\sinh\left(\frac{\Delta^{(\pm)}_{s}}{T}\right) -
s_{\perp}\sinh\left(\frac{\mu^{(\mp)}_{s}}{T}\right)}
{\cosh\left(\frac{\Delta^{(\pm)}_{s}}{T}\right)+
\cosh\left(\frac{\mu^{(\mp)}_{s}}{T} \right)}+\sum_{n=1}^{\infty}
\frac{2\Delta^{(\pm)}_{s}\sinh\left(\frac{E_{ns}^{\pm}}{T}\right)
\kappa(\sqrt{n}\,\epsilon_{B},\Lambda)}
{E_{ns}^{\pm}\left[\cosh\left(\frac{E_{ns}^{\pm} }{T}\right)+
\cosh\left(\frac{\mu^{(\mp)}_{s}}{T} \right)\right]},
\label{f1reg}
\end{equation}
where $\kappa(x,\Lambda)$ is a smooth cutoff function defined by
\begin{equation}
\kappa(x,\Lambda)=\frac{\sinh\left({\Lambda}/{\delta\Lambda}\right)}
{\cosh\left({x}/{\delta\Lambda}\right)+\cosh\left({\Lambda}/{\delta\Lambda} \right)}
\label{kappa}
\end{equation}
with $\Lambda=5000$~K and $\delta\Lambda=\Lambda/20=250$~K.
The value of  $\Lambda$ corresponds to an approximate point of the
high-energy cut-off, and the value of $\delta\Lambda$ gives the extent
of the smearing region in either direction from $\Lambda$. (Note
that the energy scale $\Lambda$ is about the same as the energy of the
$n=4$ Landau level at  $B=35~\mbox{T}$.)

{One should emphasize that the specific choice of the cutoff
energy scale $\Lambda$ has little effect on the qualitative as well as
quantitative results of our analysis, provided the dynamical energy scales
$A$ and $M=A/(1-\lambda)$ are kept fixed (see the discussion in the end
of this subsection). Here we assume that the value of the cutoff is
sufficiently large to avoid the reduction of the phase space relevant for
the quasiparticle dynamics at the $n=0$ and $n=1$ LLs.}

Because of the cutoff function $\kappa(x,\Lambda)$ the sum over $n$ on the
right hand side of Eq.~(\ref{f1reg}) is rapidly convergent. In the numerical
calculations, therefore, a sufficiently good accuracy may be achieved by keeping
a finite number of terms in the sum. The optimum choice for the maximum
value of index $n$ is $n_{\rm max}=\left[14\Lambda^2/\epsilon_{B}^2\right]$,
where the square brackets mean the integer number nearest to the result in
the brackets. This choice is large enough to insure a high precision and, at
the same time, it is small enough to make the calculation fast.

While the $f_2$-function in Eq.~(\ref{f2}) is finite, for consistency we redefine
it in the same way as function $f_1$ by smoothly cutting off the contributions
of large-$n$ LLs,
\begin{equation}
f_2\left(\Delta^{(\pm)}_{s},\mu^{(\mp)}_{s}\right)=
\frac{s_{\perp}\sinh\left(\frac{\Delta^{(\pm)}_{s}}{T}\right) -
\sinh\left(\frac{\mu^{(\mp)}_{s}}{T}\right)}
{\cosh\left(\frac{\Delta^{(\pm)}_{s}}{T}\right)+
\cosh\left(\frac{\mu^{(\mp)}_{s}}{T} \right)}
-\sum_{n=1}^{\infty}\frac{2\sinh\left(\frac{\mu^{(\mp)}_{s}}{T}\right)
\kappa(\sqrt{n}\,\epsilon_{B},\Lambda)}
{\cosh\left(\frac{E_{ns}^{\pm}}{T}\right)+ \cosh\left(\frac{\mu^{(\mp)}_{s}}{T}
\right)},
\label{f2reg}
\end{equation}
where $\kappa(x,\Lambda)$ is defined in Eq.~(\ref{kappa}). The numerical
result for the sum in $f_2$ is also approximated by dropping the terms with
$n>n_{\rm max}$ where $n_{\rm max}$ is given above.

By analyzing the solutions to Eqs.~(\ref{Deltas}) and (\ref{mus}) at very
low temperatures, we reproduce all the analytic solutions derived in
Appendix~\ref{B}. For the choice of the magnetic field $B=35~\mbox{T}$
the values of the two dynamical energy parameters $A$ and $M$ are given
by
\begin{equation}
A \approx  98~\mbox{K},\qquad
M \approx  122~\mbox{K}.
\end{equation}
As is easy to check, these correspond to the dimensionless coupling
$\lambda\approx 0.196$. Here one should keep in mind that the
smooth-cutoff regularization used in our numerical calculations is not
the same as in the analytical calculations [see, for example,
Eq.~(\ref{integral-form}) in Appendix~\ref{B}.] Despite this difference,
all analytical results agree very well even quantitatively with the
corresponding numerical ones when expressed in terms of $A$
and $M$ parameters.

\subsection{Plateau $\nu = 0$}
\label{4.1}

The plateau $\nu = 0$ is connected with a range of electron
chemical potentials in the vicinity of the Dirac neutral point
with $\mu_0 = 0$. In this case the $S1$ solution with singlet
Dirac masses of opposite sign for spin up and spin down
quasiparticles, see Eq.~(\ref{i}), is most favorable energetically
and therefore is the ground state solution, provided $\mu_0
<2A+Z$ {(other solutions related to the Dirac neutral point are
discussed in Subsec. \ref{4.4} below)}.

{From} dispersion relation (\ref{LLLenergylevels}), we find that
while $\omega_{+} =-\mu_0+Z+M+A$ is positive for spin up
states, $\omega_{-}= -\mu_0-Z-M-A$ is negative for spin down
states, i.e., the LLL is half filled (the energy spectrum in this
solution is $\sigma$ independent). Therefore there is a nonzero
spin gap $\Delta{E}_{0} = \omega_{+} - \omega_{-}$ associated with
the $\nu = 0$ plateau. The value of this gap is $\Delta{E}_{0} =
2(Z+A)+2M$.

\begin{figure}
\begin{center}
\includegraphics[width=.45\textwidth]{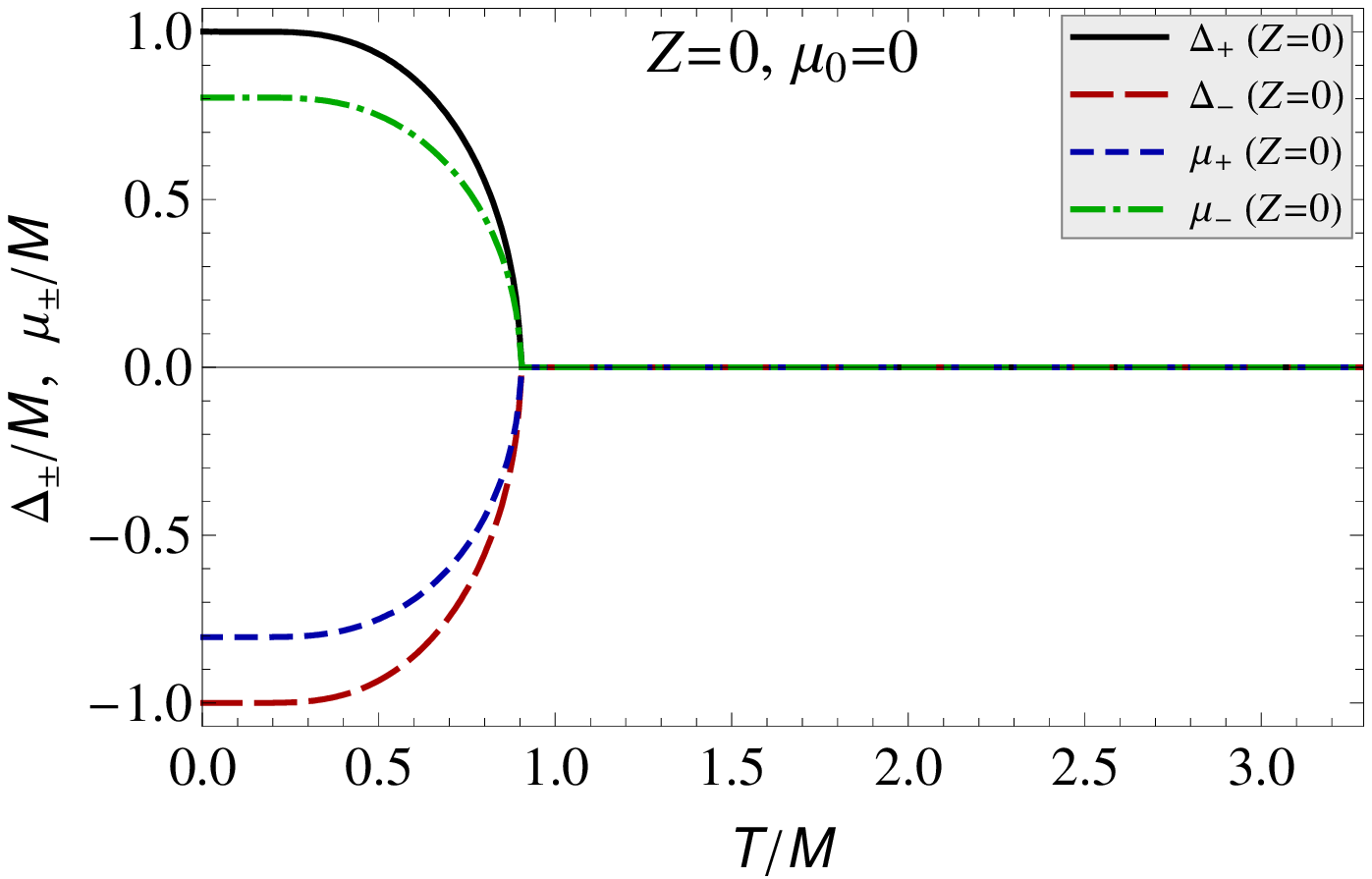}
\hspace{.04\textwidth}
\includegraphics[width=.45\textwidth]{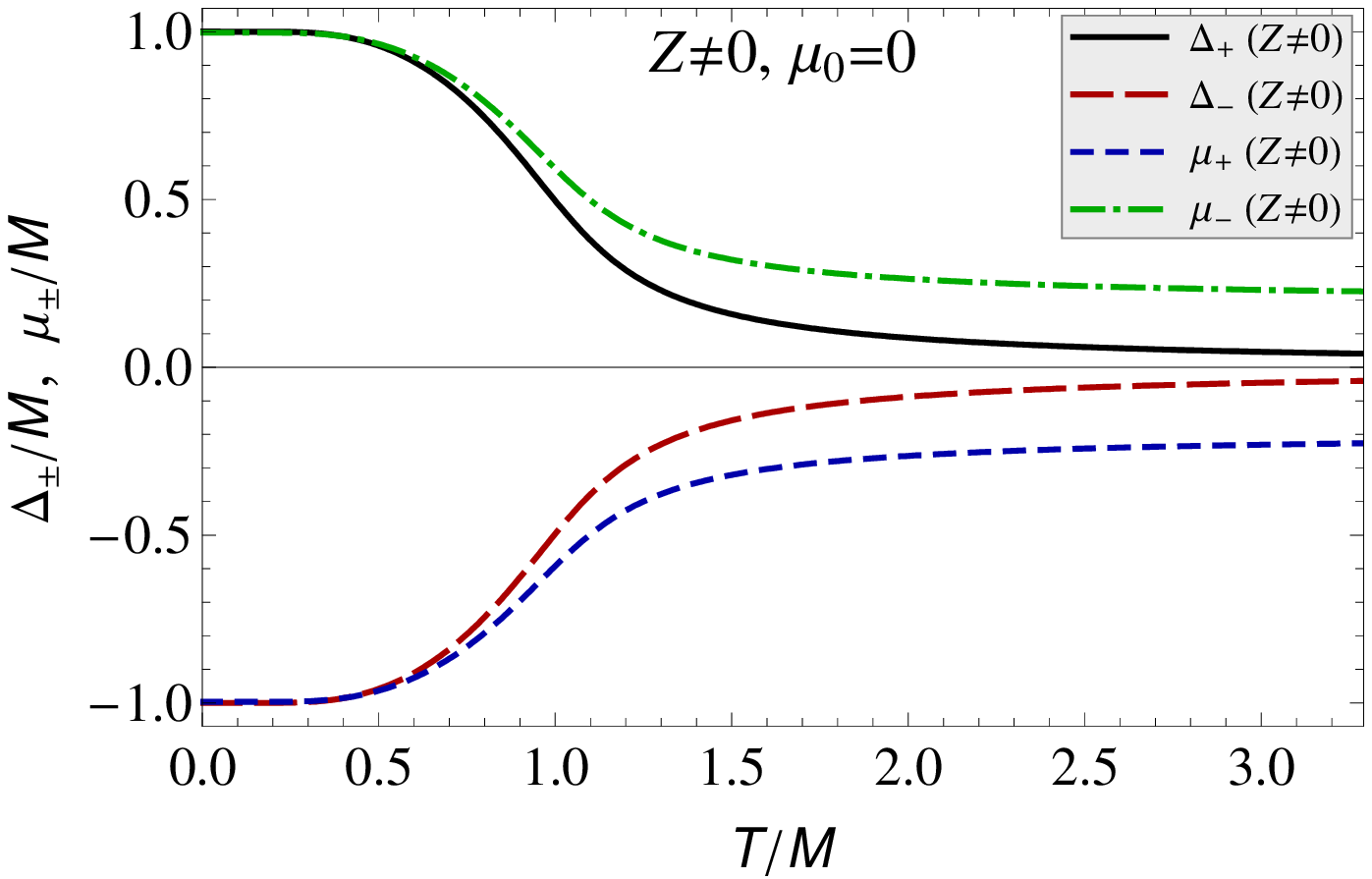}
\caption{Temperature dependence of the nontrivial order parameters
in the $\nu=0$ QH state, described by the $S1$ solution. The
results in a model with a vanishing Zeeman energy ($Z=0$) are
shown in the left panel, and the results in a realistic model with
a nonzero Zeeman energy ($Z\neq 0$) are shown in the right panel.
Note that $\tilde{\mu}_{\pm}=\tilde{\Delta}_{\pm}=0$ in both
cases. The values of the temperature and the order parameters are
given in units of the dynamical scale $M$.}
\label{fig.order_pars_mu0_vs_T}
\end{center}
\end{figure}

While no exact symmetry is broken in the state described by the
$S1$ solution, the explicit spin symmetry breaking by the Zeeman
term $Z$ is strongly enhanced by the dynamical contribution $M+A$.
In this case, it is appropriate to talk about the dynamical
symmetry breaking of the approximate spin symmetry. This is also
evident from studying the temperature dependence of the MC and QHF
order parameters in Fig.~\ref{fig.order_pars_mu0_vs_T}. In the two
panels, we compare the results in the models with the exact (left
panel) and approximate (right panel) spin symmetry. In the first
case we take $Z=0$ and see that the spontaneous spin-symmetry
breaking occurs at low temperatures. The symmetry is restored at
about $T\approx 0.9 M$ in a typical second order phase transition
(recall that we work in the mean-field approximation).
{In the second case,} a nonzero Zeeman energy term
($Z\approx 23.51~\mbox{K}$ at $B=35~\mbox{T}$) breaks the spin
symmetry explicitly and its restoration is impossible even at very
high temperatures. However, even in this latter case, there is a
well pronounced crossover (around $T\approx 0.9 M$) between the
regimes of low and high temperatures, which can be quantified by
the relative strength of the bare Zeeman and dynamical
contributions.

\begin{figure}
\begin{center}
\includegraphics[width=.45\textwidth]{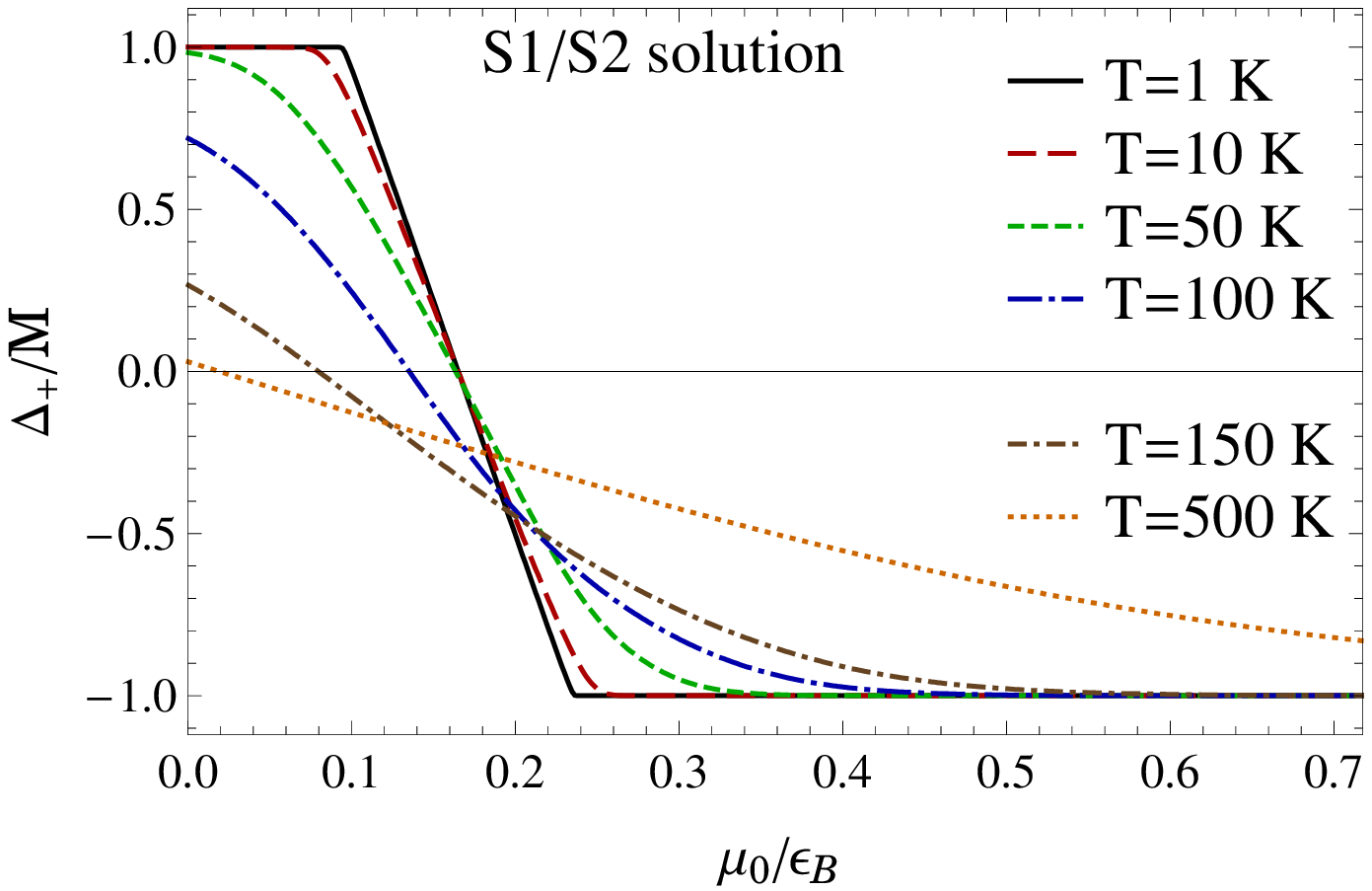}
\hspace{.04\textwidth}
\includegraphics[width=.43\textwidth]{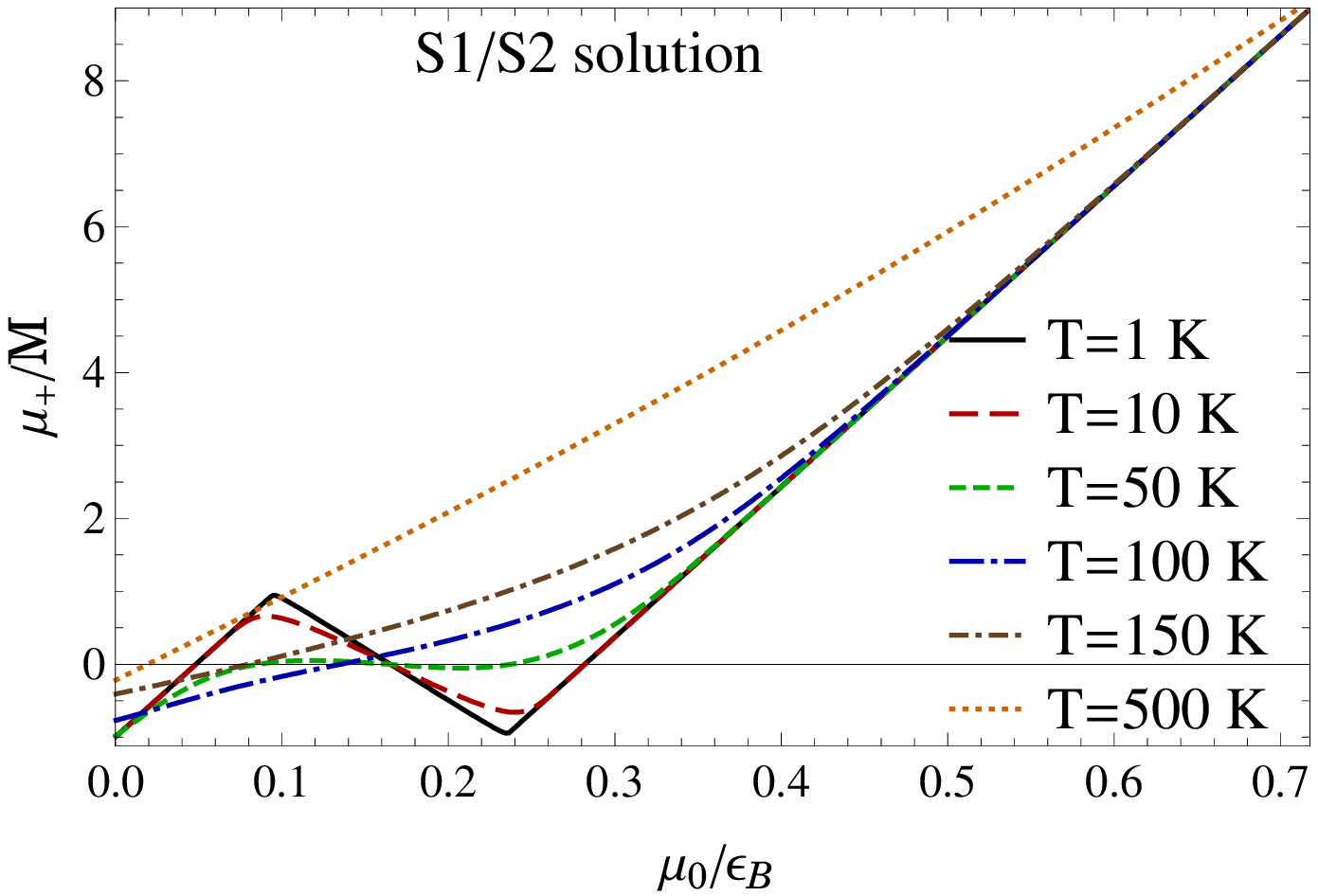}\\
\includegraphics[width=.45\textwidth]{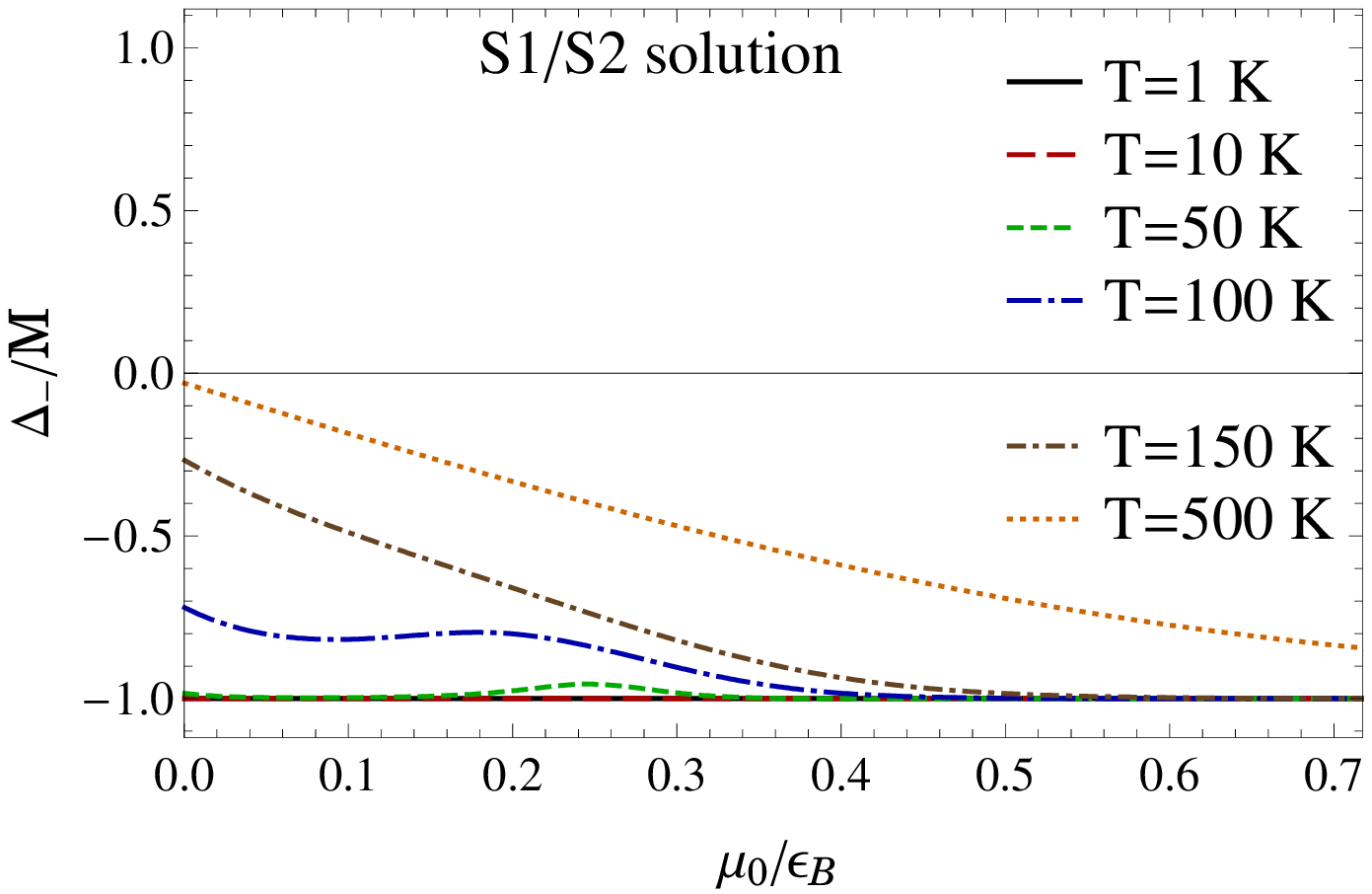}
\hspace{.04\textwidth}
\includegraphics[width=.43\textwidth]{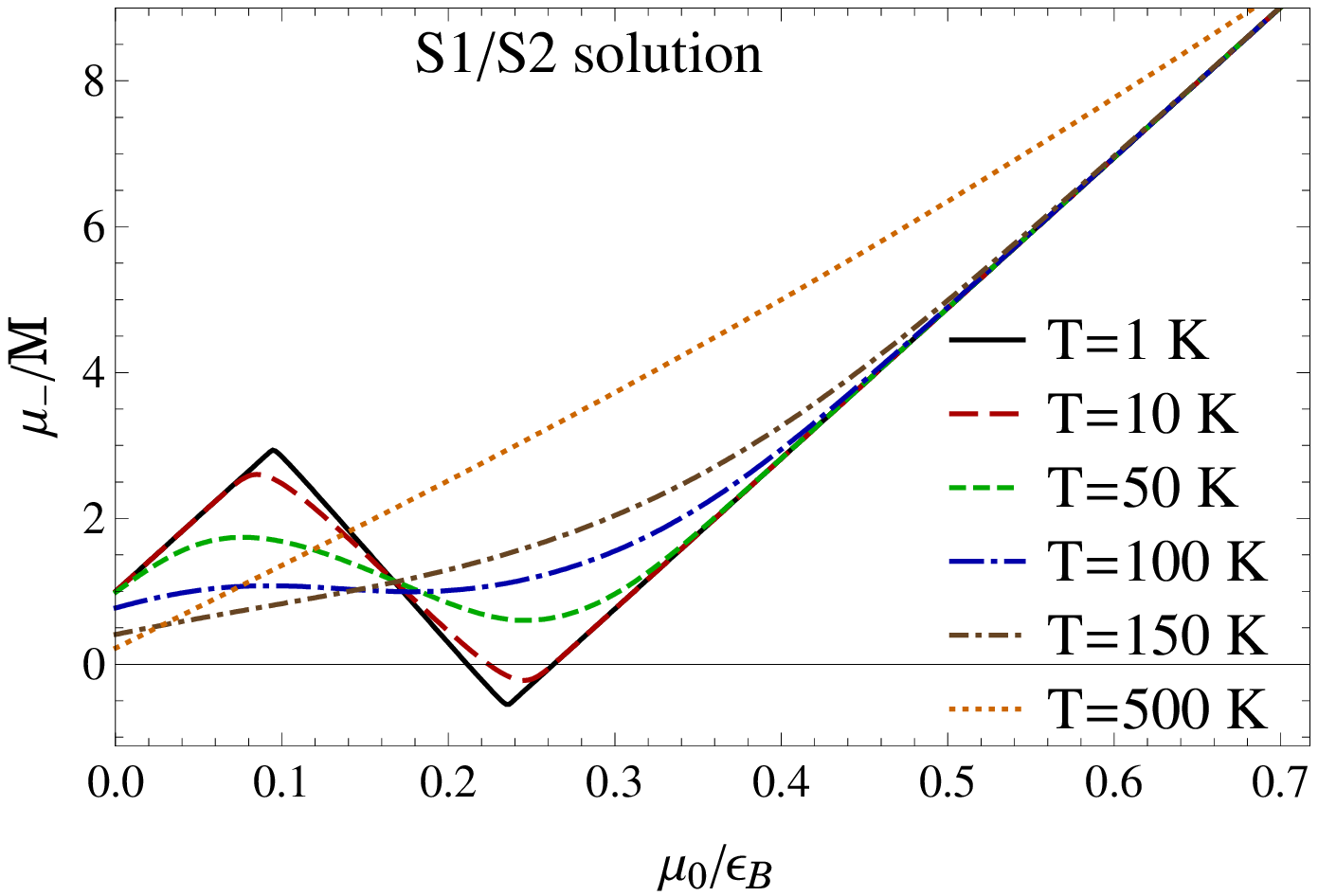}
\caption{Order parameters for the singlet solution $S1/S2$ as functions
of the electron chemical potential $\mu_0$ for several different values
of temperature.}
\label{fig.S_vs_mu_T}
\end{center}
\end{figure}

The order parameters for the solution $S1$ versus the electron chemical potential
$\mu_0$ are shown in Fig.~\ref{fig.S_vs_mu_T} for several different values of the
temperature. At $T=0$ this solution is the ground state for
$\mu_0\lesssim 0.09  \epsilon_{B}$. At sufficiently low temperature, the main
qualitative feature of this solution is that the singlet Dirac masses for spin-up
and spin-down quasiparticles have opposite signs, $\Delta_{+}= -\Delta_{-}$.
This defines the configuration of the MC order parameters that is formally
invariant under the time reversal symmetry. (Of course, the time reversal
symmetry is still explicitly broken by the external magnetic field.) As the
temperature increases, the approximate relation $\Delta_{+}\approx -\Delta_{-}$
may hold at $\mu_0\approx 0$, but deviations from such a relation grow
with increasing $\mu_0$.

It should be emphasized that the solution $S1$ is continuously connected with
the solution $S2$ responsible for the $\nu=2$ QH plateau, see Subsec.~\ref{4.3}
below. At low temperatures, the intermediate branch between the $S1$ and $S2$
solutions is metastable. At high temperatures, however, it becomes stable and
the qualitative difference between the two solutions disappears.

{The conclusion that the $\nu=0$ state is related to the
spin gap agrees with the scenario in Ref.~\onlinecite{Abanin2006PRL} and
the experiments reported in Refs.~\onlinecite{Jiang2007,Abanin2007PRL}.
The fact established in the present paper that both $\mu_3$ and the singlet
Dirac mass $\Delta_3$ contribute to the gap $\Delta{E_0}$ is
noticeable. As was already pointed out in Sec.~\ref{3}, unlike the
case of an infinite plane, in graphene samples with boundaries,
the parameters $\mu_3$ and $\Delta_3$ are independent on the LLL.
As will be discussed in Sec.~\ref{6}, this fact could have
important consequences for the dynamics of edge states.}

In conclusion, the following comment is
in order. As one can see in the right panel in Fig.\ref{fig.V.eff.T1}, besides
the $S1$ solution, there is another, triplet ($T$), solution around the Dirac
neutral point. In the $T$ solution, given in Eq.(\ref{1stgroup}) in Appendix \ref{B},
both spin up and spin down quasiparticle states have a triplet Dirac mass.
Calculating the difference of the free energy densities for these two
solutions, one finds that $\delta\Omega=\Omega_{S1}-\Omega_{T}=- Z|eB|/\pi\hbar c$.
Therefore, it is the Zeeman term which makes the $S1$ solution more favorable:
without it, the $S1$ and $T$ solutions would correspond to two degenerate ground states.
{It would be interesting to figure out the role of the small
on-site repulsion interaction terms \cite{Alicea2006PRB,LS,Herbut2006,Aleiner2007} mentioned in
Subsec. \ref{2.2} in choosing the genuine ground state in the
present dynamics.}

\subsection{Plateau $\nu = 1$}
\label{4.2}

{As was pointed out in Subsec.~\ref{4.0}}, for larger
$\mu_0$ the hybrid $H1$ solution (\ref{ii}), with a triplet Dirac
mass for spin up quasiparticles and a singlet Dirac mass for spin
down quasiparticles, is most favorable. It is the ground state for
$2A+Z < \mu_0 < 6A+Z$. As one can easily check by using
Eq.~(\ref{LLLenergylevels}), while now $\omega_{+}^{(+)} > 0$, the
energies $\omega_{+}^{(-)}$ and $\omega_{-}^{(+)}=
\omega_{-}^{(-)}$ are negative. Consequently, the LLL is now
three-quarter filled and, therefore, the gap $\Delta{E}_{1} =
\omega_{+}^{(+)} - \omega_{+}^{(-)} = 2(M + A)$ corresponds to the
$\nu = 1$ plateau. Notably, the Zeeman term does not enter the
value of the gap. Unlike the $\nu = 0$ state, therefore, the gap
in the $\nu=1$ state is directly related to the spontaneous
breakdown of the flavor symmetry $SU(2)_{+}$.

The last point regarding the nature of the ground state described
by the $H1$ solution has important consequences for the physical
properties of the $\nu=1$ QH state. {Since the coupling
constant $G_{\rm int}$ in the present model is proportional to
$1/\epsilon_B$ (see Subsec.~\ref{2.1}), Eq.~(\ref{Apar}) implies
that the dynamical parameters $A$ and $M$, and therefore the gap
$\Delta{E}_{1}$, scale with the magnetic field as
$\sqrt{|eB_{\perp}|}$. This fact agrees with the dependence of the
activation energy in the $\nu=1$ state observed in
Ref.~\onlinecite{Jiang2007}}.

{The critical temperature at which the $SU(2)_+$ symmetry is restored,
i.e., when the triplet parameters $\tilde{\mu}_+$ and $\tilde{\Delta}_+$ vanish,
is $T_c \simeq 0.9 M \simeq 110 K$. The restoration is described by a
conventional second order phase} {transition}.

\begin{figure}
\begin{center}
\includegraphics[width=.45\textwidth]{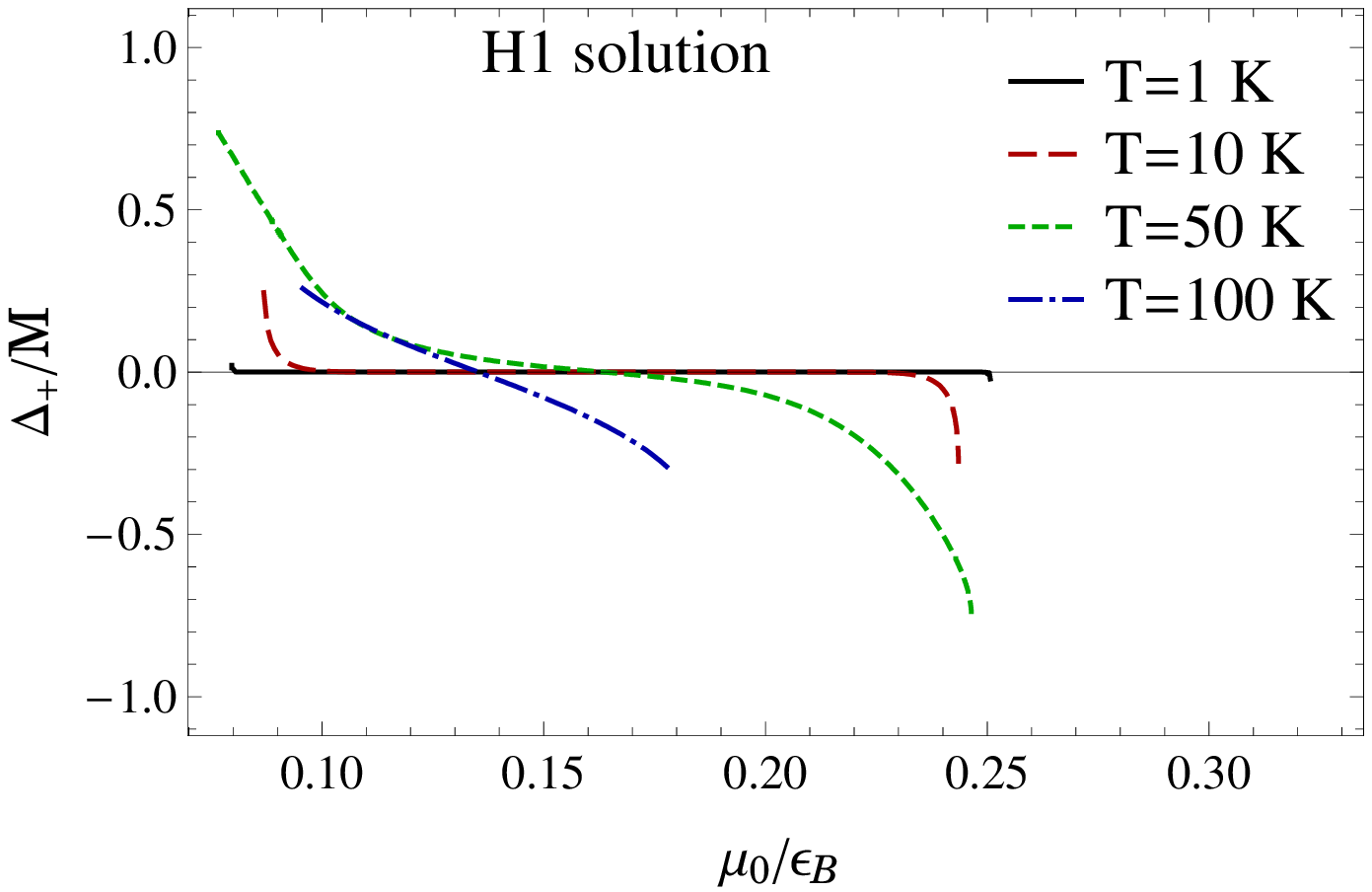}
\hspace{.04\textwidth}
\includegraphics[width=.45\textwidth]{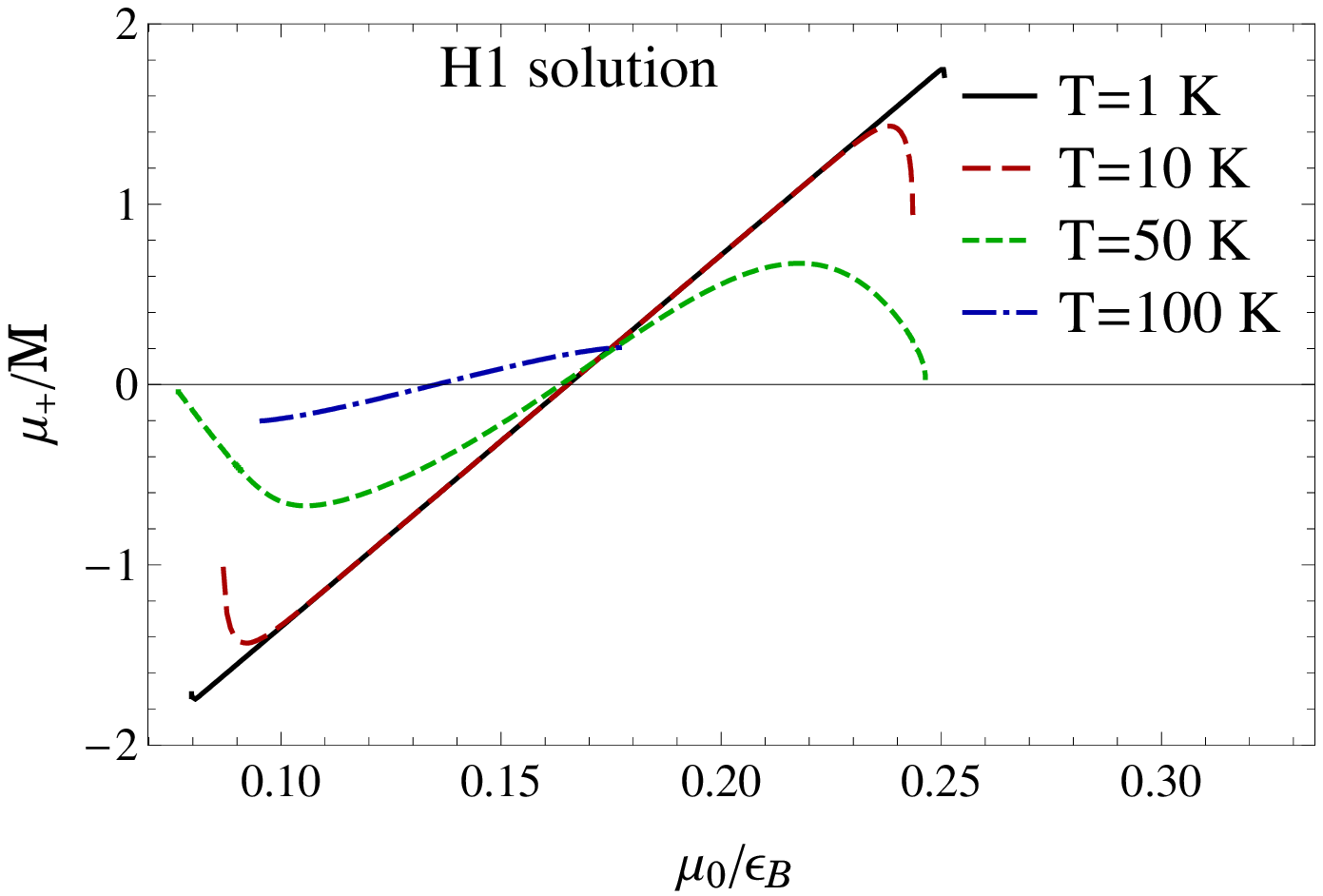}\\
\includegraphics[width=.45\textwidth]{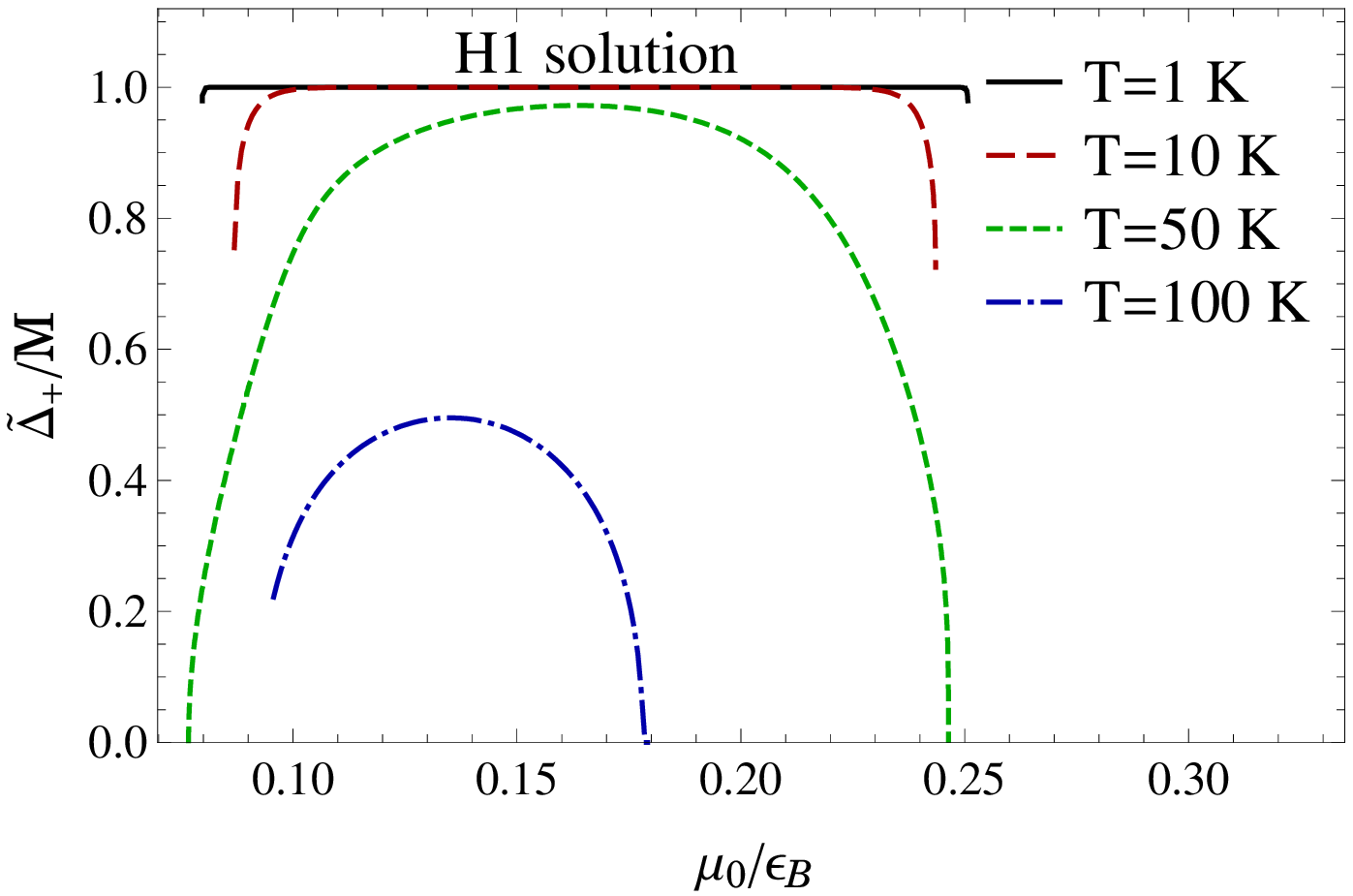}
\hspace{.04\textwidth}
\includegraphics[width=.45\textwidth]{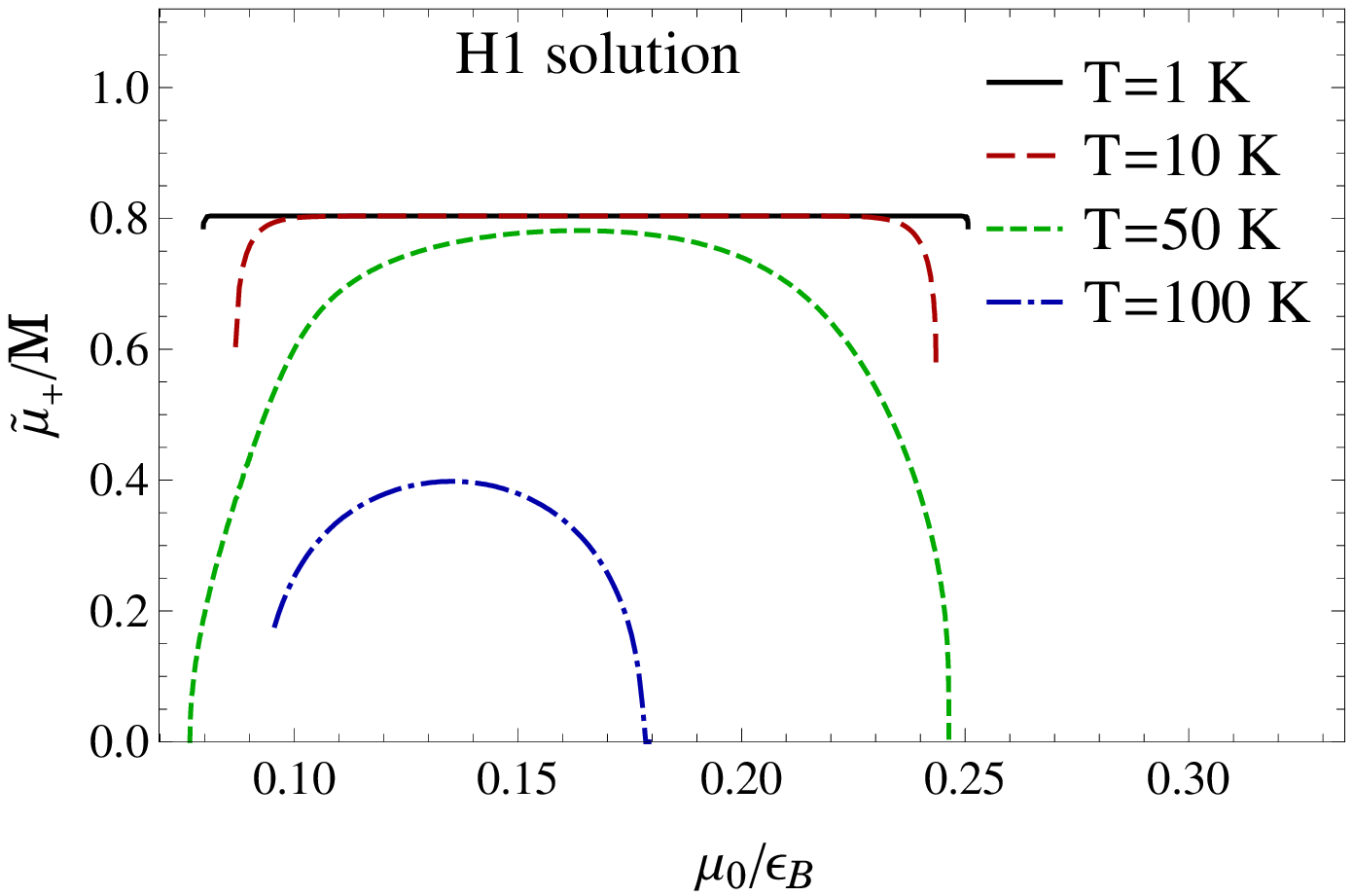}\\
\includegraphics[width=.45\textwidth]{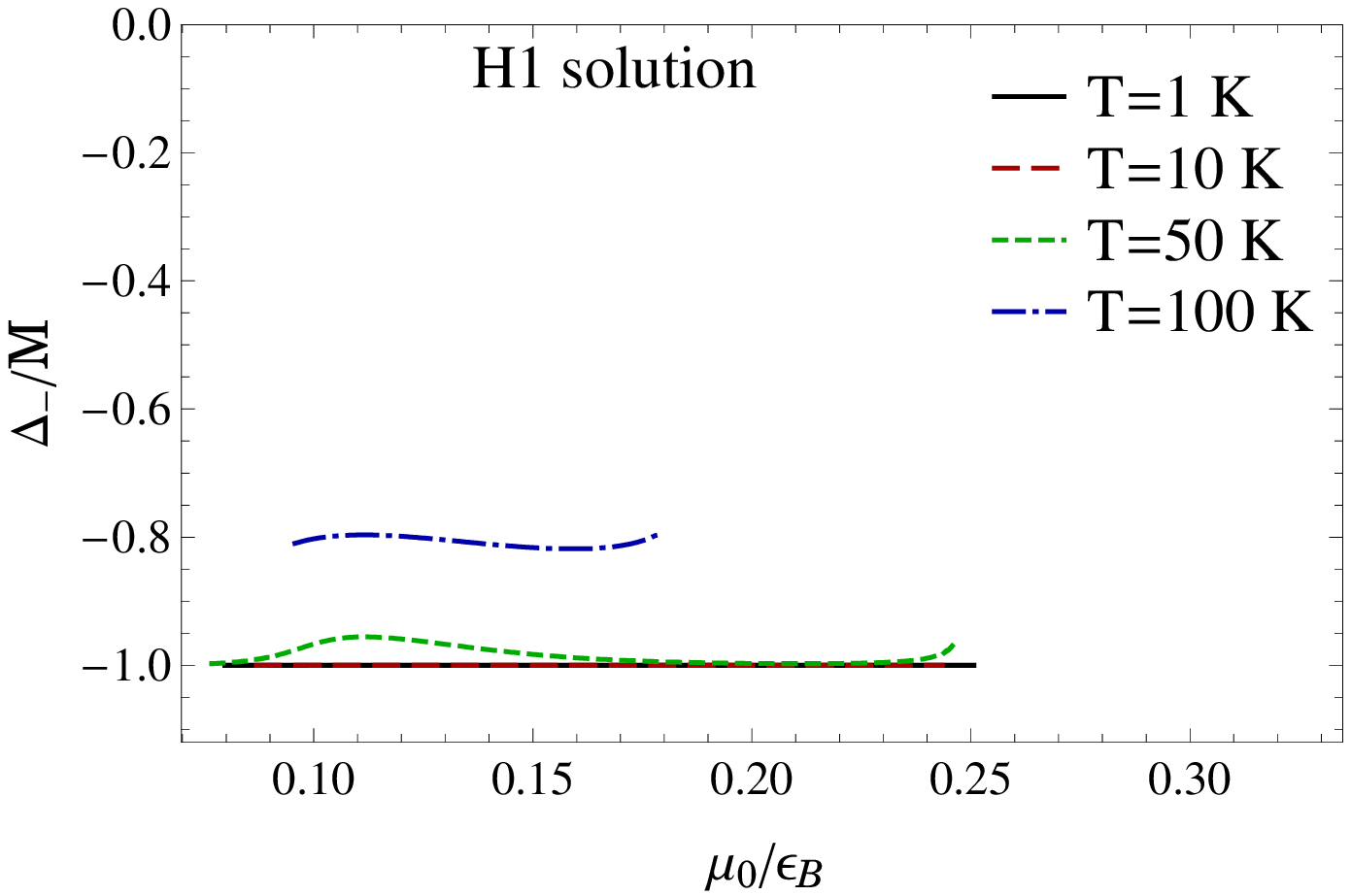}
\hspace{.04\textwidth}
\includegraphics[width=.45\textwidth]{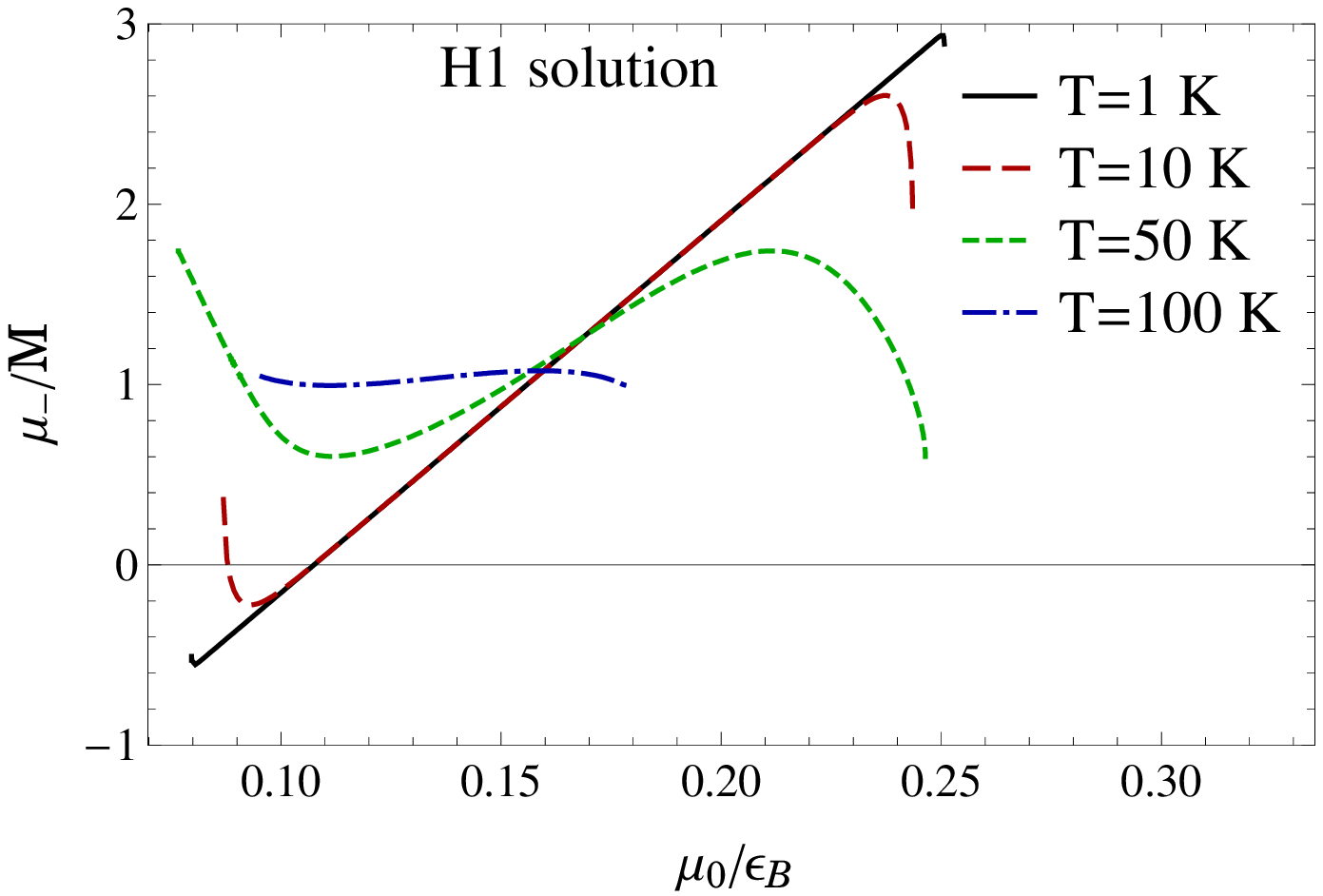}
\caption{Order parameters for the solution $H1$ as functions of the electron
chemical potential $\mu_0$ for several different values of temperature.}
\label{fig.H1_vs_mu_T}
\end{center}
\end{figure}

{The temperature dependence of the hybrid $H1$ solution is rather
interesting too. This is summarized in Fig.~\ref{fig.H1_vs_mu_T} where the
nontrivial order parameters and chemical potentials are shown for several
values of the temperature in the range from $1~$K to $100~$K.
One of the most spectacular features of this
dependence is a revival
of the singlet mass $\Delta_+$ at finite temperature shown in
Fig.~\ref{fig.H1_vs_mu_T} (recall that it vanishes at zero temperature).
This phenomenon is intimately connected with the general conclusion
in Sec.~\ref{3} that at a $\it fixed$ value of spin $s$ and {\it any} value of
temperature, there are no nontrivial solutions of the gap equation with the
both masses $\Delta_{s}$ and $\tilde{\Delta}_{s}$ being zero. Indeed,
{at $T > T_c$, when
the triplet mass $\tilde{\Delta}_+$ vanishes, the absence of
the $\Delta_+$  would contradict this conclusion}
(note that as Fig.~\ref{fig.H1_vs_mu_T} shows, the revival of this mass
occurs even at subcritical $T$). Note also that in the case of
spin down quasiparticles,
the triplet parameters $\tilde{\mu}_-$ and $\tilde{\Delta}_{-}$ are identically
zero but the singlet mass $\Delta_{-}$ remains nonzero at all temperatures.}

{These results are obtained in the mean field
approximation and for the Hamiltonian $H_{\rm tot}$ (\ref{tot}),
which is symmetric under the $U(2)_{+} \times U(2)_{-}$. However,
as was already pointed out in Sec.~\ref{2.2} above, this symmetry is not
exact for the Hamiltonian on the graphene lattice. In that case,
it is replaced by {$U(1)_{+} \times Z_{2+} \times U(1)_{-}
\times Z_{2-}$, where the elements of the
discrete group $Z_{2\pm}$ are
$\gamma^5 \otimes P_{\pm} + I_{4}\otimes P_{\mp}$ and the unit matrix.
It is important
that unlike a spontaneous breakdown of continuous symmetries, a
spontaneous breakdown of the discrete symmetry $Z_{2\pm}$, with
the order parameters $\langle {\bar{\Psi} P_{\pm}\Psi} \rangle$ and
$\langle\Psi^{\dagger}\gamma^3\gamma^{5}P_{\pm}\Psi\rangle$, is not
forbidden by the Mermin-Wagner theorem at finite temperatures in a
planar system.\cite{MW} This point strongly suggests that there
exists a genuine phase transition in temperature related to the
$\nu = 1$ state in graphene.}

\subsection{Plateau $\nu = 2$}
\label{4.3}

At zero temperature, the $S2$ solution (\ref{iii}) with equal
singlet Dirac masses for spin up and spin down states is most
favorable for $\mu_0 > 6A+Z$. It is easy to check from
Eq.~(\ref{LLLenergylevels}) that both $\omega_{+}$ and
$\omega_{-}$ are negative in this case, i.e., the LLL is
completely filled. This solution corresponds to the $\nu = 2$
plateau when the value of the electron chemical potential is in
the gap between the LLL and the $n=1$ LL.

The nonzero temperature results for the order parameters of the solution $S2$
versus the electron chemical potential $\mu_0$ are shown in Fig.~\ref{fig.S_vs_mu_T}.
At $T=0$ this solution is the ground state for $\mu_0\gtrsim 0.24 \epsilon_{B}$.
As we see, even at high temperatures, the MC order parameters
satisfy the same approximate relation, $\Delta_{+}\approx \Delta_{-}$. Such a
configuration breaks neither spin nor sublattice-valley symmetry of graphene.

\subsection{Metastable solutions on LLL}
\label{4.4}

As was already pointed above,
in addition to the three {stable} solutions $S1$, $H1$, and $S2$,
describing the $\nu=0$,  $\nu=\pm 1$,  and $\nu=\pm 2$ QH plateaus, the
numerical analysis of the gap equations reveals other, metastable, solutions.

One of such solutions is the $T$ solution with nonzero {\em triplet} Dirac masses
{for both spin up and spin down quasiparticles.
In the model of graphene used in this
paper, the explicit analytical form of this solution is given in Eq.~(\ref{1stgroup})
in Appendix~\ref{B}. Note that because there is a contribution of the bare
Zeeman term $Z\propto eB$ in the gap $\Delta{E_1}$ for this solution, the corresponding
activation energy in the $\nu = 1$ state scales with $eB$ differently from
the $\sqrt{|eB|}$ law in the hybrid $H1$ solution.}

In addition to the triplet solution, there exist also metastable hybrid ($H2$) and
singlet ($S3$) solutions. Their free energy densities are shown in
Fig.~\ref{fig.V.eff.T1} together with the energy densities of the other solutions.
As seen, neither
{the $H2$ solution nor the $S3$ one} can have sufficiently low free energy
density to become the genuine ground state.

The following remark is in order.
Unlike all the other solutions, the solutions $H2$ and $S3$
cannot be found analytically at $T=0$, see Appendix~\ref{B}. By making use of the
numerical analysis, we find that these two extra solutions are such that $\mu_{s}^{(\pm)}
\approx \pm E_{0s}^{\mp}$. At exactly zero temperature, it is problematic to get such
solutions analytically because Eqs.~(\ref{E1-a})--(\ref{E4-a}) contain undetermined
values of the step functions, e.g., $\theta(|\mu_{s}^{(\pm)}|-E_{0s}^{\mp})$. In
contrast, at a nonzero temperature, the step functions are replaced by smooth
expressions, see Eqs.~(\ref{finiteT1}) and (\ref{finiteT2}), and numerical solutions
with $\mu_{s}^{(\pm)}\approx \pm E_{0s}^{\mp}$ are easily found. The order
parameters for the solutions $H2$ and $S3$ are shown in Fig.~\ref{fig:H2&S3}.

\begin{figure}
\begin{center}
\includegraphics[width=.45\textwidth]{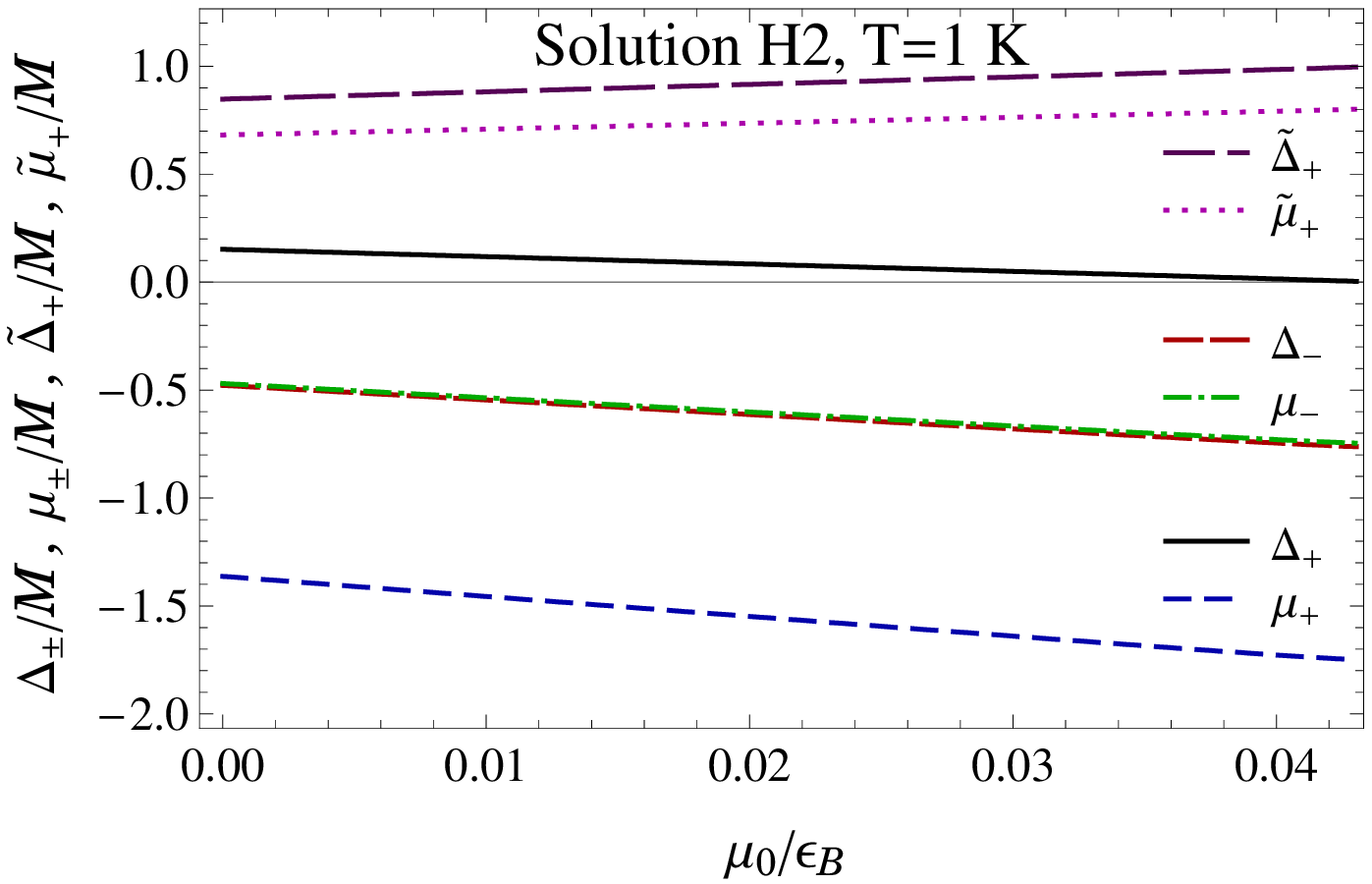}
\hspace{.04\textwidth}
\includegraphics[width=.45\textwidth]{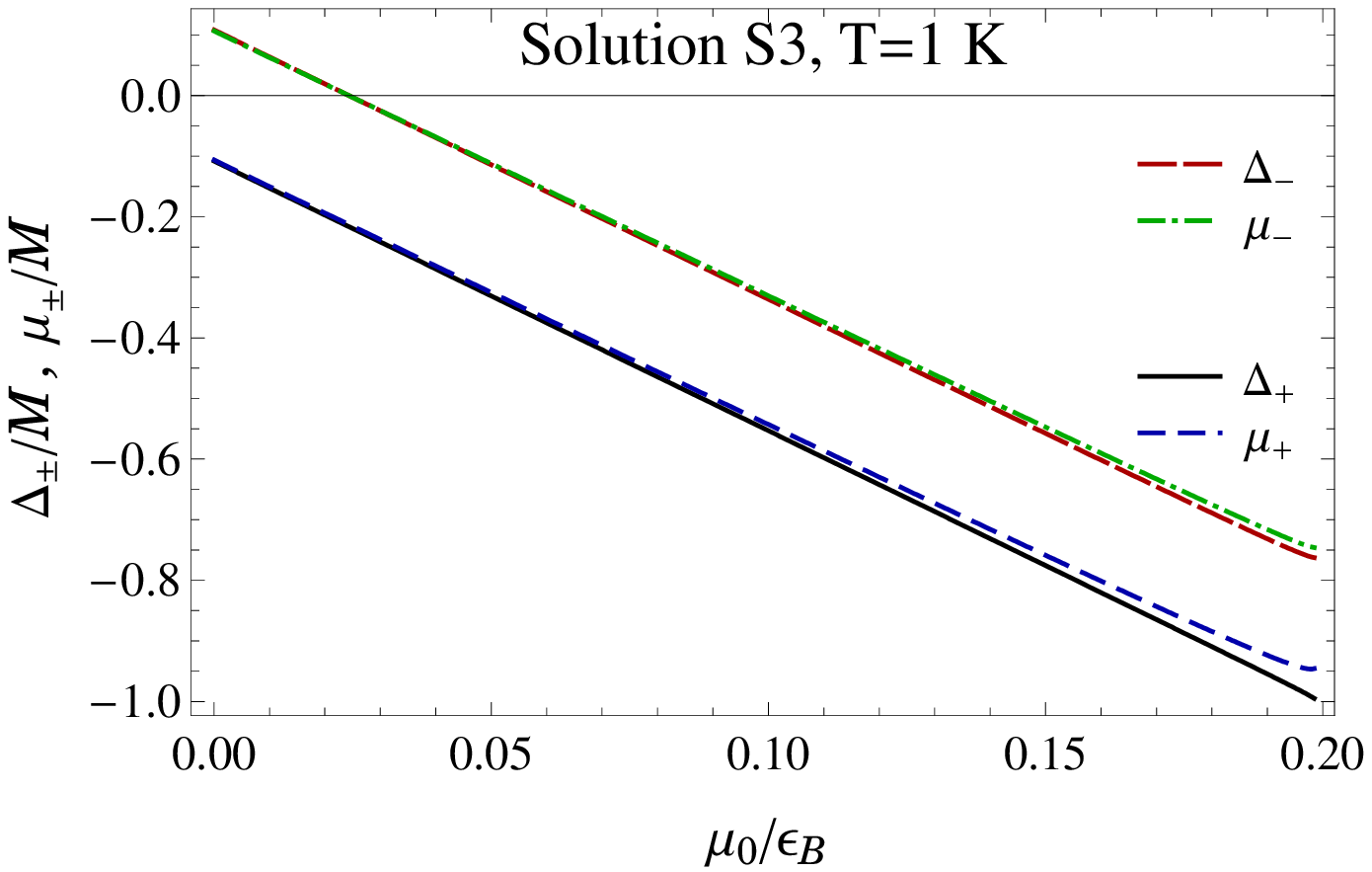}
\caption{Nontrivial order parameters of metastable solutions $H2$ (left panel)
and $S3$ (right panel) as functions of the electron chemical potential $\mu_0$.
In calculation, the temperature is taken nonzero but small, $T=1$~K.
The values of the electron chemical potential are given in units of the
Landau energy scale $\epsilon_B$, while the order parameters are
given in units of the dynamical scale $M$.}
\label{fig:H2&S3}
\end{center}
\end{figure}

\section{Dynamics on $n=1$ LL}
\label{5}

In the previous section, we analyzed solutions of the gap equations under the
condition that only states on the LLL can be filled, $|\mu_{s} \pm \tilde{\mu}_{s}|
\ll \epsilon_{B}=\sqrt{2\hbar|eB_{\perp}|v^{2}_F/c}$. Since all the dynamically
generated parameters are much less than $\epsilon_{B}$, this condition implies
that the bare chemical potential $\mu_0$ also has to satisfy a similar inequality,
$\mu_0 \ll \epsilon_{B}$. In this section, we extend the analysis by considering
the dynamics with $\mu_0$ being of the order of the Landau scale $\epsilon_{B}$,
i.e., we study the regime when quasiparticle states on the first Landau level,
$n=1$ LL, can be filled.

\subsection{Analytic solutions at $T=0$}
\label{5.1}

We will start from the gap
equations at zero temperature, which are given in Eqs.~(\ref{E1-a})--(\ref{E4-a}) in
Sec.~\ref{3}. In order to get their solutions for $\mu_0 \sim \epsilon_{B}$, we will follow
the same steps of the analysis as in Appendix~\ref{B} for the LLL. The corresponding
analysis for the $n=1$ LL, including the calculation of the free energy density for the
solutions, is done in Appendix~\ref{D}. It is shown there
that the following five stable solutions are realized (see the end of Subsec.~\ref{D3}):

\begin{itemize}
\item[(f-i)] The singlet type solution (f-I--f-I)
{(here $f$ stands for {\it first}; the nomenclature
used for the $n=1$ LL solutions is defined in Appendix \ref{D}):}

\begin{equation}
\begin{split}
& \tilde{\Delta}_{+}=\tilde{\mu}_{+}=0,
\qquad
\mu_{+} = \bar{\mu}_{+} - 7A,
\qquad
\Delta_{+}=-s_{\perp} M,
\\
& \tilde{\Delta}_{-}=\tilde{\mu}_{-}=0,
\qquad
\mu_{-} = \bar{\mu}_{-} - 7A,
\qquad
\Delta_{-}=-s_{\perp} M
\end{split}
\label{f-i}
\end{equation}
coincides with the solution S2 given by Eq.~(\ref{iii}) 
in the analysis of the LLL. It takes place for
$6A+Z<\mu_0 < 7A+\sqrt{\epsilon_{B}^2 +M^2}-Z$, and its free energy
density is
\begin{equation}
\Omega=-\frac{|eB_{\perp}|}{2\pi \hbar c}\left(M+2\mu_0-7A+h\right),
\end{equation}
where $h$ is given in Eq.~(\ref{h}). According to Subsec.~\ref{4.3}, this solution
corresponds to the regime with the filled LLL and the empty $n=1$ LL and is
connected with the $\nu = 2$ plateau.

\item[(f-ii)] The hybrid type solution (f-I--f-II)
\begin{equation}
\begin{split}
& \tilde{\Delta}_{+}=\tilde{\mu}_{+}=0,\qquad  \mu_{+} = \bar{\mu}_{+} -
11A,\qquad  \Delta_{+}=-s_{\perp}\,M,
\\
& \tilde{\Delta}_{-}=\frac{M-M_1}{2},\qquad \tilde{\mu}_{-}=-
As_{\perp},\qquad  \mu_{-} = \bar{\mu}_{-} - 10A,\qquad
\Delta_{-}=-s_{\perp}\, \frac{M+M_1}{2},
\end{split}
\label{f-ii}
\end{equation}
{with $M_1$ given in Eq.~(\ref{m2}) in Appendix \ref{D},}
takes place for $9A+\sqrt{\epsilon_{B}^2+M_1^2}-Z < \mu_0 <
11A+\sqrt{\epsilon_{B}^2 +M^2}-Z$, and its free energy density is
\begin{equation}
\Omega=-\frac{|eB_{\perp}|}{2\pi \hbar c}\left(\frac{3M+M_1}{4}+
3\mu_0-15A-\epsilon_{B}+Z+\frac{3h+h_1}{4}\right),
\end{equation}
where $h_1$ is given in Eq.~(\ref{h2}). {As is shown in Subsec.~\ref{5.2} below, 
this solution corresponds to the $\nu=3$ plateau.}

\item[(f-iii)] The singlet type solution (f-I--f-III)
\begin{equation}
\begin{split}
& \tilde{\Delta}_{+}=\tilde{\mu}_{+}=0,\qquad \mu_{+} = \bar{\mu}_{+} -
15A,\qquad \Delta_{+}=-s_{\perp}\,M,
\\
& \tilde{\Delta}_{-}=\tilde{\mu}_{-}=0,\qquad \mu_{-} = \bar{\mu}_{-} -
13A,\qquad  \Delta_{-}=-s_{\perp}\,M_1
\end{split}
\label{f-iii}
\end{equation}
is realized for $13A+\sqrt{\epsilon_{B}^2 +M_1^2}-Z < \mu_0 <
15A+\sqrt{\epsilon_{B}^2 +M^2}+Z$, and its free energy
density is
\begin{equation}
\Omega=-\frac{|eB_{\perp}|}{2\pi \hbar c}\left(\frac{M+M_1}{2}+4\mu_0-27A-
2\epsilon_{B}+2Z+\frac{h+h_1}{2}\right).
\end{equation}
{As is discussed in Subsec.~\ref{5.2}, this solution corresponds to the $\nu=4$ plateau.}
\item[(f-iv)] The hybrid type solution (f-II--f-III)
\begin{equation}
\begin{split}
& \tilde{\Delta}_{+}=\frac{M-M_1}{2},\qquad \tilde{\mu}_{+}=-
As_{\perp},\qquad  \mu_{+} = \bar{\mu}_{+} - 18A,\qquad
\Delta_{+}=-s_{\perp}\,\frac{M+M_1}{2},
\\
& \tilde{\Delta}_{-}=\tilde{\mu}_{-}=0,\qquad  \mu_{-} = \bar{\mu}_{-} -
17A,\qquad \Delta_{-}=-s_{\perp}\,M_1
\end{split}
\label{f-iv}
\end{equation}
takes place for $17A+\sqrt{\epsilon_{B}^2+M_1^2}+Z < \mu_0 < 19A+
\sqrt{\epsilon_{B}^2 +M^2}+Z$, and its free energy
density is
\begin{equation}
\Omega=-\frac{|eB_{\perp}|}{2\pi \hbar c}\left(\frac{3M_1+M}{4}+
5\mu_0-43A-3\epsilon_{B}+Z+\frac{3h_1+h}{4}\right).
\end{equation}
{This solution corresponds to the $\nu=5$ plateau (see 
Subsec.~\ref{5.2}).}
\item[(f-v)] The singlet type solution (f-III--f-III)
\begin{equation}
\begin{split}
& \tilde{\Delta}_{+}=\tilde{\mu}_{+}=0,\qquad  \mu_{+} = \bar{\mu}_{+} -
21A,\qquad \Delta_{+}=-s_{\perp}\,M_1,
\\
& \tilde{\Delta}_{-}=\tilde{\mu}_{-}=0,\qquad  \mu_{-} = \bar{\mu}_{-} -
21A,\qquad  \Delta_{-}=-s_{\perp}\,M_1
\end{split}
\label{f-v}
\end{equation}
is realized for $\mu_0 > 21A+\sqrt{\epsilon_{B}^2 +M_1^2}+Z$, and its free energy
density is
\begin{equation}
\Omega=-\frac{|eB_{\perp}|}{2\pi \hbar c}\left(M_1+6\mu_0-63A-
4\epsilon_{B}+h_1\right).
\end{equation}
\end{itemize}
{This solution corresponds to the $\nu=6$ plateau connected with the gap
between the filled $n=1$ LL and the empty $n=2$ LL.}

It should be emphasized that the above analytical solutions do not cover the
whole range of the values of the electron chemical potential around the $n=1$ LL. In
particular, there are no analytical solutions found in the following four intervals:
\begin{eqnarray}
\label{forbidden_region1}
7A+\sqrt{\epsilon_{B}^2+M^2}-Z<\mu_0 < 9A+\sqrt{\epsilon_{B}^2 +M_1^2}-Z,\\
\label{forbidden_region2}
11A+\sqrt{\epsilon_{B}^2 +M^2}-Z<\mu_0 < 13A+\sqrt{\epsilon_{B}^2 +M_1^2}-Z,\\
\label{forbidden_region3}
15A+\sqrt{\epsilon_{B}^2+M^2}+Z<\mu_0 < 17A+\sqrt{\epsilon_{B}^2 +M_1^2}+Z,\\
\label{forbidden_region4}
19A+\sqrt{\epsilon_{B}^2+M^2}+Z<\mu_0 < 21A+\sqrt{\epsilon_{B}^2+M_1^2}+Z.
\end{eqnarray}
The difficulty in finding analytical solutions at $T=0$ on these
intervals is related to the ambiguities in the definition of some
step functions in gap equations (\ref{E1-a})--(\ref{E4-a}). The
same problem, albeit in a weaker form, was also encountered in the
analysis of dynamics at the LLL {(see
Subsec.~\ref{4.4})}. As in that case, we remove the ambiguities by
considering a nonzero temperature case. The results at $T=0$ can
then be obtained by taking the limit $T\to 0$. The details of our
numerical analysis are given in the next subsection.

\subsection{Numerical analysis, $n=1$ LL}
\label{5.2}

By {performing} a nonzero temperature analysis numerically, we find that the solutions
(f-i), (f-iii), and (f-v), found analytically, are in fact continuously connected. They
are parts of a more general solution $S$ (here $S$ stands for {\em singlet})
that exists at all values of $\mu_0$. At small and intermediate values of $\mu_0$,
this solution includes solutions $S1$ and $S2$, see Fig.~\ref{fig.S_vs_mu_T}.
At larger values of $\mu_0$, relevant for the dynamics of $n=1$ LL, the solution
$S$ is shown in Fig.~\ref{fig.sol.S.f-i-iii-v}.

As seen in Fig.~\ref{fig.sol.S.f-i-iii-v}, the solution $S$
consists of five pieces defined on five adjacent intervals of
$\mu_0$. Three of them are the analytical solutions (f-i),
(f-iii), and (f-v), as defined in the previous subsection. Their
intervals of existence are $\mu_0/\epsilon_{B}\lesssim 1.27$,
$1.5\lesssim\mu_0/\epsilon_{B}\lesssim1.6$ and
$\mu_0/\epsilon_{B}\gtrsim1.83$, respectively. These intervals are
in agreement with the analytical results if one takes $M_1\approx
111~\mbox{K}$, or in terms of the Landau energy scale,
$M_1=4.42\times10^{-2}\epsilon_{B}$. The other two pieces of the
solution $S$ extend the singlet-type analytical solution to the
intermediate intervals.

\begin{figure}
\begin{center}
\includegraphics[width=.45\textwidth]{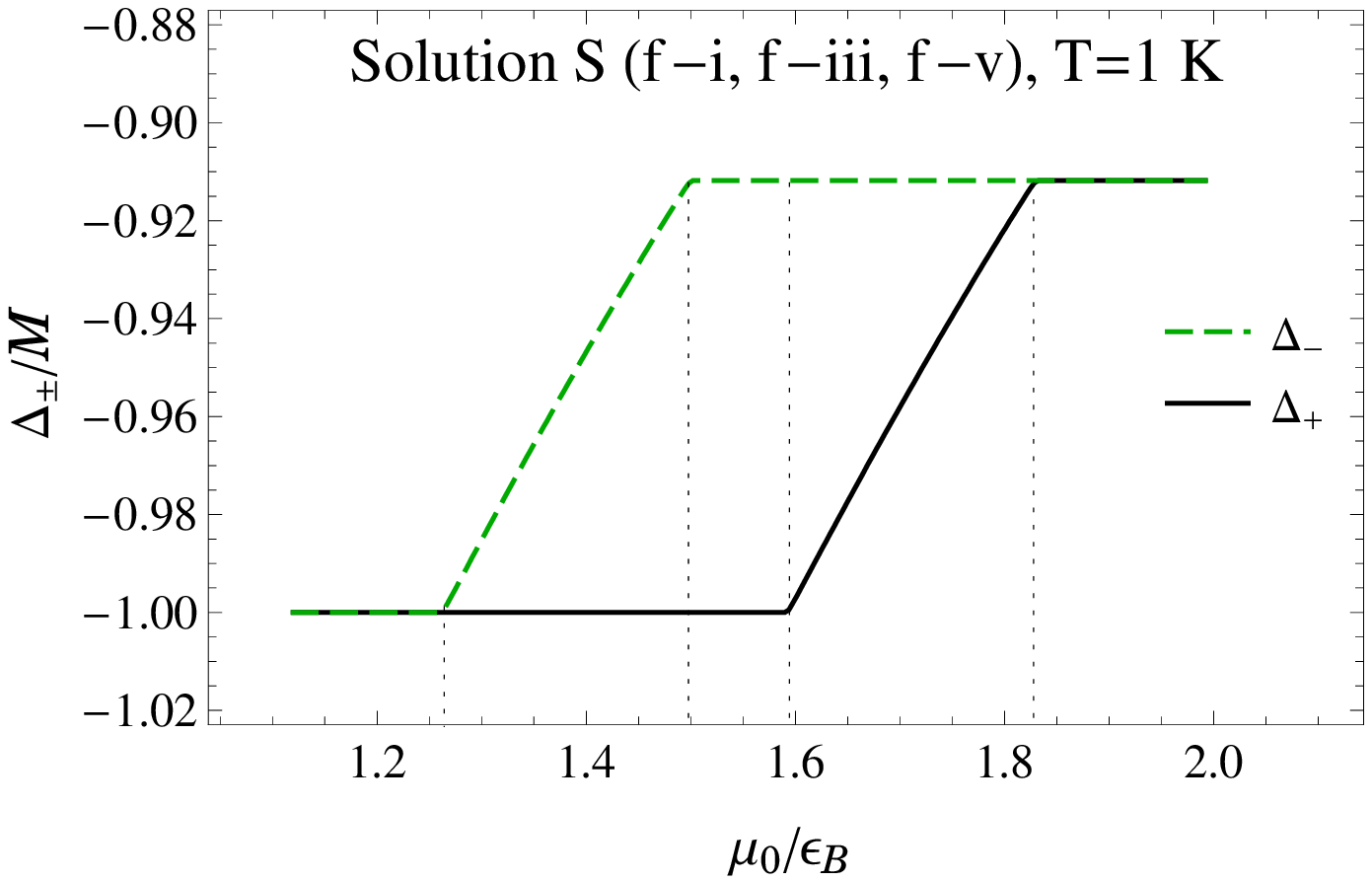}
\hspace{.04\textwidth}
\includegraphics[width=.43\textwidth]{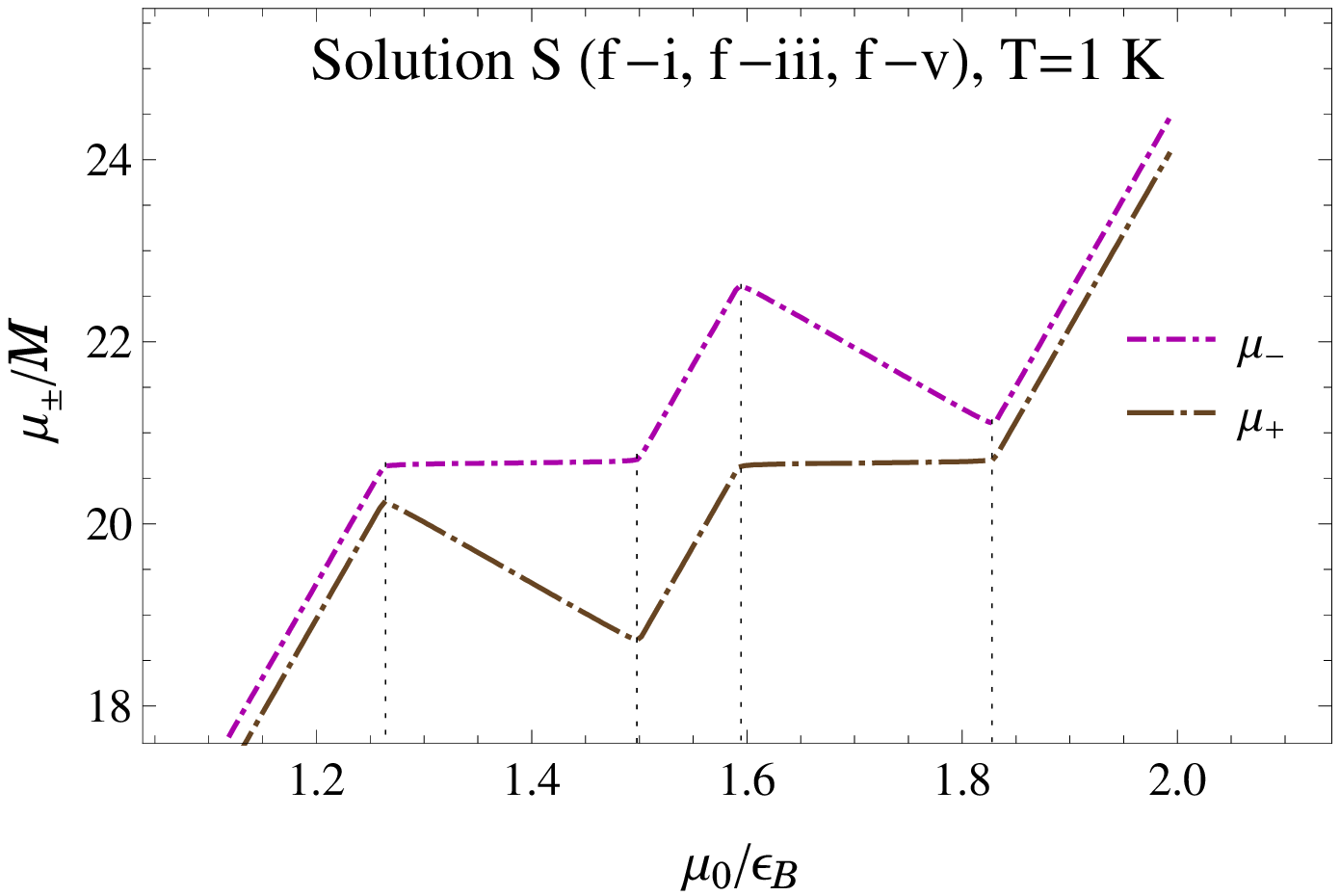}
\caption{Nontrivial order parameters of the $S$-type numerical solution
that contains the analytical solutions (f-i), (f-iii) and (f-v) as parts,
connected by two intermediate solutions.}
\label{fig.sol.S.f-i-iii-v}
\end{center}
\end{figure}

At $T=0$ the solution $S$ describes the ground state in exactly
the same regions of validity that are found analytically for
solutions (f-i), (f-iii), and (f-v) in the previous subsection.
This can be concluded from the energy consideration: among all
numerical solutions the parts of the solution $S$ have the lowest
free energy density there. {Analyzing the quasiparticle
spectra by using the dispersion relation in Eq.~(\ref{higherLLs}),
we find that the solutions (f-i), (f-iii), and (f-v) describe the
$\nu=2$, $\nu=4$, and $\nu=6$ QH states, respectively.}

{From} the symmetry viewpoint, none of the three parts of the
singlet solution break any exact symmetries in the model. However,
the part (f-iii) of the solution, describing the $\nu=4$ QH state,
corresponds to a quasi-spontaneous breakdown of the $U(4)$ symmetry
down to the $U(2)_{+} \times U(2)_{-}$. {Indeed, by using Eq.
(\ref{higherLLs}), one can check that { the LLL is half filled and
the energy gap} between the pairs of the pseudospin degenerate
spin-up and spin-down states of the $n=1$ LL is given by
$\Delta{E}_{4}\simeq 2(Z+A)+(M^{2}-M_{1}^{2})/2\epsilon_{B}$. As we
see, the spin splitting by the Zeeman term $2Z$ is strongly enhanced
by the dynamical contribution $2A$.}

This is somewhat similar to the enhancement of the spin splitting
in the $\nu=0$ QH state, discussed in Subsec.~\ref{4.1}.
{However, there is an important qualitative difference
between the cases of the LLL and the $n=1$ LL: It is only the
dynamical contribution to the chemical potentials (but not the
Dirac masses) that substantially affects the splitting in the
$\nu=4$ QH state. Indeed, the dynamical contribution due to the
Dirac masses in the gap $\Delta{E}_{4}$, i.e.,
$(M^{2}-M_{1}^{2})/2\epsilon_{B}$, is very small because $M\simeq
M_{1}\ll \epsilon_{B}$.) As a result, the gap $\Delta E_4$ is
substantially smaller than the LLL spin gap $\Delta E_0$ ($\Delta
E_4 \lesssim \Delta E_0/2$).}

Because of having nonvanishing triplet order parameters in the
extended hybrid solutions (f-ii) and (f-iv), the flavor $U(2)_{+}
\times U(2)_{-}$ symmetry of graphene is partially broken in the
corresponding ground states. {By using dispersion
relation (\ref{higherLLs}) in the analysis of the quasiparticle
spectra, we find that these solutions describe the $\nu=3$ and
$\nu=5$ plateaus corresponding to the quarter and three-quarter
filled $n=1$ LL, respectively.} In the case of the extended
solution (f-ii), the spin-down flavor subgroup $SU(2)_{-}\subset
U(2)_{-}$ is broken down to $U(1)_{-}$, while the spin-up flavor
subgroup $U(2)_{+}$ is intact. Similarly, in the case of the
extended solution (f-iv), the spin-up flavor subgroup $SU(2)_{+}
\subset U(2)_{+}$ is broken down to $U(1)_{+}$, while the
spin-down flavor subgroup $U(2)_{-}$ is intact. {Up to
small corrections due nonzero Dirac masses, the energy gaps
$\Delta E_3$ and $\Delta E_5$ associated with the (f-ii) and
(f-iv) solutions are equal to $2A$. Note that these gaps are
{substantially} smaller than the LLL gap $\Delta E_1$ ($\Delta E_3,
\Delta E_5 \lesssim \Delta E_1/2$).}

The analytical hybrid solutions (f-ii) and (f-iv) get continuous extensions to the
left and to the right from their regions of validity found analytically in the previous
subsection. In fact, they extend all the way to cover the neighboring ``forbidden"
regions defined in Eqs.~(\ref{forbidden_region1})--(\ref{forbidden_region4}).
The first two ``forbidden" interval are covered by the extension of the solution
(f-ii) to the interval
$7A+\sqrt{\epsilon_{B}^2+M^2}-Z<\mu_0 < 13A+\sqrt{\epsilon_{B}^2 +M_1^2}-Z$.
The non-trivial Dirac masses and chemical potentials for this numerical solution
are shown in Fig.~\ref{fig.sol.H.f-ii}. The last two ``forbidden" intervals, see
Eqs.~(\ref{forbidden_region3}) and (\ref{forbidden_region4}), are covered by
the extension of the solution (f-iv) to the interval
$15A+\sqrt{\epsilon_{B}^2 +M^2}+Z<\mu_0 < 21A+\sqrt{\epsilon_{B}^2 +M_1^2}+Z$.
The non-trivial parameters for this solution are shown in Fig.~\ref{fig.sol.H.f-iv}.

\begin{figure}
\begin{center}
\includegraphics[width=.45\textwidth]{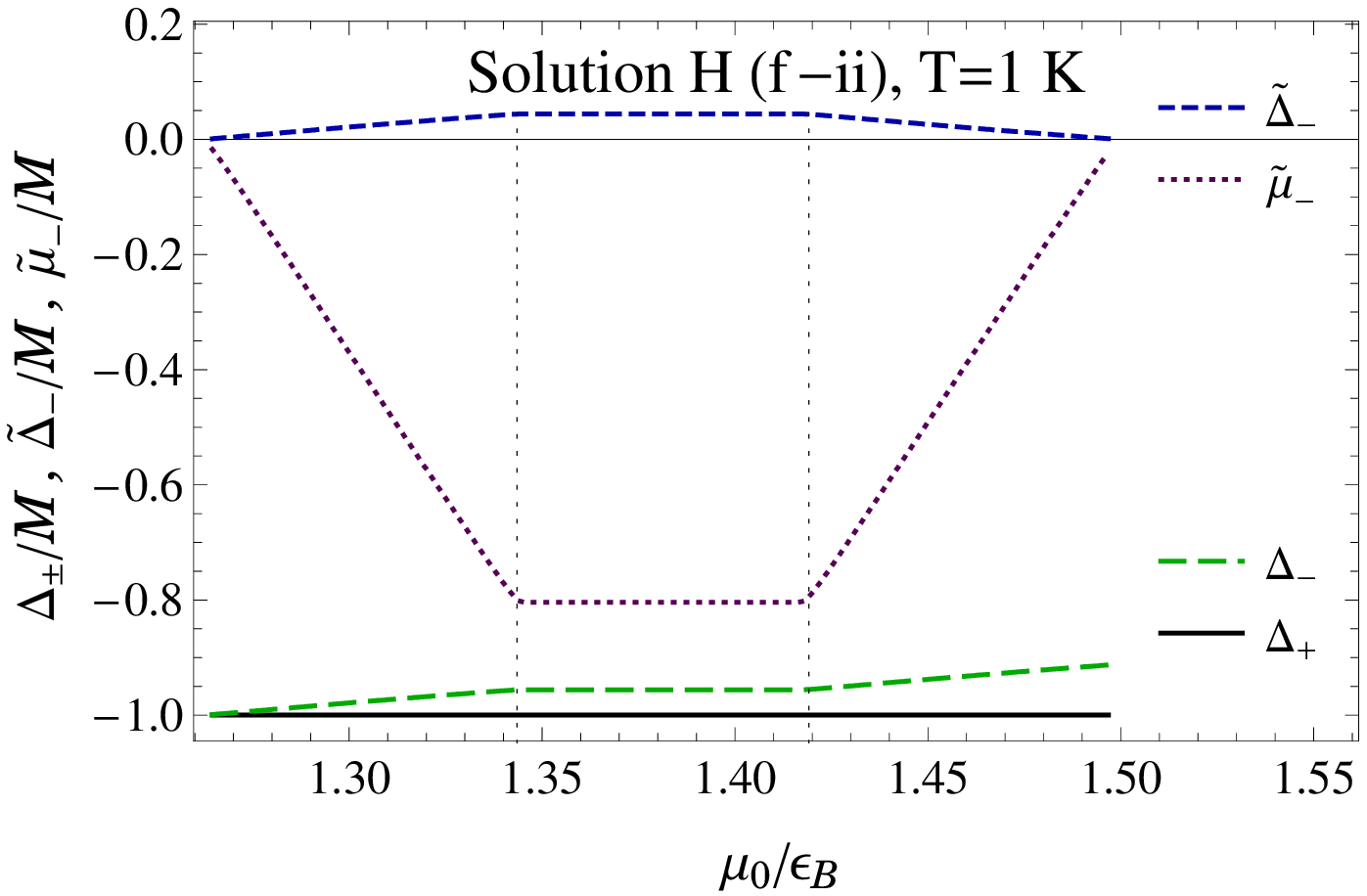}
\hspace{.04\textwidth}
\includegraphics[width=.45\textwidth]{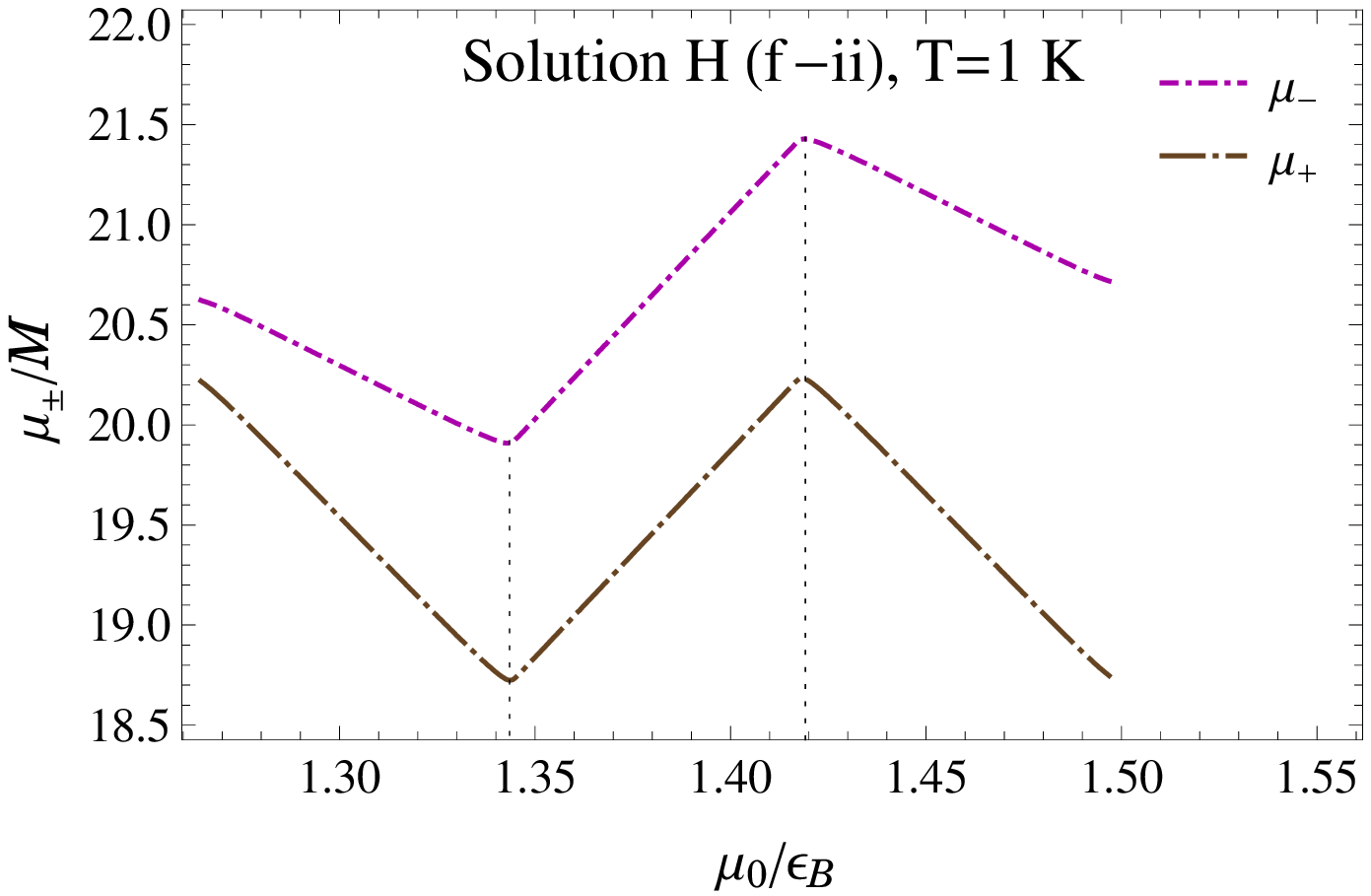}
\caption{Nontrivial order parameters of the extended hybrid solution (f-ii)
which determines the ground state for the $\nu=3$ QH plateau in graphene.}
\label{fig.sol.H.f-ii}
\end{center}
\end{figure}

\begin{figure}
\begin{center}
\includegraphics[width=.45\textwidth]{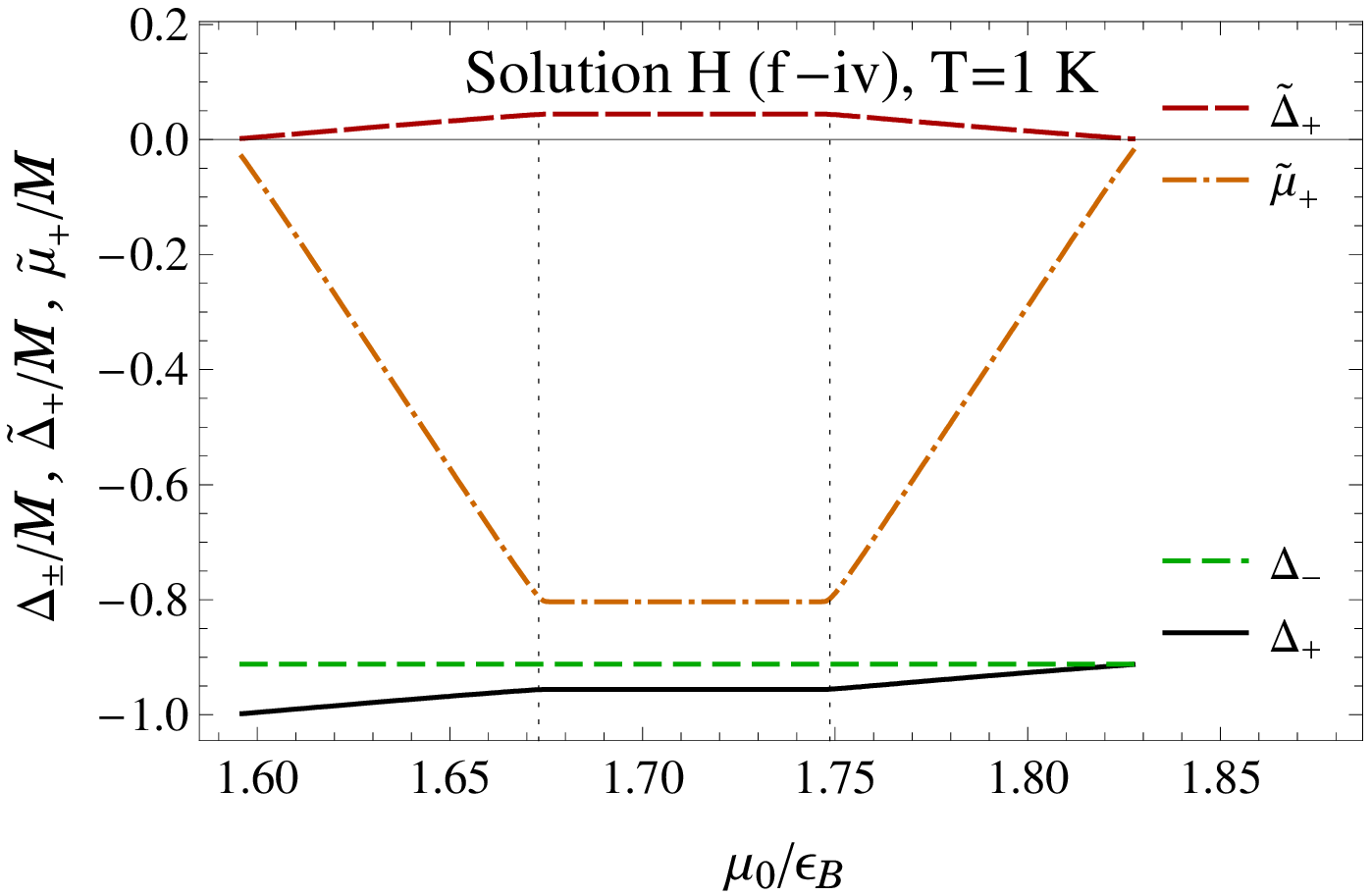}
\hspace{.04\textwidth}
\includegraphics[width=.45\textwidth]{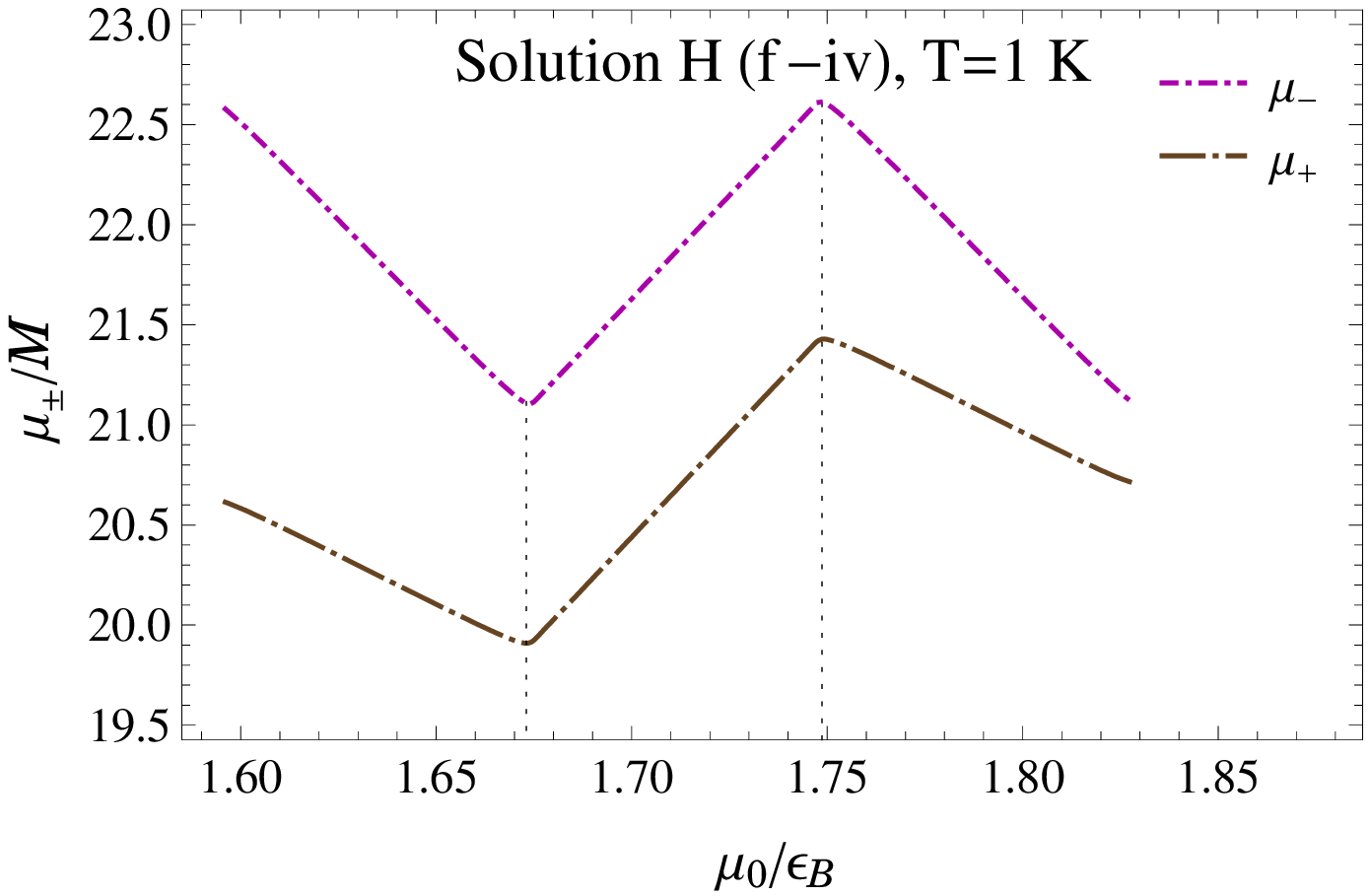}
\caption{Nontrivial order parameters of the extended hybrid solution (f-iv)
which determines the ground state for the $\nu=5$ QH plateau in graphene. }
\label{fig.sol.H.f-iv}
\end{center}
\end{figure}

In fact, the extended solutions (f-ii) and (f-iv) are the ground
states in their whole regions of existence. {This is seen
in Fig.~\ref{fig.V.eff.LL1.T1}, where we plot the difference
between the free energy density of the hybrid type solutions and
the singlet one. The results for the extended hybrid solutions
(f-ii) and (f-iv) are shown by the solid line and the long-dashed
line, respectively.}

In Fig.~\ref{fig.V.eff.LL1.T1} we also show the results for
another hybrid solution that was found numerically. It exists in
the interval of $\mu_0$ that could potentially be relevant for the
$\nu=4$ QH state. However, its free energy density is higher than
that for the solution $S$, and therefore it is unstable.

\begin{figure}
\begin{center}
\includegraphics[width=.45\textwidth]{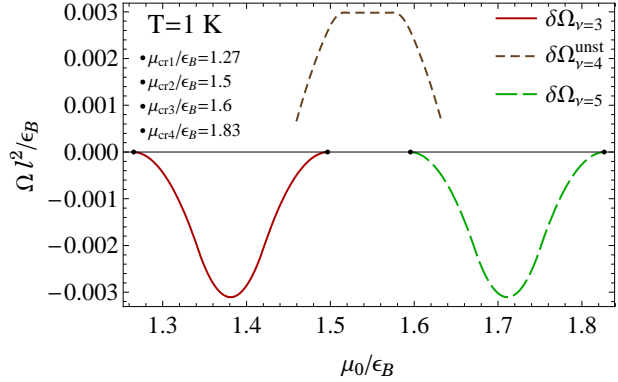}
\caption{The difference between the free energy density of three hybrid
type solutions and the free energy density of the $S$-type solution in
the range of $\mu_0$, associated with the dynamics of the $n=1$ LL. In
calculations, the temperature is taken nonzero but small, $T=1$~K. The
values of the free energy density and the electron chemical potential
are given in the same units as in Fig~\ref{fig.V.eff.T1}.}
\label{fig.V.eff.LL1.T1}
\end{center}
\end{figure}
With increasing the temperature, we find that the extended hybrid
solutions (f-ii) and (f-iv) responsible for the $\nu=3$ and
$\nu=5$ QH states gradually vanish. Their regions of existence
shrink and their free energy densities approach the free energy
density of the singlet solution $S$. At temperatures above
$T^{(\nu=3)}_{\rm cr} \simeq T^{(\nu=5)}_{\rm cr} \simeq 0.4 M
\simeq T^{(\nu=1)}_{\rm cr}/2$, they cease to exist altogether,
and the ground state is described by the singlet solution which
does not break any exact symmetries of the model.

\section{Discussion: phase diagram, experiment, disorder, and edge states}
\label{6}

{By summarizing the numerical results for the ground
states at different temperatures, we obtain the phase diagram of
graphene in the plane of temperature $T$ and electron chemical
potential $\mu_0$ shown in Fig.~\ref{fig-phase-diag}. The areas
highlighted in green correspond to hybrid solutions with a lowered
symmetry in the ground state. These regions are separated from the
rest of the diagram by phase transitions. At the boundary of the
$\nu=1$ region, the transition is of first order at low
temperatures and of second order at higher temperatures. The
transitions to/from the QH $\nu=3$ and $\nu=5$ states are of
second order. It should be kept in mind, however, that here the
analysis is done in the mean-field approximation and in a model
with a simplified contact interaction. Therefore, the predicted
types of the phase transitions may not be reliable. In particular,
the contributions of collective excitations, which are beyond the
mean-field approximation, may change the transitions to first
order type. Also, the types of the transitions may be affected by
the inclusion of disorder and a more realistic long-range Coulomb
interaction. Despite the model limitations, we still expect that
Fig.~\ref{fig-phase-diag} correctly represents the key qualitative
features of the phase diagram of graphene at least in the case of
the highest quality samples.}

\begin{figure}
\begin{center}
\includegraphics[width=.45\textwidth]{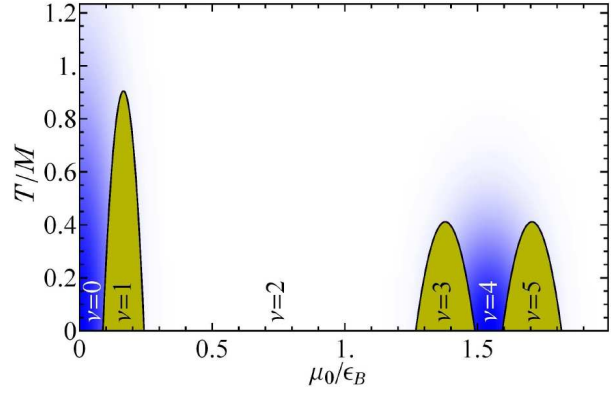}
\caption{Schematic phase diagram of graphene in the plane of temperature
and electron chemical potential. The values of chemical potential are
given in units of the Landau energy scale $\epsilon_{B}$, and the values
of temperature are given in units of the dynamical scale $M$.}
\label{fig-phase-diag}
\end{center}
\end{figure}

In Fig.~\ref{fig-phase-diag} the regions highlighted in blue
correspond to the ground states with a quasi-spontaneous breakdown
of the spin symmetry. In the case of the LLL and the $n=1$ LL,
such are the $\nu=0$ and $\nu=4$ QH states, in which the
quasi-spontaneous breakdown of the approximate $U(4)$ symmetry
down to $U(2)_{+}\times U(2)_{-}$ is enhanced by dynamical
contributions. Because of the explicit breakdown by the Zeeman
term, there is no well-defined order parameter associated with
this symmetry breakdown. Also, there is no well-defined boundary
of the corresponding regions in the diagram. In the plot, this
feature is represented by the fading shades of blue at the
approximate boundaries of the $\nu=0$ and $\nu=4$ regions.

As considered in detail in Sec.~\ref{4}, the physical properties of the $\nu=0$ and
$\nu=1$ QH states are determined by the dynamics of the LLL. The corresponding values
of the gaps, $\Delta{E}_{0}=2(Z+A+M)$ and $\Delta{E}_{1}= 2(A+M)$, are largely
determined by the dynamical contributions $A$ and $M$ of about equal magnitude.
These two contributions are associated with the QHF and MC order parameters,
respectively.

The results of this study for the LLL at least qualitatively agree with the
experimental data.\cite{Zhang2006,Jiang2007} By taking the dimensionless coupling
$\lambda=4A\Lambda/(\sqrt{\pi} \epsilon_{B}^2)$ to be a free parameter
and utilizing the cutoff $\Lambda$ to be of the order of the Landau scale $\epsilon_{B}$,
we arrive at the following scaling relations: $A  \sim \lambda\sqrt{|eB_{\perp}|}$
and $M \sim \lambda\sqrt{|eB_{\perp}|}$. This implies the same type of scaling
for the gap, $\Delta{E}_1 = 2(A + M) \sim \lambda\sqrt{|eB_{\perp}|}$, associated
with the $\nu =\pm 1$ plateaus. [Recently, the square root scaling of the activation
energy in the $\nu=1$ QH state was also obtained in the large-$N$ approximation
in Ref.~\onlinecite{Herbut-scaling}.] By making use of our results, we find that the
experimental value $\Delta{E}_1 \sim 100~\mbox{K}$ for $B_{\perp}=30~\mbox{T}$
from Ref.~\onlinecite{Jiang2007} corresponds to $\lambda \sim 0.02$. This estimate,
however, should be taken with great caution: Because interactions with impurities are
ignored and no disorder of any type is accounted for in the present model, it
may not be unreasonable to assume that actual values of $\lambda$ are up to
an order of magnitude larger.

As to the $n=1$ LL, we found that there are the gaps
$\Delta{E}_{3}=\Delta{E}_{5} \simeq 2A$ and $\Delta{E}_{4}\simeq
2(Z+A)$ corresponding to the plateaus $\nu=3, 5$ and $\nu=4$,
respectively [the contributions of Dirac masses are suppressed by
a factor of order $(M/\epsilon_{B})^2$ there]. Therefore the gaps
$\Delta{E}_{3}=\Delta{E}_{5}$ and $\Delta{E}_{4}$ are mostly due
to the QHF type order parameters and are about a factor of two
smaller than the LLL gaps $\Delta{E}_{1}$ and $\Delta{E}_{0}$,
respectively. On the other hand, the experimental data yield
$\Delta{E}_{4}\simeq 2Z$, and no gaps $\Delta{E}_{3}$,
$\Delta{E}_{5}$ have been observed.\cite{Zhang2006,Jiang2007} We
believe that a probable explanation of this discrepancy is that,
unlike $Z$, the value of the {\it dynamically} generated parameter
$A$ corresponding to the $|n| \geq 1$ LLs will be strongly
reduced if a considerable broadening of higher LLs in a magnetic
field is taken into account.\cite{Gusynin2006catalysis} If so, the
gap $\Delta{E}_{4}$ will be reduced to $2Z$, while the gaps
$\Delta{E}_{3}$ and $\Delta{E}_{5}$ will become unobservable.

In order to estimate the value of a magnetic field at which the
plateaus $\nu =3$ and $5$ could become observable, one can use the
following arguments. Recently, in Ref.~\onlinecite{LL-width}, a large
width $\Gamma_1$ of $400~\mbox{K}$ was determined for the $n=1$
LL. On the other hand, the plateaus $\nu =3, 5$ could become
observable if the gaps $\Delta E_3 =\Delta E_5 \simeq 2A$
calculated in the clean limit are at least of order $\Gamma_1$ or
larger.\cite{Gusynin2006catalysis} The LLL gap $\Delta E_1 \simeq
100~\mbox{K}$ at $|B_\perp| = 30~\mbox{T}$ corresponds to 
{$\Delta E_{3,5} \lesssim 50~\mbox{K}$}. 
Then, taking a conservative estimate
$\Gamma_1 = 100~\mbox{K}$ and using $A \sim \sqrt{|eB_\perp|}$, we
conclude that to observe the $\nu =3, 5$ plateaus, the magnetic
fields should be at least as large as $B \sim 100~\mbox{T}$.

Here it is also appropriate to mention the dynamics of gapless
edge states, whose importance for the physics of the $\nu = 0$
plateau has been recently discussed in
Refs.~\onlinecite{Abanin2006PRL,Abanin2007PRL,Ong2007}.
Generalizing the analysis in Ref.~\onlinecite{Abanin2007PRL}, it
has been recently found \cite{edge_states,edge_states_long} that
for the $S1$ solution (\ref{i}) with the zigzag boundary
conditions, such states exist only when the full Zeeman energy
$(\mu_{+} - \mu_{-})/2 = Z + A$ is larger than the Dirac mass
$\Delta_{\pm}=M$ (at an armchair edge, gapless edge states exist
for any value of a singlet Dirac mass). Because of that, {for
$\lambda$ smaller than $1$}, we find from the constraint $Z >
\lambda A/(1 - \lambda)$ in the solution (i) and Eq.~(\ref{Apar})
with $\Lambda \sim \epsilon_B$ that the gapless edge states exist
when $|B_\perp| > B_{\perp}^{\rm (cr)}\sim 8\times 10^4
\lambda^4/(1-\lambda)^2$~T. Then, for the values of the
dimensionless coupling $\lambda$ in the range $0.02\lesssim
\lambda \lesssim 0.2$, we find that $0.01~\mbox{T} \lesssim
B_{\perp}^{\rm (cr)} \lesssim 200~\mbox{T}$. As we see,
$B_{\perp}^{\rm (cr)}$ is very sensitive to the choice of $\lambda$.
Therefore, in order to fix the critical value $B_{\perp}^{\rm (cr)}$
more accurately, one should first utilize more realistic models of
graphene that incorporate disorder among other
things.\cite{disorder1,disorder2} This is a topic for future
studies however.

These results are of interest in connection with the
interpretation of the $\nu=0$ Hall plateau. Indeed, the gapless
edge states should play an important role in transport properties
of graphene in a strong magnetic field. Their presence is expected
to make graphene a so-called quantum Hall metal, while their
absence should make it an
insulator.\cite{Abanin2007PRL,Abanin2006PRL} The actual
temperature dependence of the longitudinal resistivity at the
$\nu=0$  plateau in Refs.~\onlinecite{Zhang2006,Abanin2007PRL} is
consistent with the metal type. This conclusion may be disputed in
view of the recent data from Ref.~\onlinecite{Ong2007} that reveal
a clear plateau at $\nu=0$, but the temperature dependence of the
diagonal component of the resistivity signals a crossover to an
insulating state in high fields. The latter observations do not
seem to support the existence of gapless edge states.

The analysis in this paper as well as in Refs.~\onlinecite{edge_states,edge_states_long}
suggests that the conditions for the existence and absence of gapless
edge states depend sensitively on the type of the boundary conditions
and the values of QHF and MC order
parameters that characterize the nature of the corresponding QH state.
Therefore, the
dynamics of the edge states is very likely to be rich and full of surprises.

In conclusion, we have shown that the QHF and MC order parameters
in graphene are two sides of the same coin and they necessarily
coexist. This feature could have important consequences for the QH
dynamics, in particular, for edge states. The present model leads
to a reasonable and consistent description of the new QH plateaus
in graphene in strong magnetic fields. It would be desirable to
extend the present analysis to a more realistic model setup,
including the Coulomb interaction between quasiparticles, the
quasiparticle width, and various types of disorder.

\begin{acknowledgments}
Useful discussions with S.G. Sharapov are acknowledged.
V.A.M. is grateful to G.W. Semenoff for enjoyable
discussions.
The work of E.V.G and V.P.G. was supported by the
SCOPES under Project No. IB 7320-110848 of the NSF-CH,
by Ukrainian State Foundation for Fundamental Research
under the grant No. F16-457-2007, and by the National
Academy of Sciences of Ukraine under the grants No. 10/07-N
and No. II-1-07.
The work of V.A.M. was supported by the Natural
Sciences and Engineering Research Council of Canada.
He is grateful to the Aspen Center for Physics and 
the Institute for Nuclear Theory
at the University of Washington for their hospitality
and the Department of Energy for partial support during
the completion of this work.

\end{acknowledgments}

\appendix

\section{Quasiparticle propagator and the gap equation}
\label{A}

\subsection{Quasiparticle propagator: Expansion over LLs}
\label{A1}

In this Appendix, the units with $\hbar=1$ and $c=1$ are used.

The full propagator $G_{s}(u,u^\prime)$ that corresponds to
the inverse propagator in Eq.~(\ref{full-inverse}) is given by
the following expression:
\begin{eqnarray}
G_{s}(u,u^{\prime}) &=& i\langle u|\left[(i\partial_t+\mu_{s})\gamma^0
- v_F(\bm{\pi}\cdot\bm{\gamma})
+i\tilde{\mu}_{s}\gamma^1\gamma^2+i\Delta_{s}\gamma^0\gamma^1\gamma^2
-\tilde{\Delta}_{s}\right]^{-1}|u^\prime\rangle
\nonumber\\
&=&
i\langle u| \left[(i\partial_t+\mu_{s})\gamma^0 - v_F(\bm{\pi}\cdot\bm{\gamma})
+i\tilde{\mu}_{s}\gamma^1\gamma^2 -i\Delta_{s}\gamma^0\gamma^1\gamma^2
+\tilde{\Delta}_{s}\right]
\nonumber\\
&&\times \left[\left( (i\partial_t+\mu_{s})\gamma^0 -v_F(\bm{\pi}\cdot\bm{\gamma})
+i\tilde{\mu}_{s}\gamma^1\gamma^2+
i\Delta_{s}\gamma^0\gamma^1\gamma^2-\tilde{\Delta}_{s}\right) \right.
\nonumber\\
&& \times
\left.\left( (i\partial_t+\mu_{s})\gamma^0 -v_F(\bm{\pi}\cdot\bm{\gamma})
+i\tilde{\mu}_{s}\gamma^1\gamma^2 -i\Delta_{s}\gamma^0\gamma^1\gamma^2+
\tilde{\Delta}_{s}\right)\right]^{-1}|u^\prime\rangle
\nonumber\\
&=&
i\langle u|
\left[(i\partial_t+\mu_{s})\gamma^0 - v_F(\bm{\pi}\cdot\bm{\gamma})
+i\tilde{\mu}_{s}\gamma^1\gamma^2 -i\Delta_{s}\gamma^0\gamma^1\gamma^2+\tilde{\Delta}_{s}\right]
\nonumber\\
&&\times
\left[(i\partial_t+\mu_{s})^2- v_{F}^2\bm{\pi}^2
+ 2i\tilde{\mu}_{s}(i\partial_t+\mu_{s})\gamma^0\gamma^1\gamma^2
+2i\Delta_{s}\tilde{\Delta}_{s}\gamma^0\gamma^1\gamma^2\right.
\nonumber\\
&&
\left.
-ieB_{\perp}v_{F}^2\gamma^1\gamma^2+
\tilde{\mu}_{s}^2-\tilde{\Delta}_{s}^2-\Delta_{s}^2
\right]^{-1}
|u^\prime\rangle .
\label{propagator1}
\end{eqnarray}
where $u=(t,\mathbf{r})$ and $\mathbf{r}=(x,y)$.
Our aim is to get an expression for this propagator as an expansion
over LLs. For the Fourier transform in time we need to calculate
\begin{eqnarray}
G_{s}(\omega;\mathbf{r},\mathbf{r}^{\prime})=i\left[W-v_F(\bm{\pi}_{r}\cdot{\pmb\gamma})\right]
\langle\mathbf{r} |
\left({\cal M}-v_F^{2}\bm{\pi}^{2}-ieB_{\perp}v_F^{2}\gamma^1\gamma^2\right)^{-1}
| \mathbf{r}^{\prime}\rangle,
\label{propagator-FT}
\end{eqnarray}
where the matrices $W$ and ${\cal M}$ are
\begin{eqnarray}
W&=&(\omega+\mu_{s})\gamma^{0}+i\tilde{\mu}_{s}\gamma^1\gamma^2
-i\Delta_{s}\gamma^0\gamma^1\gamma^2+\tilde{\Delta}_{s},
\label{matrices-MW}\\
{\cal M} &=& (\omega+\mu_{s}+i\tilde{\mu}_{s}\gamma^0\gamma^1\gamma^2)^2
-(\tilde{\Delta}_{s}-i\Delta_{s}\gamma^0\gamma^1\gamma^2)^2.
\label{matrices-WM}
\end{eqnarray}
The operator $\bm{\pi}^{2}$ has well known eigenvalues $(2n+1)|eB_{\perp}|$ with $n=0,1,2,\dots$
and its normalized wave functions in the Landau gauge $\mathbf{A}=(0,B_{\perp}x)$ are
\begin{eqnarray}
\psi_{np}(\mathbf{r})=\frac{1}{\sqrt{2\pi l}}\frac{1}{\sqrt{2^nn!\sqrt{\pi}}}
H_n\left(\frac{x}{l}+pl\right)e^{-\frac{1}{2l^2}(x+pl^2)^2} e^{ipy},
\end{eqnarray}
where $H_{n}(x)$ are the Hermite polynomials and $l=\sqrt{\hbar c/|eB_{\perp}|}$ is the
magnetic length. These wave functions satisfy the conditions of normalizability
\begin{eqnarray}
\int d^{2}{r}\psi^{*}_{np}(\mathbf{r})\psi_{n^{\prime}p^{\prime}}(\mathbf{r})=\delta_{nn^{\prime}}
\delta(p-p^{\prime}),
\end{eqnarray}
and completeness
\begin{eqnarray}
\sum\limits_{n=0}^{\infty}\int\limits_{-\infty}^{\infty} dp\psi^{*}_{np}(\mathbf{r})
\psi_{np}(\mathbf{r}^{\prime})
=\delta(\mathbf{r}-\mathbf{r}^{\prime}).
\label{completeness}
\end{eqnarray}
Using the spectral expansion of the unit operator (\ref{completeness}), we can write
\begin{eqnarray}
\langle\mathbf{r}|\left(
{\cal M}-v_F^{2}\bm{\pi}^{2}-ieB_{\perp}v_F^{2}\gamma^1\gamma^2\right)^{-1}|
\mathbf{r}^{\prime}\rangle
&=&
\frac{1}{2\pi l^2}\exp\left(-\frac{(\mathbf{r} -\mathbf{r}^{\,\prime})^2}
{4l^2}-i\frac{(x+x^\prime)(y-y^\prime)}{2l^2}\right)\nonumber\\
&\times&\sum\limits_{n=0}^\infty\frac{1}{{\cal M}-(2n+1)v^{2}_{F}|eB_{\perp}|
-iv^{2}_{F}eB_{\perp}\gamma^{1}\gamma^{2}}
L_n\left(\frac{(\mathbf{r}-\mathbf{r}^{\,\prime})^2}{2l^2}\right),
\label{auxillary-prop}
\end{eqnarray}
where we integrated over the quantum number $p$ by making use of the formula
$7.378$ in Ref.~\onlinecite{GR},
\begin{equation}
\int\limits_{-\infty}^\infty\,e^{-x^2}H_m(x+y)H_n(x+z)dx
=2^n\pi^{1/2}m!z^{n-m}L_m^{n-m}(-2yz),
\end{equation}
assuming $m\le n$. Here $L^{\alpha}_n$ are the generalized Laguerre polynomials, and
$L_n \equiv L^{0}_n$. The matrix $iv_F^{2}eB_{\perp} \gamma^1\gamma^2$ has eigenvalues
$\pm v_F^{2}|eB_{\perp}|$, and thus one can write
\begin{equation}
\frac{L_{n}(\xi)}{{\cal M}-(2n+1)v^{2}_{F}|eB_{\perp}|-i v^{2}_{F} eB_{\perp}\gamma^{1}\gamma^{2}}
=\frac{{\cal{P_{-}}}L_{n}(\xi)}{{\cal M}-
(2n+1)v^{2}_{F}|eB_{\perp}|+v^{2}_{F}|eB_{\perp}|}+
\frac{{\cal{P_{+}}}L_{n}(\xi)}{{\cal M}-
(2n+1)v^{2}_{F}|eB_{\perp}|-v^{2}_{F}|eB_{\perp}|},
\label{A9}
\end{equation}
where the variable $\xi$ and the projectors {$\cal{P}_{\pm}$} are
\begin{eqnarray}
\xi &=&\frac{(\mathbf{r}-\mathbf{r}^{\,\prime})^2}{2l^2}, \\
{\cal{P}_{\pm}} 
&=&\frac{1}{2}\left[1\pm i\gamma^{1}\gamma^{2}\mbox{sign}(eB_{\perp})\right].
\end{eqnarray}
Now, by redefining $n\to n-1$ in the second term in Eq.~(\ref{A9}),
equality (\ref{auxillary-prop}) can be rewritten as
\begin{eqnarray}
\langle\mathbf{r}|[{\cal M}-v_F^{2}\bm{\pi}^{2}-ieB_{\perp}v_F^{2}\gamma^1\gamma^2]^{-1}|
\mathbf{r}^{\prime}\rangle=\frac{1}{2\pi l^{2}}e^{i\Phi(\mathbf{r},\mathbf{r}^{\prime})}
e^{-\xi/2}\sum\limits_{n=0}^\infty
\frac{{\cal{P}_{-}}L_{n}(\xi)+
{\cal{P}_{+}}L_{n-1}(\xi)}{{\cal M}-
2nv^{2}_{F}|eB_{\perp}|},
\end{eqnarray}
where $L_{-1}\equiv 0$  by definition and the phase
\begin{equation}
\Phi(\mathbf{r},\mathbf{r}^{\prime})=-\frac{(x+x^\prime)(y-y^\prime)}{2l^2}=
-e\int\limits_{\mathbf{r}^{\prime}}^{\mathbf{r}}dz_{i}A_{i}(z)
\label{phase}
\end{equation}
appears because in the presence
of a constant magnetic field, the commutative group of translations is
replaced by the noncommutative group of magnetic translations \cite{Zak} 
(note that the integration in Eq. (\ref{phase}) is taken along the straight line). 
This
implies that it has a universal character. By noting that
\begin{eqnarray}
\pi_{x}e^{i\Phi} &=& e^{i\Phi}\left(-i\partial_{x}-\frac{y-y^{\prime}}{2l^{2}}\right),\\
\pi_{y}e^{i\Phi} &=& e^{i\Phi}\left(-i\partial_{y}+\frac{x-x^{\prime}}{2l^{2}}\right),
\end{eqnarray}
we see that propagator (\ref{propagator-FT}) can be presented in the form of a
product of the phase factor and a translation invariant part
$\bar{G}_{s}(\omega;\mathbf{r}-\mathbf{r}^{\prime})$,
\begin{equation}
G_{s}(\omega;\mathbf{r},\mathbf{r}^{\prime})=e^{i\Phi(\mathbf{r},\mathbf{r}^{\prime})}
\bar{G}_{s}(\omega;\mathbf{r}-\mathbf{r}^{\prime}),
\end{equation}
where
\begin{eqnarray}
\bar{G}_{s}(\omega;\mathbf{r}-\mathbf{r}^{\prime})=i\left[W-v_{F}\gamma^{1}
\left(-i\partial_{x}-\frac{y-y^{\prime}}{2l^{2}}\right)-v_{F}\gamma^{2}
\left(-i\partial_{y}+\frac{x-x^{\prime}}{2l^{2}}\right)\right]\frac{e^{-\xi/2}}{2\pi l^{2}}
\sum\limits_{n=0}^\infty
\frac{{\cal{P}_{-}}L_{n}(\xi)+{\cal{P}_{+}}
L_{n-1}(\xi)}{{\cal M}-
2nv^{2}_{F}|eB_{\perp}|}.
\label{propagatorG-FT}
\end{eqnarray}
It is important to emphasize that the phase factor does not affect
the gap equation (\ref{gap}) because the latter contains the full
propagator only at $u^\prime=u$.

The Fourier transform of the translation invariant part of
propagator (\ref{propagatorG-FT}) can be evaluated by first
performing the integration over the angle,
{
\begin{equation}
\int\limits_{0}^{2\pi}d\theta e^{ikr\cos\theta}=2\pi J_{0}(kr),
\end{equation}
where $J_{0}(x)$ is the Bessel function,} and then using the
formula $7.421.1$ in Ref.~\onlinecite{GR},
\begin{equation}
 \int_0^\infty
xe^{-\frac{1}{2}\alpha x^2}L_n\left(\frac{1}{2}\beta
x^2\right)J_0(xy)dx
=\frac{(\alpha-\beta)^n}{\alpha^{n+1}}e^{-\frac{1}{2\alpha}
y^2}L_n\left( \frac{\beta y^2}{2\alpha(\beta-\alpha)}\right),
\end{equation}
valid for $y>0$ and $\mbox{Re}\,\alpha>0$. The result is given by
\begin{equation}
\bar{G}_{s}(\omega,\mathbf{k}) = ie^{-k^2 l^{2}}\sum_{n=0}^{\infty}
\frac{(-1)^nD_{ns}(\omega,\mathbf{k})}{{\cal M}-2nv^{2}_{F}|eB_{\perp}|},
\label{Dsn-new}
\end{equation}
with
\begin{eqnarray}
D_{ns}(\omega,\mathbf{k}) = 
2W\left[{\cal{P}_{-}}L_n\left(2 k^2 l^{2}\right)
-{\cal{P}_{+}}L_{n-1}\left(2 k^2 l^{2}\right)\right]
 + 4v_F(\mathbf{k}\cdot\bm{\gamma}) L_{n-1}^1\left(2 k^2 l^{2}\right),\,
L_{-1}^\alpha \equiv 0,
\label{Dn}
\end{eqnarray}
describing the $n$th Landau level contribution (compare with corresponding expression
for the standard Dirac propagator in Ref.~\onlinecite{Gusynin1995PRD}).

\subsection{Equations for Dirac masses and chemical potentials}
\label{A2}

In order to derive Eqs.~(\ref{E1-a})--(\ref{E4-a}) for masses and
chemical potentials, we
need to know the full propagator at $u^{\prime}=u$,
$G_{s}(u,u^\prime)|_{u=u^\prime} = \bar{G}_{s}(u,u)$.
As follows from Eq.~(\ref{propagatorG-FT}), it is
\begin{eqnarray}
{G}_{s}(u,u)= \int\limits_{-\infty}^{\infty}\frac{d\omega}{2\pi}\bar{G}_{s}(\omega,0)
=\frac{i}{2\pi l^{2}}\sum\limits_{n=0}^{\infty}\int\limits_{-\infty}^{\infty}\frac{d\omega}{2\pi}
W\frac{{\cal{P}_{-}}+
{\cal{P}_{+}}\theta(n-1)}{{\cal M}-2nv^{2}_{F}|eB_{\perp}|}.
\label{interaction-new}
\end{eqnarray}
In what follows, it is convenient to work with eigenvectors of the
matrices $\gamma^1\gamma^2$ and $\gamma^0$. Since
$
(\gamma^1\gamma^2)^2=-1,
$
the eigenvectors $|s_{12}\rangle$ of the matrix $\gamma^1\gamma^2$ correspond
to imaginary eigenvalues $is_{12}=\pm i$, i.e.,
\begin{equation}
\gamma^1\gamma^2|s_{12}\rangle=is_{12}|s_{12}\rangle\, .
\end{equation}
Similarly, since
$
(\gamma^0)^2=1,
$
the eigenvectors $|s_0\rangle$ of the matrix $\gamma^0$ correspond to
eigenvalues $s_0=\pm 1$, i.e.,
\begin{equation}
\gamma^0|s_0\rangle=s_0|s_0\rangle.
\end{equation}
Because $\gamma^0$ and $\gamma^1\gamma^2$ commute, we can use states
$|s_{12}s_0\rangle$ which are simultaneously eigenvectors of
$\gamma^1\gamma^2$ and $\gamma^0$ with eigenvalues $is_{12}$ and $s_0$,
respectively. The vectors $|s_{12}s_0\rangle$ form a complete basis. Therefore,
any $4\times 4$ matrix ${\cal O}$ can be represented as
\begin{equation}
{\cal O}=\sum_{s_{12}^{\prime},s_0^{\prime},s_{12},s_0} {\cal
O}_{s_{12}^{\prime}s_0^{\prime}s_{12}s_0}
|s_{12}^{\prime}s_0^{\prime}\rangle\langle s_{12}s_0|.
\label{Dm-expansion}
\end{equation}
Now, taking into account that propagator (\ref{interaction-new}) contains only the
unit, $\gamma^0$, $\gamma^1\gamma^2$, and $\gamma^0\gamma^1\gamma^2$
matrices [see Eqs.~(\ref{matrices-MW}) and (\ref{matrices-WM})], its expansion in
the form (\ref{Dm-expansion}) has only diagonal terms with $s_{12}^{\prime}=s_{12}$
and $s_0^{\prime}=s_0$. Therefore, we can rewrite it
as follows:
\begin{eqnarray}
G_{s}(u,u)&=& \frac{i}{4\pi l^{2}}\sum_{s_{12},s_0}
\int\limits_{-\infty}^{\infty} \frac{d\omega }{2\pi}
\sum_{n=0}^{\infty}
\frac{(\omega+\mu_{s}-\tilde{\mu}_{s}s_{12}s_0)s_0+\tilde{\Delta}_{s}+\Delta_{s}
s_{12}s_0}{(\omega+\mu_{s}-\tilde{\mu}_{s}s_{12}s_0)^2 -(\tilde{\Delta}_{s}+\Delta_{s}
s_{12}s_0)^2-2v^{2}_F|eB_{\perp}|n}
\nonumber\\
&&
\times \left\{1+s_{12}\mbox{sign}(eB_{\perp})
+\left[1-s_{12}\mbox{sign}(eB_{\perp})\right]\theta(n-1)\right\} \,
\left|s_{12}s_0\right\rangle
\left\langle s_{12}s_0\right|\, .
\label{interaction-A1}
\end{eqnarray}
The zeros of the denominator in the integrand define the dispersion relations
for the Landau levels. In the case of $n \ge 1$, they are given by
\begin{equation}
\omega_{ns}^{(\sigma)}=-\mu_{s}+\sigma\tilde{\mu}_{s} \pm
\sqrt{2v^{2}_F|eB_{\perp}|n+(\tilde{\Delta}_{s}+ \sigma\Delta_{s}})^2,
\label{energylevels-A}
\end{equation}
where {$\sigma\equiv s_{12}s_{0}$ (i.e., $\sigma=\pm1$)} and the
two signs in front of the square root correspond to the energy levels above and
below the Dirac point. The case of the LLL is special because the numerator in
the $n=0$ term in Eq.~(\ref{interaction-A1}) coincides with one of the zeros in the
denominator. After taking this into account, we find the following dispersion relation:
\begin{equation}
\omega^{(\sigma)}_{s}= -\mu_{s} + \sigma[\tilde{\mu}_{s}\,\mbox{sign}(eB_{\perp})
+\,\tilde{\Delta}_{s}] + \Delta_{s}\,\mbox{sign}(eB_{\perp}).
\label{LLLenergylevels-A}
\end{equation}
Note that the parameter $\sigma = \pm1$ in Eqs.~(\ref{energylevels-A}) and
(\ref{LLLenergylevels-A}) is connected with the eigenvalues of the diagonal
pseudospin matrix $\gamma_3\gamma_5$ in Eq.~(\ref{gamma35}). Indeed,
from the expression $\gamma^5 = i\gamma^0\gamma^1\gamma^2\gamma^3$, one gets
$\gamma^3\gamma^5 = i\gamma^0\gamma^1\gamma^2$, i.e., the eigenvalues
of $\gamma^3\gamma^5$ are $-s_0s_{12}$. It is now easy to check
for higher LLs that $\sigma = \pm1$ in Eq.~(\ref{energylevels-A}) corresponds
to the eigenvalues $\mp 1$ of $\gamma^3\gamma^5$. On the other hand,
as follows from Eq.~(\ref{Dsn-new}), $s_{12} = \mbox{sign}(eB_\perp)$
on the LLL, and we find that in this case $\sigma = \pm1 $ corresponds to
$\mbox{sign}(eB_\perp)\times \,(\mp 1)$, with $\mp 1$ being the eigenvalues
of $\gamma^3\gamma^5$.

Integrating over $\omega$ in Eq.~(\ref{interaction-A1}), we obtain
\begin{eqnarray}
G_{s}(u,u) & =& \frac{1}{8\pi l^2}\sum_{s_{12},s_0}
\sum_{n=0}^{\infty}
\Bigg(-s_0\,\mbox{sign}(\mu_{s}-\tilde{\mu}_{s}s_{12}s_0)\theta(|\mu_{s}-
\tilde{\mu}_{s}s_{12}s_0|-E_{ns}^{\sigma})
\nonumber\\
&&
+ \frac{(\tilde{\Delta}_{s}+\Delta_{s}
s_{12}s_0)\theta(E_{ns}^{\sigma}-|\mu_{s}-
\tilde{\mu}_{s}s_{12}s_0|)}{E_{ns}^{\sigma}}\Bigg)
\nonumber\\
&\times&
\left\{1+s_{12}\,\mbox{sign}(eB_{\perp}) +
\left[1-s_{12}\,\mbox{sign}(eB_{\perp})\right]
\theta(n-1)\right\}\,|s_{12}s_0\rangle \langle s_{12}s_0|,
\label{interaction-A3}
\end{eqnarray}
where $E_{ns}^{\sigma} = \sqrt{2v^{2}_F|eB_{\perp}|n + (\tilde{\Delta}_{s} +
\sigma\Delta_{s}})^2$.

Using this expression and the inverse bare and full propagators in
Eqs.~(\ref{inversebare}) and (\ref{full-inverse}),
we arrive at the following form of gap equation
(\ref{gap}):
\begin{eqnarray}
&& -\mu_{s} s_0 + \tilde{\mu}_{s}s_{12} + \Delta_{s} s_{12}s_0 + \tilde{\Delta}_{s} =
-\bar{\mu}_{s}  s_0
+ A\sum_{n=0}^{\infty} \Big[1+s_{12}\mbox{sign}(eB_{\perp}) +
\left(1-s_{12}\mbox{sign}(eB_{\perp})\right)\theta(n-1)\Big]
\nonumber\\
&&
\times
\left[-s_0\mbox{sign}(\mu_{s}-\tilde{\mu}_{s}s_{12}s_0)\theta(|\mu_{s}-
\tilde{\mu}_{s}s_{12}s_0|-E_{ns}^{\sigma})
+ \frac{(\tilde{\Delta}_{s}+\Delta
s_{12}s_0)\theta(E_{ns}^{\sigma}-
|\mu_{s}-\tilde{\mu}_{s}s_{12}s_0|)}{E_{ns}^{\sigma}}\right]
\nonumber\\
&&
- A\,s_0
\sum_{n=0}^{\infty}\,\sum_{s^{\prime}=
\pm}\,\sum_{s_{12}^{\prime},s_0^{\prime}}
\Big[1+s_{12}^{\prime}\mbox{sign}(eB_{\perp}) +
\left(1-s_{12}^{\prime}\mbox{sign}(eB_{\perp})\right)\theta(n-1)\Big]
\nonumber\\
&&
\times
\left[-s_0^{\prime}\mbox{sign}(\mu_{s^{\prime}}-
\tilde{\mu}_{s^{\prime}}s_{12}^{\prime}s_0^{\prime})
\theta(|\mu_{s^{\prime}}-\tilde{\mu}_{s^{\prime}}s_{12}^{\prime}s_0^{\prime}|-
E_{ns^{\prime}}^{\sigma})
+ \frac{(\tilde{\Delta}_{s^{\prime}}+\Delta_{s^{\prime}}
s_{12}^{\prime}s_0^{\prime})
\theta(E_{ns^{\prime}}^{\sigma}-|\mu_{s^{\prime}}
-\tilde{\mu}_{s^{\prime}}s_{12}^{\prime}s_0^{\prime}|)}
{E_{ns^{\prime}}^{\sigma}}\right],
\label{SD-eigen1}
\end{eqnarray}
where $A\equiv G_{\rm int}/(8\pi l^2)$.
The last term on the right-hand side of Eq.~(\ref{SD-eigen1})
proportional to $s_0$ is the Hartree contribution. Finally, multiplying
(\ref{SD-eigen1}) by $1,\, s_{12}s_0,\, s_{12}$, and $s_0$, respectively, and
taking the sum over $s_{12}$ and $s_0$, we obtain
Eqs.~(\ref{E1-a})--(\ref{E4-a}).

\section{Analytic Solutions of Gap Equation for LLL at $T=0$}
\label{B}

In order to solve Eqs.~(\ref{E1-a})-(\ref{E3-a}) for $\Delta_{s}$, $\tilde{\Delta}_{s}$,
$\tilde{\mu}_{s}$ as functions of $\mu_{s}$, note that these equations contain
$\theta$-functions whose arguments suggest that the following three cases
have to be considered:

\begin{enumerate}
\item
$|\mu_{s} \mp \tilde{\mu}_{s}| < |\tilde{\Delta}_{s} \pm \Delta_{s}|$;
\item
$|\mu_{s} - \tilde{\mu}_{s}| > |\tilde{\Delta}_{s} + \Delta_{s}|$,
$|\mu_{s} + \tilde{\mu}_{s}| < |\tilde{\Delta}_{s} - \Delta_{s}|$ or
$|\mu_{s} - \tilde{\mu}_{s}| < |\tilde{\Delta}_{s} + \Delta_{s}|$,
$|\mu_{s} + \tilde{\mu}_{s}| > |\tilde{\Delta}_{s} - \Delta_{s}|$;
\item
$|\mu_{s} \mp \tilde{\mu}_{s}| > |\tilde{\Delta}_{s} \pm \Delta_{s}|$.
\end{enumerate}

\subsection{The first case}
\label{B1}

For $|\mu_{s} \mp \tilde{\mu}_{s}| < |\tilde{\Delta}_{s} \pm \Delta_{s}|$, the gap
equations for Dirac masses take the form
\begin{eqnarray}
\tilde{\Delta}_{s} + \Delta_{s} &=& A\sum_{n=0}^{\infty}
\frac{\tilde{\Delta}_{s}+\Delta_{s}}{E_{ns}^+} \left[1+\theta(n-1)\right],
\label{s3}
\\
\tilde{\Delta}_{s} - \Delta_{s} &=& A\sum_{n=0}^{\infty}
\frac{\tilde{\Delta}_{s}-\Delta_{s}}{E_{ns}^-} \left[1+\theta(n-1)\right].
\label{s4}
\end{eqnarray}
Equations for $\tilde{\Delta}_{s} + \Delta_{s}$ and $\tilde{\Delta}_{s} - \Delta_{s}$ are
equivalent and since each equation admits both positive and negative solutions
with the same absolute value, we have
\begin{equation}
\tilde{\Delta}_{s} + \Delta_{s} = \pm (\tilde{\Delta}_{s} - \Delta_{s}).
\end{equation}
This implies that one of the following should be true
\begin{equation}
\mbox{(a)}\quad \Delta_{s} = 0 \, , \qquad  \mbox{or}\qquad
\mbox{(b)}\quad \tilde{\Delta}_{s} = 0\, .
\label{cases(a)and(b)}
\end{equation}
Then, the gap equation for the nonvanishing parameter $\tilde{\Delta}_{s}$
(or $\Delta_{s}$) takes the form
\begin{equation}
\tilde{\Delta}_{s} = A\sum_{n=0}^{\infty}
\frac{\tilde{\Delta}_{s}}{\sqrt{n\epsilon_{B}^2 + \tilde{\Delta}_{s}^2}}
\left[1+\theta(n-1)\right].
\label{gapequation}
\end{equation}
Let us first consider the case (a) and show that Eq.~(\ref{gapequation}) can be
equivalently represented in the following integral form:
\begin{equation}
\tilde{\Delta}_{s} = \frac{A\,\tilde{\Delta}_{s}}{\sqrt{\pi}}
\int_{1/\Lambda^2}^{\infty} \frac{dy}{\sqrt{y}} e^{-y\tilde{\Delta}_{s}^2}\,
\coth\left(\frac{\epsilon_{B}^2}{2}y\right),
\label{integral-form}
\end{equation}
where $\Lambda$ is a high energy cut-off up to which the low-energy
effective theory is valid. After taking into account the identity
$\mbox{coth}(\epsilon_{B}^2y/2) = 1 + 2\sum_{n=1}^{\infty} e^{-yn\epsilon_{B}^2}$,
we can integrate over $y$ in Eq.~(\ref{integral-form}) by using the
following table integral:
\begin{equation}
\int_{1/\Lambda^2}^{\infty} \frac{dy}{\sqrt{y}}
e^{-y\left(n\epsilon_{B}^2+\tilde{\Delta}_{s}^2\right)}
\simeq
\int_{0}^{\infty} \frac{dy}{\sqrt{y}}
e^{-y\left(n\epsilon_{B}^2+\tilde{\Delta}_{s}^2\right)}
=\frac{\sqrt{\pi}}{\sqrt{n\epsilon_{B}^2+\tilde{\Delta}_{s}^2}},
\end{equation}
where we replaced the lower limit of integration by $0$ because the integral is
convergent for $y \to 0$. Therefore, up to corrections suppressed by the inverse
powers of cutoff $\Lambda$, Eq.~(\ref{integral-form}) is indeed equivalent to
Eq.~(\ref{gapequation}). Then, by using the same approach as in the second
paper in Ref.~\onlinecite{Gusynin1995PRD}, we expand the result on the right hand
side of Eq.~(\ref{integral-form}) in powers of $1/\Lambda$ and arrive at the
following form of the gap equation:
\begin{equation}
\tilde{\Delta}_{s} = \lambda\tilde{\Delta}_{s}+A+\frac{2A\tilde{\Delta}_{s}}{\epsilon_{B}}
\zeta\left(\frac{1}{2},1+\frac{\tilde{\Delta}_{s}^2}{\epsilon_{B}^2}\right)
+ O\left(\lambda\frac{\tilde{\Delta}_{s}^2}{\Lambda^2}\right),
\end{equation}
where $\lambda\equiv 4A\Lambda/(\sqrt{\pi}\epsilon_{B}^2)
=G_{\rm int}\Lambda/(4\pi^{3/2}\hbar^2v_F^2)$ is the dimensionless
coupling constant and $\zeta(z,q)$ is the generalized Riemann zeta
function.\cite{GR} By assuming that the gap $\tilde{\Delta}_{s}$ is much
smaller than the Landau energy scale $\epsilon_{B}$, we find the solution
in an analytical form,
\begin{equation}
\tilde{\Delta}_{s} = M\equiv \frac{A}{1-\lambda},
\qquad
\tilde{\mu}_{s} = A\, s_{\perp} .
\label{gap-solution-1}
\end{equation}
Here, in order to get the result for the chemical potential $\tilde{\mu}_{s}$ we used
Eq.~(\ref{E3-a}). It is easy to check that the gap equation also has another solution, which
is obtained from Eq.~(\ref{gap-solution-1}) by replacing $\tilde{\Delta}_{s}$ and $\tilde{\mu}_{s}$
with $-\tilde{\Delta}_{s}$ and $-\tilde{\mu}_{s}$. However, the second solution is
equivalent to that in Eq.~(\ref{gap-solution-1}): one can see this from dispersion
relations (\ref{higherLLs}), (\ref{LLLenergylevels}) by transforming $\sigma \to -\sigma$
there. In other words, these solutions describe two degenerate ground states
connected by a $Z_{2s}$ [$\subset SU(2)_{s}$] symmetry transformation.

Turning to case (b) in Eq.~(\ref{cases(a)and(b)}), we have $\tilde{\Delta}_{s}=0$,
\begin{equation}
\Delta_{s} = \pm M
\qquad\mbox{and}\qquad
\tilde{\mu}_{s} = 0,
\label{gap-solution-2}
\end{equation}
where the last relation follows from Eq.~(\ref{E3-a}).

Finally, we would like to note that by analyzing the inequalities $|\mu_{s} \mp
\tilde{\mu}_{s}| < |\tilde{\Delta}_{s} \pm \Delta_{s}|$, one can show that solution
(\ref{gap-solution-1}) with a triplet Dirac mass exists for
\begin{equation}
|\mu_{s}| < M-A
\label{unfilled-1}
\end{equation}
and solution (\ref{gap-solution-2}) with a singlet Dirac mass exists for
\begin{equation}
|\mu_{s}| < M.
\label{unfilled-2}
\end{equation}

\subsection{The second case}
\label{B2}

There are two possibilities
$|\mu_{s} + \tilde{\mu}_{s}| < |\tilde{\Delta}_{s} -\Delta_{s}|$,
$|\mu_{s} - \tilde{\mu}_{s}| > |\tilde{\Delta}_{s} + \Delta_{s}|$ or
$|\mu_{s} - \tilde{\mu}_{s}| < |\tilde{\Delta}_{s} + \Delta_{s}|$,
$|\mu_{s} + \tilde{\mu}_{s}| > |\tilde{\Delta}_{s} - \Delta_{s}|$.
In the first case, the equations for Dirac masses take the
form
\begin{eqnarray}
\tilde{\Delta}_{s} + \Delta_{s} &=&  -As_{\perp} \mbox{sign}(\mu_{s}-\tilde{\mu}_{s}) +
2A\sum_{n=1}^{\infty}\frac{\tilde{\Delta}_{s}+\Delta_{s}}{E_{ns}^+}, \label{pf-1}
\\
\tilde{\Delta}_{s} - \Delta_{s} &=&  A\sum_{n=0}^{\infty}
\frac{\tilde{\Delta}_{s}-\Delta_{s}}{E_{ns}^-} \left[1+\theta(n-1)\right],
\label{pf-2}
\end{eqnarray}
where $s_{\perp} \equiv {\rm sgn} (eB_{\perp})$.
While the equation for $\tilde{\Delta}_{s}-\Delta_{s}$ coincides with Eq.~(\ref{s4}),
the equation for $\tilde{\Delta}_{s}+\Delta_{s}$ is slightly different from its
counterpart in Eq.~(\ref{s3}). Unlike Eq.~(\ref{s3}), the above equation for
$\tilde{\Delta}_{s}+\Delta_{s}$ does not contain the sign factor $\mbox{sign}
(\tilde{\Delta}_{s}+\Delta_{s})$ in the LLL contribution. The absence of such a factor
in Eq.~(\ref{pf-1}) means that the sign of the LLL contribution is fixed for a
given set of values of $\mu_{s}$, $\tilde{\mu}_{s}$, and $eB_{\perp}$. In turn,
this implies that Eq.~(\ref{pf-1}) [unlike the gap equation (\ref{s3})] has only
one solution whose sign is correlated with the sign of the LLL contribution.
In order to prove this, let us consider the following equation:
\begin{equation}
x =  -A + 2A\sum_{n=1}^{\infty} \frac{x}{\sqrt{n\epsilon_{B}^2+x^2}}.
\label{auxiliary1}
\end{equation}
By taking $x$ negative, we see that its absolute value $|x|$ satisfies an
equation that is equivalent to the equation for positive $\tilde{\Delta}_{s} $ that
follows from Eq.~(\ref{gapequation}). Therefore, the solution for $|x|$
coincides with the positive solution for $\tilde{\Delta}_{s}$ in (\ref{gap-solution-1}).
We can also show that Eq.~(\ref{auxiliary1}) does not have a solution for
positive $x$ by using the integral form of (\ref{auxiliary1}), i.e.,
\begin{equation}
1 = \frac{A}{\sqrt{\pi}} \int_{1/\Lambda^2}^{\infty} \frac{dy}{\sqrt{y}}
e^{-yx^2}\,\left[\coth\left(\frac{\epsilon_{B}^2}{2}y\right)-2\right],
\label{integral-form-1}
\end{equation}
where the term $-2$ is subtracted in order to get the negative LLL contribution
as in (\ref{auxiliary1}) [cf. Eq.~(\ref{integral-form})].

In order to prove that Eq.~(\ref{integral-form-1}) does not have solution, we
will use the fact that Eq.~(\ref{integral-form}) does not have a nontrivial solution
for $B_{\perp} \to 0$ in the case when the coupling constant $G_{\rm int}$ is
subcritical, i.e., $G_{\rm int} < {4\pi^{3/2}v_F^2\hbar^2}/{\Lambda}$, or
equivalently $\lambda < 1$. Note that the coupling constant should indeed
be subcritical because, as we know from experiment, there is no gap
generation at $B_{\perp}=0$. It is not difficult to prove that the right hand
side of Eq.~(\ref{integral-form-1}) is less than $\lambda$ after taking into
account that $t(\coth t - 2) < 1$ for $t>0$. Then we conclude that
Eq.~(\ref{integral-form-1}) does not have a solution for a subcritical
coupling constant $\lambda<1$. As for Eq.~(\ref{auxiliary1}), it has only one
solution which, in fact, coincides with the solution for $\tilde{\Delta}_{s} $ in
(\ref{gap-solution-1}) times $-1$. Thus, the solutions of Eqs.~(\ref{pf-1}) and (\ref{pf-2})
are
\begin{equation}
\tilde{\Delta}_{s} + \Delta_{s} = -\mbox{sign}(\mu_{s}-\tilde{\mu}_{s})\,
s_{\perp} M
\label{pf-1-solution}
\end{equation}
and
\begin{equation}
\tilde{\Delta}_{s} - \Delta_{s} = \pm (\tilde{\Delta}_{s} + \Delta_{s}).
\label{pf-2-solution}
\end{equation}
{From} the fact that the solutions for $\tilde{\Delta}_{s} + \Delta_{s}$ and
$\tilde{\Delta}_{s} - \Delta_{s}$ have the same absolute value, we conclude
that either $\tilde{\Delta}_{s} \ne 0$, $\Delta_{s}=0$ or vice versa
$\Delta_{s} \ne 0$, $\tilde{\Delta}_{s}=0$ depending on the sign in
Eq.~(\ref{pf-2-solution}). If Eq.~(\ref{integral-form-1}) had solution, there
would exist solutions with both nonzero $\tilde{\Delta}_{s}$ and $\Delta_{s}$.

Further, solution of Eq.~(\ref{E3-a}) for $\tilde{\mu}_{s}$ in the case under
consideration  takes the form
\begin{equation}
\tilde{\mu}_{s} = \frac{A}{2} [-\mbox{sign}(\mu_{s}-\tilde{\mu}_{s}) +
\mbox{sign}(\tilde{\Delta}_{s}-\Delta_{s})\,s_{\perp}]. \label{E3-a-2}
\end{equation}
Using (\ref{pf-1-solution}), it is easy to check that for the plus sign in
(\ref{pf-2-solution}) (when $\tilde{\Delta}_{s} \ne 0$ and $\Delta_{s}=0$)
$\tilde{\mu}_{s}=A\,\mbox{sign}(\tilde{\Delta}_{s})\,s_{\perp}$ and for the sign
minus in (\ref{pf-2-solution}) (when $\tilde{\Delta}_{s}=0$ and $\Delta_{s} \ne 0$)
$\tilde{\mu}_{s}=0$. In the latter case, the assumed inequalities $|\mu_{s} +
\tilde{\mu}_{s}| < |\tilde{\Delta}_{s} - \Delta_{s}|$ and $|\mu_{s} - \tilde{\mu}_{s}| >
|\tilde{\Delta}_{s} + \Delta_{s}|$ cannot be satisfied, therefore, only solution
with triplet Dirac mass $\tilde{\Delta}_{s}$ is realized
\begin{equation}
\tilde{\Delta}_{s} =
-s_{\perp}\,\mbox{sign}{(\mu_{s}-\tilde{\mu}_{s})}\,M,
\qquad
\Delta_{s}=0,
\qquad
\tilde{\mu}_{s}=A\,\mbox{sign}(\tilde{\Delta}_{s})\,s_{\perp}.
\label{p-f-s-1}
\end{equation}
In the other case $|\mu_{s} + \tilde{\mu}_{s}| > |\tilde{\Delta}_{s} -\Delta_{s}|$ and
$|\mu_{s} - \tilde{\mu}_{s}| < |\tilde{\Delta}_{s} + \Delta_{s}|$, we find the
following solution:
\begin{equation}
\tilde{\Delta}_{s} =s_{\perp}\,\mbox{sign}{(\mu_{s}+\tilde{\mu}_{s})}\, M ,
\qquad
\Delta_{s}=0,
\qquad
\tilde{\mu}_{s}=A\,\mbox{sign}(\tilde{\Delta}_{s})\,s_{\perp}.
\label{p-f-s-2}
\end{equation}
One can show that it is possible to join solutions (\ref{p-f-s-1}) and
(\ref{p-f-s-2}) into one solution with triplet Dirac mass
\begin{equation}
\tilde{\Delta}_{s} = M,
\qquad
\tilde{\mu}_{s} =A\,s_{\perp}, \qquad \Delta_{s}=0,
\label{second-unique}
\end{equation}
which exists for
\begin{equation}
M-A < |\mu_{s}| < M+A.
\end{equation}
In fact, like in the previous subsection, there is another solution, with
$\tilde{\Delta}_{s}$, $\tilde{\mu}_{s}$ replaced by $-\tilde{\Delta}_{s}$,
$-\tilde{\mu}_{s}$. However, such a solution is equivalent to solution
(\ref{second-unique}) by a $SU(2)_{s}$ (or $Z_{2s}$) symmetry transformation.

\subsection{The third case}
\label{B3}

For $|\mu_{s} \pm \tilde{\mu}_{s}| > |\tilde{\Delta}_{s} \mp \Delta_{s}|$, the
equations for Dirac masses take the form
\begin{eqnarray}
\tilde{\Delta}_{s} + \Delta_{s} &=&  -A\,\mbox{sign}(\mu_{s} - \tilde{\mu}_{s})\,
s_{\perp} + 2A\sum_{n=1}^{\infty} \frac{\tilde{\Delta}_{s}+\Delta_{s}}{E_{ns}^+},
\label{Bf1}
\\
\tilde{\Delta}_{s} - \Delta_{s} &=&  A\,\mbox{sign}(\mu_{s} + \tilde{\mu}_{s})\,
s_{\perp} + 2A\sum_{n=1}^{\infty} \frac{\tilde{\Delta}_{s}-\Delta_{s}}{E_{ns}^-}.
\label{Bf2}
\end{eqnarray}
Solutions of Eqs.~(\ref{Bf1}) and (\ref{Bf2}) are
\begin{eqnarray}
\tilde{\Delta}_{s} + \Delta_{s} &=& -s_{\perp}\,\mbox{sign}(\mu_{s}-\tilde{\mu}_{s})\,M,
\label{Bf1-solution}
\\
\tilde{\Delta}_{s} - \Delta_{s} &=&
s_{\perp}\,\mbox{sign}(\mu_{s}+\tilde{\mu}_{s})\,M.
\label{Bf2-solution}
\end{eqnarray}
Using these solutions and taking into account the inequalities
$|\mu_{s} \pm \tilde{\mu}_{s}| > |\tilde{\Delta}_{s} \mp \Delta_{s}|$, one
can check that Eq.~(\ref{E3-a}) has only the trivial solution.
Then it follows from Eqs.~(\ref{Bf1-solution}) and
(\ref{Bf2-solution}) that
\begin{equation}
\Delta_{s} = -s_{\perp}\, \mbox{sign}(\mu_{s})\, M,
\qquad
\tilde{\Delta}_{s}=\tilde{\mu}_{s}=0.
\label{Bf12-solution}
\end{equation}
Taking into account the assumed inequalities $|\mu_{s} \pm \tilde{\mu}_{s}| >
|\tilde{\Delta}_{s} \mp \Delta_{s}|$, we find that this solution with singlet Dirac
mass exists for
\begin{equation}
|\mu_{s}| >M.
\label{filled}
\end{equation}

\subsection{Final solutions for $\Delta_{s}$, $\tilde{\Delta}_{s}$,
and $\tilde{\mu}_{s}$ as functions of $\mu_{s}$}
\label{B4}

Using the results derived above and calculating the quantity $X_{s}$ in
Eq.~(\ref{Xs}) (which is needed for solving the equation for $\mu_{s}$),
we obtain the following three distinct solutions.

\begin{itemize}
\item Solution I (triplet Dirac mass).
By joining the two solutions of the same type in Eqs.~(\ref{gap-solution-1})
and (\ref{second-unique}), considered in Subsecs.~\ref{B1} and ~\ref{B2},
respectively, we arrive at the following solution:
\begin{equation}
\tilde{\Delta}_{s} = M,
\qquad
\tilde{\mu}_{s} = A\,s_{\perp},
\qquad
\Delta_{s}=0,
\qquad
X_{s}=0
\label{I}
\end{equation}
which exists over the combined range of validity $|\mu_{s}| < M+A$. Let us mention
that there is also another solution, in which $\tilde{\Delta}_{s}$ and $\tilde{\mu}_{s}$
are replaced by $-\tilde{\Delta}_{s}$ and $-\tilde{\mu}_{s}$, respectively. However,
this second solution is equivalent to that in Eq.~(\ref{I}): one can see this from
dispersion relations (\ref{higherLLs}), (\ref{LLLenergylevels}) by transforming
$\sigma \to -\sigma$ there. In other words, the two solutions are related to two
degenerate ground states connected by a $SU(2)_{s}$ (or $Z_{2s}$) flavor
transformation.

\item Solution II (singlet Dirac mass).
This is one of the two solutions in Eq.~(\ref{gap-solution-2}) from Subsec.~\ref{B1}
that corresponds to a particular choice of the sign for the singlet Dirac mass,
\begin{equation}
\Delta_{s} = s_{\perp}\, \mbox{sign}(\mu_{s})\, M, \qquad
\tilde{\Delta}_{s}=\tilde{\mu}_{s}=0,\qquad  X_{s}=4A\,\mbox{sign}(\mu_{s}).
\label{II}
\end{equation}
It exists for $|\mu_{s}| < M$.

\item Solution III (singlet Dirac mass).
This combines the remaining solution in Eq.~(\ref{gap-solution-2}) from
Subsec.~\ref{B1} with solution (\ref{Bf12-solution}) in Subsec.~\ref{B3}
to give
\begin{equation}
\Delta_{s} = -s_{\perp}\, \mbox{sign}(\mu_{s})\, M,
\qquad
\tilde{\Delta}_{s}=\tilde{\mu}_{s}=0,
\qquad
X_{s} =-4A\,\mbox{sign}(\mu_{s}).
\label{III}
\end{equation}
This solution exists for {\it all} values of $\mu_{s}$.
\end{itemize}
A noticeable point is that unlike the case with a triplet Dirac mass,
the solutions II and III, with a different sign for a singlet Dirac mass,
are different. This in particular can be seen from dispersion relation
(\ref{LLLenergylevels}). This feature is directly connected with the fact
that while the triplet mass is even under time reversal $\cal{T}$, the
singlet mass is $\cal{T}$-odd. The latter is in turn connected with the fact that
$\Delta \propto s_{\perp} = \mbox{sign}(B_{\perp})$ (recall that a magnetic
field is also $\cal{T}$-odd).

Let us also emphasize that the expressions for Dirac masses in solutions
I, II, and III are valid only for $\lambda < 1$: in the supercritical regime,
with $\lambda > 1$, a Dirac mass $\tilde{\Delta}$ is generated even with
no magnetic field.\cite{Gusynin1995PRD} Experiments clearly show that
the subcritical regime, with $\lambda < 1$, takes place in
graphene.\cite{Geim2005Nature,Kim2005Nature}  As argued
in Sec.~\ref{4} in the main text, realistic values for $\lambda$ in this model
are $\lambda \lesssim 0.2$.

\subsection{Including both spin up and spin down states}
\label{B5}

In the previous subsection, the solutions for masses and chemical
potentials were found for a fixed spin, treating the electron chemical
potential $\mu_{s}$ as a free parameter. Here we will describe full
solutions, including both spin up and spin down states. For this
purpose, we need to solve Eq.~(\ref{E4-a}) for the chemical
potentials $\mu_{\pm}$. Since the $X$ term in that equation
contains both spin up and spin down contributions,
the equations for $\mu_{+}$ and $\mu_{-}$
are now coupled and have to be solved together. As a result,
the full chemical potentials $\mu_{\pm}$ will be expressed through
the bare electron chemical potentials $\bar{\mu}_{\pm} = \mu_0 \mp Z$.

At a fixed spin, there are 3 different types of solutions for masses and
$\tilde{\mu}_{s}$ described in Subsec.~\ref{B4}. Since we can choose any of them
for each spin, there are nine possible types and, therefore, nine systems of
coupled equations for $\mu_{+}$ and $\mu_{-}$. Fortunately, noting that the
solutions for the types II-I, III-I, and III-II can be obtained from those for
I-II, I-III, and II-III by interchanging the spin subscripts $+$ and $-$ in the
latter, this number can be reduced to six coupled systems. We will analyze them
below case by case.

It will be convenient to separate these systems of equations into 3 groups. The
first group includes one system, I-I. This is the simplest case with triplet
masses $\tilde{\Delta}_{\pm}$ for both spins, when the Hartree diagram
does not contribute in the equations for $\mu_{\pm}$.
The second group consists of hybrid systems I-II and I-III, where while
fields with spin up have a triplet mass $\tilde{\Delta}_{+}$, the fields
with spin down have a singlet mass $\Delta_{-}$. The third group,
II-II, II-III, and III-III, consists of solutions with singlet masses
$\Delta_{\pm}$ only.

In the analysis, it will be assumed that the Zeeman energy $Z < A$. As argued
in Sec.~\ref{4}, this assumption is valid for magnetic fields $|B_\perp| \lesssim 45T$
used in experiments.\cite{Zhang2006,Jiang2007}

\begin{itemize}
\item
{\sl The first group: Triplet Dirac masses}

    \begin{itemize}
    \item[$\circ$] I-I.
In this simplest case, using Eq.~(\ref{I}),
we immediately find from Eq.~(\ref{E4-a}) that $\mu_{\pm} = \bar{\mu}_{\pm}$
and the solution is:
\begin{eqnarray}
\begin{split}
&
\tilde{\Delta}_{+} = M,
\qquad
\tilde{\mu}_{+} = A\,s_{\perp}\,,
\qquad
\mu_{+} = \bar{\mu}_{+},
\qquad
\Delta_{+}=0,
\\
&
\tilde{\Delta}_{-} = M,
\qquad
\tilde{\mu}_{-} = A\,s_{\perp}\,,
\qquad
\mu_{-} = \bar{\mu}_{-},
\qquad
\Delta_{-}=0.
\end{split}
\label{1stgroup}
\end{eqnarray}
It exists for
\begin{equation}
|\bar{\mu}_{+}| < A + M ,
\qquad
|\bar{\mu}_{-}| < A + M .
\end{equation}
    \end{itemize}

{The physical meaning of these constraints is clear: they imply that the LLL is neither
completely filled nor empty.}

\item
{\sl The second group: Hybrid solutions}
    \begin{itemize}
    \item[$\circ$] I-II.
By using Eqs.~(\ref{I}) and (\ref{II}), we analyze the system
of two equations
(\ref{E4-a}) for $\mu_{+}$ and $\mu_{-}$ and find that the
solution
\begin{eqnarray}
\begin{split}
&
\tilde{\Delta}_{+} = M,
\qquad
\tilde{\mu}_{+}= A\,s_{\perp}\,,
\qquad
\mu_{+} = \bar{\mu}_{+} - 4A\, \mbox{sign}(\bar{\mu}_{+}),
\qquad
\Delta_{+}=0,
\\
&
\tilde{\Delta}_{-}=\tilde{\mu}_{-}=0,
\qquad
\mu_{-} = \bar{\mu}_{-} - 3A\, \mbox{sign}(\bar{\mu}_{-}),
\qquad
\Delta_{-} = -s_{\perp}\,
\mbox{sign}(\bar{\mu}_{-})\, M
\label{I-II}
\end{split}
\end{eqnarray}
exists for
\begin{equation}
3A - M < |\bar{\mu}_{+}| < 5A +M,
\qquad
3A -M < |\bar{\mu}_{-}| < 3A,
\qquad
\mbox{sign}(\bar{\mu}_{+})\,\mbox{sign}(\bar{\mu}_{-})> 0.
\end{equation}

    \item[$\circ$] I-III.
In this case, using Eqs.~(\ref{E4-a}), (\ref{I}),
and (\ref{III}), we find the solution
\begin{eqnarray}
\begin{split}
&
\tilde{\Delta}_{+} = M,
\qquad
\tilde{\mu}_{+}= A\,s_{\perp}\,,
\qquad  \mu_{+} = \bar{\mu}_{+} - 4A\, \mbox{sign}(\bar{\mu}_{+}),
\qquad
\Delta_{+}=0,
\\
&
\tilde{\Delta}_{-}=\tilde{\mu}_{-}=0,
\qquad
\mu_{-} = \bar{\mu}_{-} - 3A\,
\mbox{sign}(\bar{\mu}_{-}),
\qquad
\Delta_{-} = -s_{\perp}\,
\mbox{sign}(\bar{\mu}_{-})\,M,
\end{split}
\label{I-III}
\end{eqnarray}
which exists for
\begin{equation}
3A - M < |\bar{\mu}_{+}| < 5A +M,
\qquad
|\bar{\mu}_{-}| > 3A,
\qquad
\mbox{sign}(\bar{\mu}_{+}) \mbox{sign}(\bar{\mu}_{-}) > 0.
\end{equation}
    \end{itemize}

\item
{\sl The third group: Singlet Dirac masses}

    \begin{itemize}
    \item[$\circ$] II-II.
Using Eq.~(\ref{II}) and analyzing equations (\ref{E4-a}) for
$\mu_{+}$ and $\mu_{-}$, we find the solution
\begin{eqnarray}
\begin{split}
&
\tilde{\Delta}_{+}=\tilde{\mu}_{+}=0,
\qquad
\mu_{+} = \bar{\mu}_{+} - 7A\, \mbox{sign}(\bar{\mu}_{+}),
\qquad
\Delta_{+}=-s_{\perp}\, \mbox{sign}(\bar{\mu}_{+})M,
\\
&
\tilde{\Delta}_{-}=\tilde{\mu}_{-}=0,
\qquad
\mu_{-} = \bar{\mu}_{-} - 7A\, \mbox{sign}(\bar{\mu}_{-}),
\qquad
\Delta_{-}=-s_{\perp}\, \mbox{sign}(\bar{\mu}_{-})M,
\end{split}
\label{II-II-1}
\end{eqnarray}
which exists for
\begin{equation}
7A - M < |\bar{\mu}_{+}| < 7A,
\qquad
7A -M < |\bar{\mu}_{-}| < 7A ,
\qquad
\mbox{sign}(\bar{\mu}_{+}) \mbox{sign}(\bar{\mu}_{-}) > 0.
\end{equation}
[Formally, there is also another solution,
\begin{eqnarray}
\begin{split}
&
\tilde{\Delta}_{+}=\tilde{\mu}_{+}=0,
\qquad
\mu_{+} = \bar{\mu}_{+} -
A\, \mbox{sign}(\bar{\mu}_{+}),
\qquad
\Delta_{+}=s_{\perp}\, \mbox{sign}(\bar{\mu}_{+})\, M,
\\
&
\tilde{\Delta}_{-}=\tilde{\mu}_{-}=0,
\qquad
\mu_{-} = \bar{\mu}_{-} -
A\, \mbox{sign}(\bar{\mu}_{-}),
\qquad
\Delta_{-}=s_{\perp}\, \mbox{sign}(\bar{\mu}_{-})\, M ,
\end{split}
\label{II-II-2}
\end{eqnarray}
which exists for
\begin{equation}
A < |\bar{\mu}_{+}| < A+M,
\qquad
A < |\bar{\mu}_{-}| < A+M,
\qquad
\mbox{sign}(\bar{\mu}_{+}) \mbox{sign}(\bar{\mu}_{-}) < 0.
\end{equation}
However, because of the latter inequalities, it is easy to check that
this solution does not satisfy the condition $Z < A$ and therefore is not
realized for magnetic fields $|B_\perp| \lesssim 45T$.]

\item[$\circ$] II-III.
As in the previous case, there are two solutions.
The first solution, II-III-1,
\begin{eqnarray}
\begin{split}
&
\tilde{\Delta}_{+}=\tilde{\mu}_{+}=0,
\qquad
\mu_{+} = \bar{\mu}_{+} - A\,\mbox{sign}(\bar{\mu}_{+}),
\qquad
\Delta_{+}=s_{\perp}\, \mbox{sign}(\bar{\mu}_{+}) \, M,
\\
&
\tilde{\Delta}_{-}=\tilde{\mu}_{-}=0,
\qquad
\mu_{-} = \bar{\mu}_{-} + A\, \mbox{sign}(\bar{\mu}_{+}),
\qquad
\Delta_{-}=-s_{\perp}\, \mbox{sign}(\bar{\mu}_{+})\, M
\end{split}
\label{II-III-1}
\end{eqnarray}
exists for
\begin{equation}
A < |\bar{\mu}_{+}| <A+M,
\qquad
\bar{\mu}_{-}\,
\mbox{sign}(\bar{\mu}_{+})> -A.
\end{equation}

The second solution, II-III-2,
\begin{eqnarray}
\begin{split}
&
\tilde{\Delta}_{+}=\tilde{\mu}_{+}=0,
\qquad
\mu_{+} = \bar{\mu}_{+} - 7A\,\mbox{sign}(\bar{\mu}_{+}),
\qquad
\Delta_{+}=-s_{\perp}\, \mbox{sign}(\bar{\mu}_{+}) \, M,
\\
&
\tilde{\Delta}_{-}=\tilde{\mu}_{-}=0,
\qquad
\mu_{-} = \bar{\mu}_{-} - 7A\, \mbox{sign}(\bar{\mu}_{-}),
\qquad
\Delta_{-}=-s_{\perp}\, \mbox{sign}(\bar{\mu}_{-}) \, M
\end{split}
\label{II-III-2}
\end{eqnarray}
exists for
\begin{equation}
7A - M < |\bar{\mu}_{+}| < 7A,
\qquad
\bar{\mu}_{-}\,
\mbox{sign}(\bar{\mu}_{+}) > 7A\,.
\end{equation}

\item[$\circ$] III-III.
There are three solutions in this case.
The first solution, III-III-1,
\begin{eqnarray}
\begin{split}
&
\tilde{\Delta}_{+}=\tilde{\mu}_{+}=0,
\qquad
\mu_{+} = \bar{\mu}_{+} -7A\,\mbox{sign}(\bar{\mu}_{+}),
\qquad
\Delta_{+}=-s_{\perp}\, \mbox{sign}(\bar{\mu}_{+})\, M,
\\
&
\tilde{\Delta}_{-}=\tilde{\mu}_{-}=0,
\qquad
\mu_{-} = \bar{\mu}_{-} -
7A\, \mbox{sign}(\bar{\mu}_{-}),
\qquad
\Delta_{-}=-s_{\perp}\, \mbox{sign}(\bar{\mu}_{-})\, M
\end{split}
\label{III-III-1}
\end{eqnarray}
exists for
\begin{equation}
|\bar{\mu}_{+}| > 7A,
\qquad
|\bar{\mu}_{-}| > 7A,
\qquad
\mbox{sign}(\bar{\mu}_{+})\, \mbox{sign}(\bar{\mu}_{-}) > 0.
\end{equation}

The second solution, III-III-2, is
\begin{eqnarray}
\begin{split}
&
\tilde{\Delta}_{+}=\tilde{\mu}_{+}=0,
\qquad
\mu_{+} = \bar{\mu}_{+} +A,
\qquad
\Delta_{+}=-s_{\perp}\,M,
\\
&
\tilde{\Delta}_{-}=\tilde{\mu}_{-}=0,
\qquad
\mu_{-} = \bar{\mu}_{-} -A,
\qquad
\Delta_{-}=s_{\perp}\,M.
\end{split}
\label{III-III-2}
\end{eqnarray}
It is realized for
\begin{equation}
\bar{\mu}_{+} > -A,
\qquad
\bar{\mu}_{-} < A\,.
\end{equation}

The third solution, III-III-3,
\begin{eqnarray}
\begin{split}
&
\tilde{\Delta}_{+}=\tilde{\mu}_{+}=0,
\qquad
\mu_{+} = \bar{\mu}_{+} -A,
\qquad
\Delta_{+}=s_{\perp}\, M,
\\
&
\tilde{\Delta}_{-}=\tilde{\mu}_{-}=0,
\qquad
\mu_{-} = \bar{\mu}_{-} +A,
\qquad
\Delta_{-}=-s_{\perp}\,M
\end{split}
\label{III-III-3}
\end{eqnarray}
takes place for
\begin{equation}
\bar{\mu}_{+} < A,
\qquad
\bar{\mu}_{-} > -A\,,
\end{equation}
i.e., in fact, it is obtained from the second solution by interchanging
spin subscripts $+$ and $-$.
    \end{itemize}
\end{itemize}

\subsection{Dependence of solutions on electron chemical potential
$\mu_0$ and free energy density of their ground states}
\label{B6}

The process of filling LLs is described by varying the electron chemical
potential $\mu_0$. Therefore, it will be convenient to express the intervals of
the existence of the solutions found in the previous subsection in terms of
$\mu_0$. Henceforth we will consider $\mu_0 \geq 0$. (Dynamics with negative
$\mu_0$ is related by electron-hole symmetry and will not be discussed
separately.)

Some intermediate results of the analysis in this subsection will depend on
whether the inequality $M > 2Z$ or $M < 2Z$ is satisfied. We will consider both
these cases and indicate explicitly which inequality is satisfied for a particular
solution. If nothing will be said, this means that the corresponding results are
valid in both cases. Fortunately, the final results do not depend on whether
$M > 2Z$ or $M < 2Z$.

\begin{table*}
\caption{
\label{tab:intervals}
Intervals of the existence of solutions, relevant for the dynamics in the LLL at $T=0$.}
\begin{ruledtabular}
\begin{tabular}{lll}
 & $M>2Z$  & $M<2Z$ \\[1mm]
\hline\hline
I-I  &  {$ 0 \le \mu_0 < M + A - Z $}  & {$ 0 \le \mu_0 < M + A - Z $} \\[1mm]
\hline
I-II & $ M + A + Z < \mu_0 < 3A-Z $ & no solution\\[1mm]
\hline
II-I  & {$3A-M+Z < \mu_0 < 3A+Z $}  & {$3A-M+Z < \mu_0 < 3A+Z $}\\[1mm]
\hline
I-III & $ 3A-Z < \mu_0 < 5A+M+Z $ & $ 3A-M+Z < \mu_0 < 5A+M+Z $ \\[1mm]
\hline
III-I  &  {$ 3A+Z < \mu_0 < 5A+M-Z $} & {$ 3A+Z < \mu_0 < 5A+M-Z $}\\[1mm]
\hline
II-II & $ 7A-M+Z < \mu_0 < 7A-Z $ & no solution \\[1mm]
\hline
II-III-1  &   {$ A+Z < \mu_0 < M+A+Z $} & {$ A+Z < \mu_0 < M+A+Z $}\\[1mm]
\hline
II-III-2 & $ 7A-Z < \mu_0 < 7A+Z $ & $ 7A-M+Z < \mu_0 < 7A+Z $ \\[1mm]
\hline
III-II-1 &  {$ A-Z < \mu_0 < M+A-Z $}    & {$ A-Z < \mu_0 < M+A-Z $} \\[1mm]
\hline
III-II-2 &  {no solution}  & {no solution}\\[1mm]
\hline
III-III-1 &  {$ 7A+Z < \mu_0 $}  & {$ 7A+Z < \mu_0 $}\\[1mm]
\hline
III-III-2 &  {$ 0 < \mu_0 < A-Z $}  & {$ 0 < \mu_0 < A-Z $}\\[1mm]
\hline
III-III-3 &  {$ 0 < \mu_0 < A+Z $}  & {$ 0 < \mu_0 < A+Z $}\\[1mm]
\end{tabular}
\end{ruledtabular}
\end{table*}

Taking into account that $\bar{\mu}_{\pm} = \mu_0 \mp Z$, we find the intervals
of existence for solutions. These are given in Table~\ref{tab:intervals}. Using
this information, we see that some solutions may coexist. The list of coexisting
solutions for a set of non-overlapping intervals of $\mu_0$ is summarized in
Table~\ref{tab:mu0intervals}. [We assume that $Z > M-A\equiv {A}\lambda/(1-\lambda)$
which is likely to be satisfied because, as will be shown in Sec.~\ref{4}, realistic
values for $\lambda$ in this model are relatively small, $\lambda\lesssim 0.2$.]

\begin{table*}
\caption{
\label{tab:mu0intervals}
The list of solutions that coexist in a set of non-overlapping intervals
of $\mu_0$, relevant for the dynamics in the LLL at $T=0$. The solutions
with the lowest free energy density are marked by stars.}
\begin{ruledtabular}
\begin{tabular}{llll}
\# &Interval & $M>2Z$  & $M<2Z$ \\[1mm]
\hline
\hline
1& $0 \le \mu_0 < A-Z$  & I-I, III-III-2, III-III-3$^{\star}$ &  I-I, III-III-2, III-III-3$^{\star}$ \\[1mm]
\hline
2& $A-Z < \mu_0 < A+Z$ & I-I, III-II, III-III-3$^{\star}$ & I-I, III-II, III-III-3$^{\star}$  \\[1mm]
\hline
3& $A+Z < \mu_0 < M+A-Z$ & I-I, III-II, II-III-1$^{\star}$  & I-I, III-II, II-III-1$^{\star}$ \\[1mm]
\hline
4& $M+A-Z < \mu_0 <3A-M+Z$ & II-III-1$^{\star}$ &  II-III-1$^{\star}$ \\[1mm]
\hline
5& $3A-M+Z < \mu_0 < 2A+Z$ & I-II, II-I, II-III-1$^{\star}$  &  I-III, II-I, II-III-1$^{\star}$ \\[1mm]
\hline
6& $2A+Z < \mu_0 < M+A+Z$ & I-II$^{\star}$, II-I, II-III-1 &  I-III$^{\star}$, II-I, II-III-1 \\[1mm]
\hline
7& $M+A+Z < \mu_0 < 3A-Z$ & I-II$^{\star}$, II-I  & I-III$^{\star}$, II-I \\[1mm]
\hline
8& $3A-Z < \mu_0 < 3A+Z$ & I-III$^{\star}$, II-I  &  I-III$^{\star}$, II-I \\[1mm]
\hline
9& $3A+Z < \mu_0 < 5A+M-Z$ & I-III$^{\star}$, III-I  & I-III$^{\star}$, III-I  \\[1mm]
\hline
10& $5A+M-Z < \mu_0 < 7A-M+Z$ &  I-III$^{\star}$ &  I-III$^{\star}$ \\[1mm]
\hline
11& $7A-M+Z < \mu_0 < 6A+Z$ & I-III$^{\star}$, II-II & I-III$^{\star}$, II-III-2  \\[1mm]
\hline
12& $6A+Z < \mu_0 < 5A+M+Z$ & I-III, II-II$^{\star}$ & I-III, II-III-2$^{\star}$   \\[1mm]
\hline
13& $5A+M+Z < \mu_0 < 7A-Z$ & II-II$^{\star}$  & II-III-2$^{\star}$ \\[1mm]
\hline
14& $7A-Z < \mu_0 < 7A+Z$ & II-III-2$^{\star}$  & II-III-2$^{\star}$ \\[1mm]
\hline
15& $7A+Z < \mu_0$ &  III-III-1$^{\star}$  &  III-III-1$^{\star}$ \\[1mm]
\end{tabular}
\end{ruledtabular}
\end{table*}

Thus, there are several coexistent solutions on different intervals of $\mu_0$.
In order to find the most stable solution among them, we have to calculate the free
energy density $\Omega$ of the ground states corresponding to these solutions.
To facilitate this, we first calculate the free energy densities of the fixed spin
solutions I, II and III considered in Subsec.~\ref{B4} by using expression
(\ref{eff-pot}) for $\Omega$ derived in Appendix~\ref{C}. The results are
\begin{eqnarray}
\mbox{solution I:} &\qquad &
\Omega_{\mbox{\scriptsize I}}=-\frac{|eB_{\perp}|}{4\pi \hbar c}\left[M+A+h\right],\\
\mbox{solution II:} &\qquad &
\Omega_{\mbox{\scriptsize II}}=-\frac{|eB_{\perp}|}{4\pi \hbar c}\left[M -
(\mu+\bar{\mu})\mbox{sign}(\mu)+h\right],\\
\mbox{solution III:} &\qquad &
\Omega_{\mbox{\scriptsize III}}=-\frac{|eB_{\perp}|}{4\pi\hbar c}\left[M +
(\mu+\bar{\mu})\mbox{sign}(\mu)+h\right],
\end{eqnarray}
where $h$ is the higher LLs contribution, defined by
\begin{equation}
h \equiv \sum_{n=1}^{\infty}
\frac{2M^4}{\sqrt{n \epsilon_{B}^2+M^2}
\left(\sqrt{n \epsilon_{B}^2+M^2}+
\sqrt{n}\, \epsilon_{B}\right)^2}
\simeq
\frac{M^4}{2\epsilon_{B}^3}
\left[
\zeta\left(\frac32\right)-\zeta\left(\frac52\right) \frac{M^2}{\epsilon_{B}^2}
+O\left(\frac{M^4}{\epsilon_{B}^4}\right)
\right],
\label{h}
\end{equation}
where $\zeta (x)$ is the Riemann zeta function.
On the right hand side we used the expansion in powers of $(M/\epsilon_{B})^2$.
When keeping only the first two terms in the expansion, we find that the result
deviates by less than $1\%$ from the exact one for $M\lesssim 0.4 \epsilon_{B}$.
Note that the above contribution from higher LLs is the same for all solutions. Therefore,
it is only the LLL contribution that is relevant for choosing the lowest free energy
density.

It is not difficult now to calculate the free energy densities for
all the solutions. In Table~\ref{tab:mu0intervals}, the solutions
that have the lowest values of $\Omega$ and thus correspond to the
ground states in the given intervals of $\mu_0$ are marked by stars.
As for the explicit expression for the energy density in the
ground state, it reads
\begin{eqnarray}
\Omega &=& -\frac{|eB_{\perp}|}{2\pi\hbar c}\left[M+A+2Z+h\right],
\quad\mbox{for}\quad 0<\mu_0 < 2A+Z,\\
\Omega &=& -\frac{|eB_{\perp}|}{2\pi\hbar c}\left[M-A+Z+h+\mu_0\right],
\quad\mbox{for}\quad 2A+Z < \mu_0 < 6A+Z,\\
\Omega &=& -\frac{|eB_{\perp}|}{2\pi\hbar c}\left[M-7A+h+2\mu_0\right],
\quad\mbox{for}\quad 6A+Z < \mu_0 ,
\end{eqnarray}
Using now the explicit form of the solutions obtained in Subsec.~\ref{B5}, we
can significantly reduce the number of the cases. As result, we conclude that
only the following three solutions are realized:
\vspace{3mm}

\begin{itemize}
\item[(i)] The solution with singlet Dirac masses for both spin up and spin
down:
\begin{equation}
\begin{split}
& \tilde{\Delta}_{+}=\tilde{\mu}_{+}=0,\qquad \mu_{+}=\bar{\mu}_{+}-A,\qquad
\Delta_{+}=s_{\perp}M,
\\
& \tilde{\Delta}_{-}=\tilde{\mu}_{-}=0,\qquad \mu_{-}=\bar{\mu}_{-}+ A,\qquad
\Delta_{-}=-s_{\perp}M.
\end{split}
\label{i-app}
\end{equation}
It is the most favorable for $0 \le \mu_0 < 2A+Z$.\cite{footnote2}
We will call it the $S1$ solution, which is one of several solutions with
nonvanishing  {\em singlet} Dirac masses.

\item[(ii)] The hybrid solution with a triplet Dirac mass for spin up and a singlet
mass for spin down:
\begin{equation}
\begin{split}
& \tilde{\Delta}_{+} = M,\qquad \tilde{\mu}_{+}=As_{\perp},\qquad \mu_{+} =
\bar{\mu}_{+} - 4A,\qquad \Delta_{+}=0,
\\
& \tilde{\Delta}_{-}=\tilde{\mu}_{-}=0,\qquad \mu_{-}=\bar{\mu}_{-}-3A,\qquad
\Delta_{-}=-s_{\perp}M.
\end{split}
\label{ii-app}
\end{equation}
It is the most favorable for $2A+Z \le \mu_0 < 6A+Z$. We will call it the
$H1$ solution.

\item[(iii)] The solution with equal singlet masses for both spin up and spin
down:
\begin{equation}
\begin{split}
& \tilde{\Delta}_{+}=\tilde{\mu}_{+}=0,\qquad \mu_{+}=\bar{\mu}_{+}-7A,\qquad
\Delta_{+}=-s_{\perp}M,
\\
& \tilde{\Delta}_{-}=\tilde{\mu}_{-}=0,\qquad  \mu_{-}=\bar{\mu}_{-}- 7A,\qquad
\Delta_{-}=-s_{\perp}M.
\end{split}
\label{iii-app}
\end{equation}
It is the most favorable for $\mu_0 > 6A+Z $. We will call it the $S2$ solution.
\end{itemize}

\section{Free Energy Density}
\label{C}

In this Appendix, the units with $\hbar =1$ and $c = 1$ are used.

In order to calculate a free energy density $\Omega$, it is
convenient to use the Baym-Kadanoff formalism (the effective
action formalism for composite operators) developed in
Ref.~\onlinecite{potential} (see in particular the last paper there). In
the mean field approximation that we use, the corresponding
effective action $\Gamma$ has the following form:
\begin{eqnarray}
\Gamma(G)={-i}\,\mbox{Tr}\left[\mbox{Ln} G^{-1} +S^{-1}G-1\right]
+\frac{ G_{int}}{2}\int d^{3}x\left\{\mbox{tr}
\left[\gamma^{0}G(x,x)\gamma^{0}G(x,x)\right]
-\left(\mbox{tr}\left[\gamma^{0}G(x,x)\right]\right)^{2}\right\},
\label{potential}
\end{eqnarray}
where the trace, the logarithm, and the product $S^{-1}G$ are
taken in the functional sense, and $G = \mbox{diag}(G_{+},
G_{-})$. The free energy density $\Omega$ is expressed through
$\Gamma$ as $\Omega = -\Gamma/TV$, where $TV$ is a space-time
volume. The stationarity condition $\delta\Gamma(G)/\delta{G}=0$
leads to the gap equation (\ref{gap}). On its solutions 
we have
\begin{equation}
\Gamma= -i\,\mbox{Tr}\left[\mbox{Ln} G^{-1}
+\frac{1}{2}\left(S^{-1}G-1\right)\right]. \label{omega}
\end{equation}
Henceforth we will use the symmetric gauge with
$\mathbf{A}(\mathbf{r}) = (-B_{\perp}y/2, B_{\perp}x/2)$. Then, as
was shown in Appendix~\ref{A}, the Green's function
$G_{s}(u,u^\prime)$, with $u = (t,\mathbf{r})$, has the form:
\begin{equation}
G_{s}(u,u^{\prime})=e^{i\Phi(u,u^\prime)}\bar G_{s}(u-u^\prime),
\label{form}
\end{equation}
where $\Phi(u,u^\prime)=-e \mathbf{r}\cdot\mathbf{A}(\bf
r^\prime)$ is the Schwinger phase in the symmetric gauge.

Because of the translation invariance in time, we have
\begin{equation}
G_{s}(u,u^{\prime})=\int\limits_{-\infty}^{\infty}\frac{d\omega}{2\pi}\,
e^{-i\omega(t-t^\prime)}G_{s}(\omega;\mathbf{r},
\mathbf{r}^\prime).
\end{equation}
Then the effective action $\Gamma$ can be rewritten as
\begin{equation}
\Gamma =
-i\,T\int\limits_{-\infty}^{\infty}\frac{d\omega}{2\pi}\mbox{Tr}\left[\ln
G^{-1}(\omega)
+\frac{1}{2}\left(S^{-1}(\omega)G(\omega)-1\right)\right],
\label{effpot-integral-in-omega}
\end{equation}
where
\begin{eqnarray}
G_{s}^{-1}(\omega;\mathbf{r},\mathbf{r}^\prime)
&=&-i\left[(\omega+\mu_{s})\gamma^0-v_{F}(\bm{\pi}\cdot\bm{\gamma})
+i\tilde{\mu}_{s}\gamma^1\gamma^2+i\Delta_{s}\gamma^0\gamma^1\gamma^2-
\tilde{\Delta}_{s}\right] \delta(\mathbf{r}-\mathbf{r}^\prime ),\\
S_{s}^{-1}(\omega;\mathbf{r},\mathbf{r}^\prime )
&=&-i\left[(\omega+\bar\mu_{s})\gamma^0-v_{F}(\bm{\pi}\cdot\bm{\gamma})\right]
\delta(\mathbf{r}-\mathbf{r}^\prime ).
\end{eqnarray}
In Eq.~(\ref{effpot-integral-in-omega}) the functional operation
$\mbox{Tr}$ includes now the integration over the space
coordinates only and the trace over matrix indices.

Integrating by parts the logarithm term in
Eq.~(\ref{effpot-integral-in-omega}) and omitting the irrelevant
surface term (independent of the physical parameters), we arrive
at the expression
\begin{equation}
\Gamma=-iT\int\limits_{-\infty}^{\infty}\frac{d\omega}{2\pi}\mbox{Tr}
\left[-\omega\frac{\partial G^{-1}(\omega)}
{\partial\omega}\,G(\omega)+\frac{1}{2}\left( S^{-1}(\omega)
\,G(\omega)-1\right)\right] \label{effpot2}
\end{equation}
with
\begin{equation}
\frac{\partial G^{-1}(\omega)}{\partial\omega} =
-i\gamma^{0}\delta(\mathbf{r}-\mathbf{r}^\prime).
\end{equation}
Substituting now expression (\ref{form}) for the Green's function
in $\Gamma$, one can see that the Schwinger phase goes away and we
get
\begin{eqnarray}
\Gamma=-iTV\int\limits_{-\infty}^{\infty}\frac{d\omega}
{2\pi}\mbox{tr} \left[{i}\gamma^{0}\omega\bar
G(\omega;0)+\frac{1}{2}\left(-i \left[(\omega+\bar\mu)\gamma^0
-v_{F}(\bm{\pi}\cdot\bm{\gamma})\right]\bar{G}(\omega;\mathbf{r})|_{r=0}-
\delta(0)\right)\right].
\end{eqnarray}

Dividing $\Gamma$ by the space-time volume $TV$, we find the free
energy density:
\begin{eqnarray}
\Omega &=& i\int\limits_{-\infty}^{\infty} \frac{d\omega}{2\pi}
\int\frac{d^{2}k}{(2\pi)^{2}}
\mbox{tr}\left\{{i}\omega\gamma^{0}\bar{G}(\omega,\mathbf{k})+
\frac{1}{2}\left(-i\left[(\omega+\bar{\mu})\gamma^{0}
-v_{F}(\mathbf{k}\cdot\bm{\gamma})\right]\bar{G}(\omega,\mathbf{k})-
1\right)\right\}\nonumber\\
&=&-\int\limits_{-\infty}^{\infty}\frac{d\omega}{4\pi}\int
\frac{d^{2}k}{(2\pi)^{2}}
\mbox{tr}\left\{\left[(\omega-\bar{\mu})\gamma^{0}+v_{F}(\mathbf{k}\cdot\bm{\gamma})\right]
\bar{G}(\omega,\mathbf{k})+i\right\}, \label{Omega-formal-div}
\end{eqnarray}
where the propagator $\bar{G}_{s}(\omega,\mathbf{k})$ is given in
Eq.~(\ref{Dsn-new}) in Appendix~\ref{A}. By making use of its
explicit form, we can calculate the following two integrals that
contribute to the free energy density,
\begin{eqnarray}
\int\frac{d^{2}k}{(2\pi)^{2}}\,\gamma^{0}\bar{G}_{s}(\omega,\mathbf{k})
&=&\frac{i}{4\pi l^2}\sum\limits_{n=0}^{\infty}
\frac{\left(\omega+\mu_{s}+i\tilde{\mu}_{s}\gamma^0\gamma^1\gamma^2-
i\Delta_{s}\gamma^1\gamma^2+ \tilde{\Delta}_{s}\gamma^0\right)P_{n}}
{(\omega+\mu_{s}+i\tilde{\mu}_{s}\gamma^0\gamma^1\gamma^2)^2-
(\tilde{\Delta}_{s}-i\Delta_{s}\gamma^0\gamma^1
\gamma^2)^2-2v_{F}^2|eB_{\perp}|n},\\
\int\frac{d^{2}k}{(2\pi)^{2}} v_{F}(\mathbf{k}\cdot\bm{\gamma})
\bar{G}_{s}(\omega,\mathbf{k}) &=&
 \frac{i}{\pi l^2}\sum\limits_{n=0}^{\infty}\frac{v_{F}^2|eB_{\perp}|n\,\theta(n-1)}
{(\omega+\mu_{s}+i\tilde{\mu}_{s}\gamma^0\gamma^1\gamma^2)^2-
(\tilde{\Delta}_{s}-i\Delta_{s}\gamma^0\gamma^1
\gamma^2)^2-2v_{F}^2|eB_{\perp}|n},
\end{eqnarray}
where
\begin{equation}
P_{n}=1-i\gamma^1\gamma^2\mbox{sign}(eB_{\perp})
 +\left[1+i\gamma^1\gamma^2\mbox{sign}(eB_{\perp})\right]\theta(n-1).
\end{equation}
In the calculation, we used formula 7.414.7 from Ref.~\onlinecite{GR},
i.e.,
\begin{equation}
\int_{0}^{\infty}e^{-at}t^{\alpha}L^{\alpha}_{n}(t)dt =
\frac{\Gamma(\alpha+n+1)(a-1)^{n}}{n!a^{\alpha+n+1}}, \quad
\mbox{Re}\,\alpha>-1,\, \mbox{Re}\,a>0.
\end{equation}
By dropping an infinite divergent term which is independent of the
physical parameters, from Eq.~(\ref{Omega-formal-div}) we derive
the following expression for the free energy density:
\begin{equation}
\Omega=-\frac{i}{(4\pi l)^2}\sum_{s=\pm}
\int\limits_{-\infty}^{\infty}d\omega\, \mbox{tr}_D
\sum\limits_{n=0}^{\infty}
\frac{(\omega-\bar{\mu}_{s})\left[\omega+\mu_{s}+i\tilde{\mu}_{s}\gamma^0\gamma^1\gamma^2
-i\Delta_{s}\gamma^1\gamma^2+\tilde{\Delta}_{s}\gamma^0\right]P_{n}+
4v_{F}^2|eB_{\perp}|n\theta(n-1)}
{(\omega+\mu_{s}+i\tilde{\mu}_{s}\gamma^0\gamma^1\gamma^2)^2-
(\tilde{\Delta}_{s}-i\Delta_{s}\gamma^0\gamma^1
\gamma^2)^2-2v_{F}^2|eB_{\perp}|n} .
\end{equation}
Here the trace $\mbox{tr}_D$ is taken over the Dirac indices.

The free energy density $\Omega$ is a function of
$\tilde{\Delta}_{s}$, $\tilde{\mu}_{s}$, $\mu_{s}$, $\Delta_{s}$,
$\bar{\mu}_{s}$, and $B_{\perp}$. Normalizing $\Omega$ by
subtracting its value at
$\tilde{\Delta}_{s}=\tilde{\mu}_{s}=\mu_{s}=\Delta_{s}=\bar{\mu}_{s}=0$, we
obtain:
\begin{eqnarray}
\Omega &=&-\frac{i}{(4\pi l)^2}\sum_{s=\pm}
\sum\limits_{n=0}^{\infty}\int\limits_{-\infty}^{\infty}d\omega\,
\mbox{tr}_D\left[
\frac{(\omega-\bar{\mu}_{s})[\omega+\mu_{s}+i\tilde{\mu}_{s}\gamma^0\gamma^1\gamma^2
-i\Delta_{s}\gamma^1\gamma^2+\tilde{\Delta}_{s}\gamma^0]P_{n}+
4v_{F}^2|eB_{\perp}|n\theta(n-1)}
{(\omega+i\epsilon\mbox{sign}({\omega})
+\mu_{s}+i\tilde{\mu}_{s}\gamma^0\gamma^1\gamma^2)^2-
(\tilde{\Delta}_{s}-i\Delta_{s}
\gamma^0\gamma^1\gamma^2)^2-2v_{F}^2|eB_{\perp}|n}\right.\nonumber\\
&-&\left.\frac{\omega^{2}P_{n}+4v_{F}^2|eB_{\perp}|n\theta(n-1)}
{\left(\omega+i\epsilon\mbox{sign}
({\omega})\right)^{2}-2v_{F}^2|eB_{\perp}|n}
\right].\nonumber\\
\end{eqnarray}
One can check that for
$\tilde{\mu}_{s}=\Delta_{s}=\mu_{s}=\bar{\mu}_{s}=B_{\perp}=0$ and
$\tilde{\Delta}_{+}=\tilde{\Delta}_{-}=\tilde{\Delta}$ this
expression reduces to
\begin{eqnarray}
\Omega(\tilde{\Delta},0,0,0,0,0)=
-\frac{\tilde{\Delta}^{4}}{4\pi}\int\limits_{0}^{\infty}
\frac{dx}{\sqrt{\tilde{\Delta}^{2}+x}\left(\sqrt{\tilde{\Delta}^{2}+x}
+\sqrt{x}\right)^{2}}=-\frac{\tilde{\Delta}^{3}}{6\pi},
\end{eqnarray}
which coincides with the known expression for the vacuum energy
density in $2+1$ dimension.\cite{Gorbar2003PLA}

Finally, integrating over $\omega$ and taking trace, we find the
following expression for the free energy density:
\begin{eqnarray}
&&\Omega=-\frac{1}{8\pi l^2}\sum_{s=\pm} \Bigg\{\!
\left[\mu_{s}+\bar{\mu}_{s}-\tilde{\mu}_{s}-(\tilde{\Delta}_{s}+\Delta_{s})
\mbox{sign}(eB_{\perp})\right] \mbox{sign}(\mu_{s}-\tilde{\mu}_{s})
\theta(|\mu_{s}-\tilde{\mu}_{s}|-|\tilde{\Delta}_{s}+\Delta_{s}|)
\nonumber\\
&&+\left[\tilde{\Delta}_{s}+\Delta_{s}-(\mu_{s}+\bar{\mu}_{s}-
\tilde{\mu}_{s})\mbox{sign}(eB_{\perp})\right]\,\mbox{sign}(\tilde{\Delta}_{s}+\Delta_{s})
\theta(|\tilde{\Delta}_{s}+\Delta_{s}|-|\mu_{s}-\tilde{\mu}_{s}|)
\nonumber\\
&&\left. +2\sum_{n=1}^{\infty} \left[ \left[
(\mu_{s}+\bar{\mu}_{s}-\tilde{\mu}_{s})\mbox{sign}(\mu_{s}-\tilde{\mu}_{s})-
2\epsilon_{B}\sqrt{n}\right] \theta
\left(|\mu_{s}-\tilde{\mu}_{s}|-E_{ns}^{+} \right) +
\frac{(\tilde{\Delta}_{s}+\Delta_{s})^4\theta(E_{ns}^+-|\mu_{s}-\tilde{\mu}_{s}|)}
{E_{ns}^+(E_{ns}^{+}+\epsilon_{B}\sqrt{n})^2}\right]
\right.\nonumber\\
&&+\left[\tilde{\mu}_{s} \to -\tilde{\mu}_{s},\,\Delta_{s} \to -
\Delta_{s},\,\mbox{sign}(eB_{\perp}) \to
-\mbox{sign}(eB_{\perp})\right] \Bigg\}, \label{eff-pot}
\end{eqnarray}
where $E_{ns}^{\pm}=\sqrt{n\epsilon_{B}^2+(\tilde{\Delta}_{s} \pm
\Delta_{s})^2}$ and $\epsilon_{B}=\sqrt{2v_{F}^2|eB_{\perp}|}$.

\section{Analytic Solutions of Gap Equation for $n=1$ LL at $T=0$}
\label{D}

\subsection{Fixed spin}
\label{D1}

In Appendix~\ref{B}, we analyzed solutions of the gap equations under
the condition that only states on the LLL can be filled,
$|\mu_{s} \pm \tilde{\mu}_{s}| \ll \epsilon_{B}=\sqrt{2\hbar|eB_{\perp}|v^{2}_F/c}$.
Since all the dynamically generated parameters are much less than $\epsilon_{B}$,
this condition implies that the bare chemical potential $\mu_0$ also has
to satisfy $\mu_0 \ll \epsilon_{B}$ in that case.

In this section, we will consider the case when $\mu_0$ is of the
order of the Landau scale $\epsilon_{B}$, i.e., we will study the dynamics
when states on the first Landau level, $n=1$ LL, can be filled.
The gap equations are given in Eqs.~(\ref{E1-a})--(\ref{E4-a})
in Sec.~\ref{3}. In order to get their solutions for $\mu_0 \sim \epsilon_{B}$,
we will follow the steps in the analysis in Appendix~\ref{B}. The equations
for the dynamical parameters $\tilde{\Delta}_{s}$, $\Delta_{s}$, and
$\tilde{\mu}_{s}$ form independent systems of equations for each
spin. From these systems, we can find their solutions as functions
of $\mu_{s}$. We obtain the following three solutions.

\begin{itemize}
\item
Solution f-I. This solution corresponds to the case with
$|\mu_{s}-\tilde{\mu}_{s}|<\sqrt{\epsilon_{B}^2 +(\tilde{\Delta}_{s}+\Delta_{s})^2}$
and
$|\mu_{s}+\tilde{\mu}_{s}|<\sqrt{\epsilon_{B}^2 +(\tilde{\Delta}_{s}-\Delta_{s})^2}$.
It is
\begin{equation}
\Delta_{s}=-s_{\perp}\,\mbox{sign}(\mu_{s})\,M,
\qquad
\tilde{\Delta}_{s}=\tilde{\mu}_{s}=0.
\label{f-I}
\end{equation}
This solution exists for $M < |\mu_{s}| < \sqrt{\epsilon_{B}^2
+M^2}$. Actually, it is exactly the same as solution (\ref{Bf12-solution})
considered in Subsec. 3 of Appendix~\ref{B}. For positive $\mu_{s}$, this
solution corresponds to a state with the completely filled LLL and the empty
$n=1$ LL. With increasing $\mu_{s}$, this solution exists up to the point where
the first LL starts to fill, which is defined by the upper limit of the above inequality
for $\mu_{s}$. Recall that $X_{s}=-4A\,\mbox{sign}(\mu_{s})$ for such a solution.

\item
Solution f-II. This solution is realized when the inequalities
$|\mu_{s}-\tilde{\mu}_{s}|<\sqrt{\epsilon_{B}^2 +(\tilde{\Delta}_{s}+\Delta_{s})^2}$
and
$|\mu_{s}+\tilde{\mu}_{s}| > \sqrt{\epsilon_{B}^2 +(\tilde{\Delta}_{s}-\Delta_{s})^2}$
are satisfied. In this solution, all three dynamical parameters $\Delta_{s}$, $\tilde{\Delta}_{s}$,
and $\tilde{\mu}_{s}$ are nonzero:
\begin{equation}
\tilde{\Delta}_{s} = -s_{\perp}\mbox{sign}(\mu_{s})\frac{M-M_1}{2},
\qquad
\Delta_{s} = -s_{\perp}\, \mbox{sign}(\mu_{s})\frac{M_1+M}{2},
\qquad
\tilde{\mu}_{s}=\mbox{sign}(\mu_{s})\,A.
\label{f-II}
\end{equation}
Here $M_1$ satisfies the following equation:
\begin{equation}
1 = \frac{A}{\sqrt{\pi}}
\int_{1/\Lambda^2}^{\infty}
\frac{dy}{\sqrt{y}}
e^{-yM_1^2}\left[\coth\left(\frac{\epsilon_{B}^2}{2}y\right) -
2\exp\left(-y\epsilon_{B}^2\right)\right].
\label{m2-equation}
\end{equation}
Note that the last term in the square brackets of the integrand appears because
the $n=1$ LL contribution is absent in the equation for $\tilde{\Delta} - \Delta$
[cf. Eq.~(\ref{integral-form}) where all LLs are included].

Utilizing the analysis in the second paper in Ref.~\onlinecite{Gusynin1995PRD},
we arrive at the following gap equation for $M_1$:
\begin{equation}
1= \lambda + \frac{A}{M_1} + \frac{2A}{\epsilon_{B}}
\zeta\left(\frac{1}{2},1+\frac{M_1^2}{\epsilon_{B}^2}\right)
-\frac{2A}{\sqrt{\epsilon_{B}^2 +M_1^2}}+O\left(\lambda\frac{M_1^2}{\Lambda^2}\right),
\end{equation}
where $\zeta(z,q)$ is the generalized Riemann zeta function.\cite{GR}
In the subcritical regime ($\lambda < 1$) its solution is given by
\begin{equation}
M_1 \simeq \frac{A}{1-\lambda+2\left[1-\zeta(1/2)\right]A/\epsilon_{B}}.
\label{m2}
\end{equation}
Since the last term in the denominator is positive, we have $M_1 < M$ that is
consistent with the fact that the equation for $\tilde{\Delta} - \Delta$ misses the
contribution of the $n=1$ LL.

This solution exists for $\sqrt{\epsilon_{B}^2 +M_1^2} - A <
|\mu_{s}| < \sqrt{\epsilon_{B}^2 +M^2} + A$. One can check that the
corresponding parameter $X_{s}$ is $X_{s}=-8A\,\mbox{sign}(\mu_{s})$. As in the
case of the LLL (see Subsec.~\ref{B4}), there is another solution with $\tilde{\Delta}_{s}$,
$\tilde{\mu}_{s}$ replaced by $-\tilde{\Delta}_{s}$, $-\tilde{\mu}_{s}$, which
takes place for $|\mu_{s}-\tilde{\mu}_{s}|> \sqrt{\epsilon_{B}^2
+(\tilde{\Delta}_{s}+\Delta_{s})^2}$ and $|\mu_{s}+\tilde{\mu}_{s}| <
\sqrt{\epsilon_{B}^2 + (\tilde{\Delta}_{s}-\Delta_{s})^2}$. These two
solutions are equivalent: one can see this from dispersion relations
(\ref{higherLLs}), (\ref{LLLenergylevels}) by transforming $\sigma \to
-\sigma$ there, i.e., as in the case of the LLL
solution I (\ref{I}), these solutions
are related to two degenerate ground states connected by a
$SU(2)_{s}$ (or $Z_{2s}$) flavor transformation.

\item
Solution f-III. This solution corresponds to the case with
$|\mu_{s}-\tilde{\mu}_{s}|> \sqrt{\epsilon_{B}^2 +(\tilde{\Delta}_{s}+\Delta_{s})^2}$
and
$|\mu_{s}+\tilde{\mu}_{s}| >\sqrt{\epsilon_{B}^2 +(\tilde{\Delta}_{s}-\Delta_{s})^2}$.
Its explicit form reads
\begin{equation}
\tilde{\Delta}_{s}=\tilde{\mu}_{s}=0,\qquad \Delta_{s} =
-s_{\perp}\, \mbox{sign}(\mu_{s}) M_1.
\label{f-III}
\end{equation}
This solution takes place for $|\mu_{s}| > \sqrt{\epsilon_{B}^2
+M_1^2}$ and the corresponding $X_{s}$ is $X_{s}=-12A\,\mbox{sign}(\mu_{s})$.

\end{itemize}

\subsection{Including both spin up and spin down}
\label{D2}

In Subsec.~\ref{D1}, the solutions for the dynamical parameters $\Delta_{s}$,
$\tilde{\Delta}_{s}$, and $\tilde{\mu}_{s}$ at fixed spin were described. Since $X$
contains contribution of fields of both spins, the equations for chemical
potentials $\mu_{+}$ and $\mu_{-}$ for fields of different spin are coupled and
have to be solved together. Since we can choose any of the found three
solutions for masses at a fixed spin, we should solve 9 systems of coupled
equations for $\mu_{+}$ and $\mu_{-}$. Like in the case of the LLL, it is enough to
consider only 6 systems. The simplest case is the solution f-I--f-I because it
corresponds to the case of completely filled LLL, which was  already considered
in Subsec.~\ref{B5}. We have

\begin{itemize}
\item
f-I--f-I solution is given by
\begin{equation}
\begin{split}
& \tilde{\Delta}_{+}=\tilde{\mu}_{+}=0,
\qquad
\mu_{+} = \bar{\mu}_{+} - 7A\,\mbox{sign}(\bar{\mu}_{+}),
\qquad
\Delta_{+}=-s_{\perp}\, \mbox{sign}(\bar{\mu}_{+}) M,
\\
& \tilde{\Delta}_{-}=\tilde{\mu}_{-}=0,
\qquad
\mu_{-} = \bar{\mu}_{-} - 7A\,\mbox{sign}(\bar{\mu}_{-}),
\qquad
\Delta_{-}=-s_{\perp}\,\mbox{sign}(\bar{\mu}_{-}) M.
\end{split}
\label{f-I-f-I}
\end{equation}
It exists when
$\mbox{sign}(\bar{\mu}_{+})\, \mbox{sign}(\bar{\mu}_{-}) > 0$
and
\begin{equation}
7A+M < |\bar{\mu}_{+}| < 7A+\sqrt{\epsilon_{B}^2+M^2},
\qquad
7A+M < |\bar{\mu}_{-}| < 7A+\sqrt{\epsilon_{B}^2+ M^2}.
\end{equation}
It coincides with the solution III-III-1 in Eq.~(\ref{III-III-1}) in Subsec.~\ref{B5},
except for having a different lower limit for $|\bar{\mu}_{\pm}|$. The latter is connected
with the point that while the solution III in Eq.~(\ref{III}) exists for all values of $\mu_{s}$,
the solution f-I in Eq.~(\ref{f-I}) exists only for $|\mu_{s}| > M$. This is because,
according to the analysis in Subsec.~\ref{B4}, the solution III is a combination
of two solutions: the solution (\ref{Bf12-solution}), which is equivalent to the solution f-I,
and one of the two solutions in Eq.~(\ref{gap-solution-2}).

\item
f-I--f-II solution is given by
\begin{equation}
\begin{split}
& \tilde{\Delta}_{+}=\tilde{\mu}_{+}=0,
\qquad
\mu_{+} = \bar{\mu}_{+} -11A\,\mbox{sign}(\bar{\mu}_{+}),
\qquad
\Delta_{+}=-s_{\perp}\,\mbox{sign}(\bar{\mu}_{+}) M,
\\
& \tilde{\Delta}_{-}=\frac{M-M_1}{2},
\qquad
\tilde{\mu}_{-}=-As_{\perp},
\qquad
\mu_{-} = \bar{\mu}_{-} -10A\,\mbox{sign}(\bar{\mu}_{-}),
\qquad
\Delta_{-}=-s_{\perp}\,\mbox{sign}(\bar{\mu}_{-}) \frac{M+M_1}{2}.
\end{split}
\label{f-I-f-II}
\end{equation}
It exists when
$\mbox{sign}(\bar{\mu}_{+})\, \mbox{sign}(\bar{\mu}_{-}) > 0$
and
\begin{equation}
11A+M < |\bar{\mu}_{+}| < 11A+\sqrt{\epsilon_{B}^2 +M^2},\qquad
9A+\sqrt{\epsilon_{B}^2 +M_1^2}
<|\bar{\mu}_{-}| < 11A+\sqrt{\epsilon_{B}^2 +M^2}.
\end{equation}

\item
f-I--f-III solution reads
\begin{equation}
\begin{split}
& \tilde{\Delta}_{+}=\tilde{\mu}_{+}=0,\qquad  \mu_{+} = \bar{\mu}_{+} -
15A\,\mbox{sign}(\bar{\mu}_{+}),\qquad \Delta_{+}=-s_{\perp}\,
\mbox{sign}(\bar{\mu}_{+}) M,
\\
& \tilde{\Delta}_{-}=\tilde{\mu}_{-}=0,\qquad  \mu_{-} = \bar{\mu}_{-} - 13A\,
\mbox{sign}(\bar{\mu}_{-}), \qquad  \Delta_{-}=-s_{\perp}\,
\mbox{sign}(\bar{\mu}_{-}) M_1,
\end{split}
\label{f-I-f-III}
\end{equation}
and takes place when
$\mbox{sign}(\bar{\mu}_{+})\, \mbox{sign}(\bar{\mu}_{-}) > 0$
and
\begin{equation}
15A+M < |\bar{\mu}_{+}| < 15A+\sqrt{\epsilon_{B}^2 +M^2},
\qquad
13A+\sqrt{\epsilon_{B}^2 +M_1^2} <|\bar{\mu}_{-}|.
\end{equation}

\item
f-II--f-II solution is given by
\begin{equation}
\begin{split}
& \tilde{\Delta}_{+}=\frac{M-M_1}{2},
\qquad
\tilde{\mu}_{+}=-As_{\perp},
\qquad
\mu_{+} = \bar{\mu}_{+} -14A\,\mbox{sign}(\bar{\mu}_{+}),
\qquad
\Delta_{+}=-s_{\perp}\,\mbox{sign}(\bar{\mu}_{+}) \frac{M+M_1}{2},
\\
& \tilde{\Delta}_{-}=\frac{M-M_1}{2},\qquad \tilde{\mu}_{-}=-
As_{\perp},\qquad  \mu_{-} = \bar{\mu}_{-} -
14A\,\mbox{sign}(\bar{\mu}_{-}),\qquad  \Delta_{-}=-s_{\perp}\,
\mbox{sign}(\bar{\mu}_{-}) \frac{M+M_1}{2},
\end{split}
\label{f-II-f-II}
\end{equation}
and exists when
$\mbox{sign}(\bar{\mu}_{+})\, \mbox{sign}(\bar{\mu}_{-}) > 0$
and
\begin{equation}
13A+\sqrt{\epsilon_{B}^2 +M_1^2} < |\bar{\mu}_{\pm}| <
15A+\sqrt{\epsilon_{B}^2 +M^2}.
\end{equation}

\item
f-II--f-III solution is given by
\begin{equation}
\begin{split}
& \tilde{\Delta}_{+}=\frac{M-M_1}{2},\qquad \tilde{\mu}_{+}=
-As_{\perp},\qquad  \mu_{+} = \bar{\mu}_{+} -
18A\,\mbox{sign}(\bar{\mu}_{+}),\qquad \Delta_{+}=-s_{\perp}\,
\mbox{sign}(\bar{\mu}_{+}) \frac{M+M_1}{2},
\\
& \tilde{\Delta}_{-}=\tilde{\mu}_{-}=0,\qquad  \mu_{-} = \bar{\mu}_{-} - 17A\,
\mbox{sign}(\bar{\mu}_{-}),\qquad  \Delta_{-}=-s_{\perp}\,
\mbox{sign}(\bar{\mu}_{-}) M_1,
\end{split}
\label{f-II-f-III}
\end{equation}
and takes place when
$\mbox{sign}(\bar{\mu}_{+})\, \mbox{sign}(\bar{\mu}_{-}) > 0$
and
\begin{equation}
17A+\sqrt{\epsilon_{B}^2 +M_1^2} < |\bar{\mu}_{+}| <
19A+\sqrt{\epsilon_{B}^2+M^2},
\qquad
|\bar{\mu}_{-}| > 17A+ \sqrt{\epsilon_{B}^2 +M_1^2}.
\end{equation}

\item
f-III--f-III solution is given by
\begin{equation}
\begin{split}
& \tilde{\Delta}_{+}=\tilde{\mu}_{+}=0,\qquad  \mu_{+} = \bar{\mu}_{+} - 21A\,
\mbox{sign}(\bar{\mu}_{+}),\qquad  \Delta_{+}=-s_{\perp}\,
\mbox{sign}(\bar{\mu}_{+}) M_1,
\\
& \tilde{\Delta}_{-}=\tilde{\mu}_{-}=0,\qquad  \mu_{-} = \bar{\mu}_{-} - 21A\,
\mbox{sign}(\bar{\mu}_{-}),\qquad  \Delta_{-}=-s_{\perp}\,
\mbox{sign}(\bar{\mu}_{-}) M_1,
\end{split}
\label{f-III-f-III}
\end{equation}
and exists when
$\mbox{sign}(\bar{\mu}_{+})\, \mbox{sign}(\bar{\mu}_{-}) > 0$
and
\begin{equation}
|\bar{\mu}_{+}| > 21A+\sqrt{\epsilon_{B}^2 +M_1^2},
\qquad
|\bar{\mu}_{-}|> 21A+\sqrt{\epsilon_{B}^2 +M_1^2}.
\end{equation}

\end{itemize}

\subsection{Dependence of solutions on $\mu_0$ and their
free energy density energy}
\label{D3}

\begin{table*}
\caption{
\label{tab:1LL.intervals}
Intervals of the existence of solutions, relevant for the dynamics in the $n=1$ LL at $T=0$.}
\begin{ruledtabular}
\begin{tabular}{ll}
f-I--f-I & $7A+M+Z < \mu_0 < 7A+\sqrt{\epsilon_{B}^2 +M^2}-Z$\\[1mm]
\hline
f-I--f-II & $9A+\sqrt{\epsilon_{B}^2 +M_1^2}-Z < \mu_0 <
11A+\sqrt{\epsilon_{B}^2 +M^2}-Z$ \\[1mm]
\hline
f-II--f-I & $9A+\sqrt{\epsilon_{B}^2 +M_1^2}+Z < \mu_0 <
11A+\sqrt{\epsilon_{B}^2 +M^2}-Z$ \\[1mm]
\hline
f-I--f-III & $13A+\sqrt{\epsilon_{B}^2 +M_1^2}-Z < \mu_0 <
15A+\sqrt{\epsilon_{B}^2 +M^2}+Z$ \\[1mm]
\hline
f-III--f-I & $13A+\sqrt{\epsilon_{B}^2 +M_1^2}+Z < \mu_0 <
15A+\sqrt{\epsilon_{B}^2 +M^2}-Z$ \\[1mm]
\hline
f-II--f-II & $13A+\sqrt{\epsilon_{B}^2 +M_1^2}+Z < \mu_0 <
15A+\sqrt{\epsilon_{B}^2 +M^2}-Z$ \\[1mm]
\hline
f-II--f-III & $17A+\sqrt{\epsilon_{B}^2 +M_1^2}+Z < \mu_0 <
19A+\sqrt{\epsilon_{B}^2 +M^2}+Z$ \\[1mm]
\hline
f-III--f-II  & $17A+\sqrt{\epsilon_{B}^2 +M_1^2}+Z < \mu_0 <
19A+\sqrt{\epsilon_{B}^2 +M^2}-Z$ \\[1mm]
\hline
f-III--f-III  & $\mu_0 > 21A + \sqrt{\epsilon_{B}^2 +M_1^2} + Z$ \\
\end{tabular}
\end{ruledtabular}
\end{table*}

Using the solutions found in the previous subsection, we find that the
intervals of their existence in terms of $\mu_0$ for $\mu_0 \ge 0$
(dynamics with negative $\mu_0$ is related by the electron-hole
symmetry and will not be discussed separately). These are given in
Table~\ref{tab:1LL.intervals}. By making use of this information, we
can also determine the complete set of non-overlapping intervals of
$\mu_0$ and the solutions that (co-)exist on such intervals. This is
summarized in Table~\ref{tab:1LL.mu0intervals}.

\begin{table*}
\caption{
\label{tab:1LL.mu0intervals}
The list of solutions that coexist in a set of non-overlapping intervals
of $\mu_0$, relevant for the dynamics in the $n=1$ LL at $T=0$. The
solutions with the lowest free energy density are marked by stars.}
\begin{ruledtabular}
\begin{tabular}{lll}
\# &Interval & Solution(s) \\[1mm]
\hline
\hline
1 &
$7A+M+Z < \mu_0 < 7A+\sqrt{\epsilon_{B}^2 +M^2}-Z$  & f-I--f-I$^{\star}$ \\[1mm]
\hline
2 &
$9A+\sqrt{\epsilon_{B}^2+M_1^2}-Z < \mu_0
<9A+\sqrt{\epsilon_{B}^2 +M_1^2}+Z$& f-I--f-II$^{\star}$ \\[1mm]
\hline
3 &
$9A+\sqrt{\epsilon_{B}^2 +M_1^2}+Z < \mu_0
<11A+\sqrt{\epsilon_{B}^2 +M^2}-Z$ & f-I--f-II$^{\star}$, f-II--f-I \\[1mm]
\hline
4 &
$13A+\sqrt{\epsilon_{B}^2 +M_1^2}-Z < \mu_0 <
13A+\sqrt{\epsilon_{B}^2 +M_1^2}+Z$ &  f-I--f-III$^{\star}$ \\[1mm]
\hline
5 &
$13A+\sqrt{\epsilon_{B}^2 +M_1^2}+Z < \mu_0 <
15A+\sqrt{\epsilon_{B}^2 +M^2}-Z$ &  f-I--f-III$^{\star}$, f-III--f-I, f-II--f-II \\[1mm]
\hline
6 &
$15A+\sqrt{\epsilon_{B}^2 + M^2}-Z < \mu_0 <
15A+\sqrt{\epsilon_{B}^2 +M^2}+Z$ & f-I--f-III$^{\star}$ \\[1mm]
\hline
7 &
$17A+\sqrt{\epsilon_{B}^2 +M_1^2}+Z < \mu_0 <
19A+\sqrt{\epsilon_{B}^2 +M^2}-Z$ &  f-II--f-III$^{\star}$, f-III--f-II \\[1mm]
\hline
8 &
$19A+\sqrt{\epsilon_{B}^2 +M^2}-Z < \mu_0 <
19A+\sqrt{\epsilon_{B}^2 +M^2}+Z$ & f-II--f-III$^{\star}$  \\[1mm]
\hline
9 &
$\mu_0 > 21A+\sqrt{\epsilon_{B}^2 +M_1^2}+Z$ & f-III--f-III$^{\star}$  \\
\end{tabular}
\end{ruledtabular}
\end{table*}

Thus, there are several coexistent solutions on certain intervals of $\mu_0$.
In order to define which solutions are realized, we have to calculate their free
energy densities. To facilitate this calculation, first we will calculate free energy
densities of solutions f-I, f-II, and f-III. Using the effective potential given by
Eq.~(\ref{eff-pot}), we have
\begin{eqnarray}
\mbox{solution f-I:}
& \quad &
\Omega_{\mbox{\scriptsize f-I}}
=-\frac{|eB_{\perp}|}{4\pi\hbar c}
\left[M + (\mu+\bar{\mu})\,\mbox{sign}(\mu)+h\right], \\
\mbox{solution f-II:}
& \quad &
\Omega_{\mbox{\scriptsize f-II}}
=-\frac{|eB_{\perp}|}{4\pi\hbar c}
\left(\frac{M+M_1}{2}+A+2(\mu+\bar{\mu})\,\mbox{sign}(\mu)
-2\epsilon_{B}+\frac{h+h_1}{2}\right), \\
\mbox{solution f-III:}
& \quad &
\Omega_{\mbox{\scriptsize f-III}}
=-\frac{|eB_{\perp}|}{4\pi\hbar c}
\left(M_1 + 3(\mu+\bar{\mu})\,\mbox{sign}(\mu)
-4\epsilon_{B}+h_1\right),
\end{eqnarray}
where $h$ is given in Eq.~(\ref{h}) and
\begin{equation}
h_1 \equiv \sum_{n=2}^{\infty} \frac{2M_1^4}
{\sqrt{n\epsilon_{B}^2 +M_1^2}
\left(\sqrt{n \epsilon_{B}^2+M_1^2} +
\sqrt{n}\epsilon_{B}\right)^2}
\simeq
\frac{M_1^4}{2\epsilon_{B}^3}
\left[ \zeta\left(3/2\right)-1
-\left[\zeta\left(5/2\right)-1\right] \frac{M_1^2 }{\epsilon_{B}^2}
+O\left(\frac{M_1^4}{\epsilon_{B}^4}\right)
\right].
\label{h2}
\end{equation}
Now it is not difficult to calculate free energy densities for all solutions and
determine the ground state on each interval. The solutions with the lowest
free energy density are marked by stars in Table~\ref{tab:1LL.mu0intervals}.
The explicit form of the corresponding energy densities are
\begin{eqnarray}
\mbox{f-I--f-I:} & \qquad &
\Omega=-\frac{|eB_{\perp}|}{2\pi\hbar c}\left(M+2\mu_0-7A+h\right),
\label{Omegaf-I--f-I}\\
\mbox{f-I--f-II:} & \qquad &
\Omega=-\frac{|eB_{\perp}|}{2\pi\hbar c}\left(\frac{3M+M_1}{4}+
3\mu_0-15A-\epsilon_{B}+Z+\frac{3h+h_1}{4}\right),
\label{Omegaf-I--f-II}\\
\mbox{f-I--f-III:} & \qquad &
\Omega=-\frac{|eB_{\perp}| }{2\pi\hbar c}\left(\frac{M+M_1}{2}
+4\mu_0-27A-2\epsilon_{B}+2Z+\frac{h+h_1}{2}\right),
\label{Omegaf-I--f-III}\\
\mbox{f-II--f-III:} & \qquad &
\Omega=-\frac{|eB_{\perp}|}{2\pi\hbar c}\left(\frac{3M_1+M}{4}
+5\mu_0-43A-3\epsilon_{B}+Z+\frac{3h_1+h}{4}\right),
\label{Omegaf-II--f-III}\\
\mbox{f-III--f-III:} & \qquad &
\Omega=-\frac{|eB_{\perp}|}{2\pi\hbar c}\left(M_1+6\mu_0-63A
-4\epsilon_{B}+h_1\right).
\label{Omegaf-III--f-III}
\end{eqnarray}
Therefore, the number of different solutions is reduced down to following five.

\begin{itemize}
\item[(f-i)] The solution f-I--f-I
\begin{equation}
\begin{split}
& \tilde{\Delta}_{+}=\tilde{\mu}_{+}=0,
\qquad
\mu_{+} = \bar{\mu}_{+} -7A,
\qquad
\Delta_{+}=-s_{\perp}\,M,
\\
& \tilde{\Delta}_{-}=\tilde{\mu}_{-}=0,
\qquad
\mu_{-} = \bar{\mu}_{-} -7A,
\qquad
\Delta_{-}=-s_{\perp}\, M
\end{split}
\label{Df-i}
\end{equation}
is realized for $7A+M+Z < \mu_0 < 7A+ \sqrt{\epsilon_{B}^2
+M^2}-Z$ and has free energy density in Eq.~(\ref{Omegaf-I--f-I}). This result
means that the solution S2 given by Eq.~(\ref{iii-app}) in Subsec.~\ref{B6} takes
place for $\mu_0 < 7A+ \sqrt{\epsilon_{B}^2 +M^2}-Z$.
\vspace{3mm}

\item[(f-ii)] The solution f-I--f-II
\begin{equation}
\begin{split}
& \tilde{\Delta}_{+}=\tilde{\mu}_{+}=0,\qquad \mu_{+} = \bar{\mu}_{+} -
11A,\qquad\, \Delta_{+}=-s_{\perp}\,M,
\\
& \tilde{\Delta}_{-}=\frac{M-M_1}{2},\qquad\,\tilde{\mu}_{-}=-
As_{\perp},\qquad\, \mu_{-} = \bar{\mu}_{-} - 10A,\qquad\,
\Delta_{-}=-s_{\perp}\, \frac{M+M_1}{2}
\end{split}
\label{Df-ii}
\end{equation}
takes place for $9A+\sqrt{\epsilon_{B}^2 +M_1^2}-Z < \mu_0 <
11A+\sqrt{\epsilon_{B}^2 +M^2}-Z$ and has free energy density
in Eq.~(\ref{Omegaf-I--f-II}).
\vspace{3mm}

\item[(f-iii)] The solution f-I--f-III
\begin{equation}
\begin{split}
& \tilde{\Delta}_{+}=\tilde{\mu}_{+}=0,\qquad \mu_{+} = \bar{\mu}_{+} -
15A\,,\,\qquad \Delta_{+}=-s_{\perp}\,M,
\\
& \tilde{\Delta}_{-}=\tilde{\mu}_{-}=0,\qquad \mu_{-} = \bar{\mu}_{-} -
13A,\qquad\, \Delta_{-}=-s_{\perp}\,M_1
\end{split}
\label{Df-iii}
\end{equation}
is realized for $13A+\sqrt{\epsilon_{B}^2 +M_1^2}-Z < \mu_0 <
15A+\sqrt{\epsilon_{B}^2 +M^2}+Z$ and has free energy density
in Eq.~(\ref{Omegaf-I--f-III}).
\vspace{3mm}

\item[(f-iv)] The solution f-II--f-III
\begin{equation}
\begin{split}
& \tilde{\Delta}_{+}=\frac{M-M_1}{2},\qquad\,\tilde{\mu}_{+}=-
As_{\perp},\,\qquad\, \mu_{+} = \bar{\mu}_{+} - 18A,\qquad\,
\Delta_{+}=-s_{\perp}\,\frac{M+M_1}{2},
\\
& \tilde{\Delta}_{-}=\tilde{\mu}_{-}=0,\qquad \mu_{-} = \bar{\mu}_{-} -
17A,\qquad\, \Delta_{-}=-s_{\perp}\,M_1
\end{split}
\label{Df-iv}
\end{equation}
takes place for $17A+\sqrt{\epsilon_{B}^2+M_1^2}+Z < \mu_0 < 19A+
\sqrt{\epsilon_{B}^2 +M^2}+Z$ and has free energy density
in Eq.~(\ref{Omegaf-II--f-III}).
\vspace{3mm}

\item[(f-v)] The solution f-III--f-III
\begin{equation}
\begin{split}
& \tilde{\Delta}_{+}=\tilde{\mu}_{+}=0,\qquad \mu_{+} = \bar{\mu}_{+} -
21A,\qquad  \Delta_{+}=-s_{\perp}\,M_1,
\\
& \tilde{\Delta}_{-}=\tilde{\mu}_{-}=0,\qquad  \mu_{-} = \bar{\mu}_{-} -
21A,\qquad  \Delta_{-}=-s_{\perp}\,M_1
\end{split}
\label{Df-v}
\end{equation}
is realized for $\mu_0 > 21A+\sqrt{\epsilon_{B}^2 +
M_1^2}+Z$ and has free energy density in Eq.~(\ref{Omegaf-III--f-III}).
\end{itemize}


\begin{thebibliography}{99}

\bibitem{Geim2004Science} K.S. Novoselov, A.K. Geim, S.V. Morozov, D.
Jaing, Y. Zhang, S.V. Dubonos, I.V. Grigorieva, and A.A. Firsov,
Science {\bf 306}, 666 (2004).

\bibitem{Geim2005Nature}
K.S. Novoselov, A.K. Geim, S.V. Morozov, D. Jaing, M.I. Katsnelson, I.V.
Grigorieva, S.V. Dubonos, and A.A. Firsov, Nature {\bf 438}, 197 (2005).

\bibitem{Kim2005Nature} Y. Zhang, Y.-W. Tan, H.L. St\"ormer, and P. Kim,
Nature {\bf 438}, 201 (2005).

\bibitem{Ando2002} Y.~Zheng and T.~Ando, Phys. Rev. B {\bf 65}, 245420 (2002).

\bibitem{Gusynin2005PRL} V.P.~Gusynin and S.G.~Sharapov, Phys. Rev. Lett.
{\bf 95}, 146801 (2005);
Phys. Rev. B. {\bf 73}, 245411 (2006).

\bibitem{Peres2005} N.M.R.~Peres, F.~Guinea, and A.H.~Castro Neto,
Phys. Rev. B {\bf 73}, 125411 (2006).

\bibitem{Semenoff1984PRL} G.W.~Semenoff, Phys. Rev. Lett.
{\bf 53}, 2449 (1984).

\bibitem{Haldane1988PRL} F.D.M.~Haldane, Phys. Rev. Lett. {\bf 61},
2015 (1988).

\bibitem{Khveshchenko2001PRL} D.V.~Khveshchenko,
Phys. Rev. Lett. {\bf 87}, 206401 (2001); {\it ibid.} {\bf 87}, 246802 (2001).

\bibitem{Gorbar2002PRB} E.V.~Gorbar, V.P.~Gusynin, V.A.~Miransky,
and I.A.~Shovkovy, Phys. Rev. B {\bf 66}, 045108 (2002).

\bibitem{SGB2004}
S.G.~Sharapov, V.P.~Gusynin, and H.~Beck, Phys.Rev. B {\bf 69}, 075104 (2004).

\bibitem{Luk'yanchuk2004} I.A.~Luk'yanchuk and Y.~Kopelevich,
Phys. Rev. Lett. {\bf 93}, 166402 (2004).

\bibitem{Zhang2006} Y.~Zhang, Z.~Jiang, J. P.~Small, M. S.~Purewal,
Y.-W.~Tan, M.~Fazlollahi, J.D.~Chudow, J.A.~Jaszczak, H.L.~St\"ormer, and
P.~Kim, Phys. Rev. Lett. {\bf 96}, 136806 (2006).

\bibitem{Jiang2007} Z.~Jiang, Y.~Zhang, H.L.~St\"ormer, and P.~Kim,
Phys. Rev. Lett. {\bf 99}, 106802 (2007).

\bibitem{Nomura2006PRL} K.~Nomura and A.H.~MacDonald, Phys. Rev.
Lett. {\bf 96}, 256602 (2006); K.~Yang, S.~Das Sarma, and A.H.~MacDonald,
Phys. Rev. B {\bf 74}, 075423 (2006).

\bibitem{Goerbig2006} M.O.~Goerbig, R.~Moessner, and B.~Dou\c{c}ot,
Phys. Rev. B {\bf 74}, 161407(R) (2006).

\bibitem{Alicea2006PRB} J.~Alicea and M.P.A.~Fisher, Phys. Rev. B
{\bf 74}, 075422 (2006).

\bibitem{Sheng2007} L.~Sheng, D.N.~Sheng, F.D.M.~Haldane, and
L.~Balents, Phys. Rev. Lett. {\bf 99}, 196802 (2007).

\bibitem{LS} {V. Lukose and R. Shankar, arXiv:0706.4280 [cond-mat.mes-hall].}

\bibitem{Abanin2006PRL} D.A.~Abanin, P.A.~Lee, and L.S.~Levitov,
Phys. Rev. Lett. {\bf 96}, 176803 (2006); Solid State Comm. {\bf 143}, 77
(2007).

\bibitem{Gusynin2006catalysis} V.P.~Gusynin, V.A.~Miransky, S.G.~Sharapov,
and I.A.~Shovkovy, Phys. Rev. B {\bf 74}, 195429 (2006);
arXiv:cond-mat/0612488.

\bibitem{Herbut2006} I.F.~Herbut, Phys. Rev. Lett. {\bf 97}, 146401 (2006);
Phys. Rev. B {\bf 75}, 165411 (2007);
{\it ibid} , {\bf 76}, 085432 (2007).

\bibitem{Fuchs2006} J.-N.~Fuchs and P.~Lederer, Phys. Rev. Lett.
{\bf 98}, 016803 (2007).

\bibitem{Ezawa2006} M.~Ezawa, J. Phys. Soc. Jpn. {\bf 76} (2007) 094701;
Physica E {\bf 40}, 269 (2007).

\bibitem{Yang2007} K.~Yang, Solid State Comm. {\bf 143}, 27 (2007).

\bibitem{Fogler1995} M.M.~Fogler and B.I.~Shklovskii,
Phys. Rev. B {\bf 52}, 17366 (1995).

\bibitem{Arovas1999} D.P.~Arovas, A.~Karlhelde, and D.~Lilliehook,
Phys. Rev. B {\bf 59}, 13147 (1999);
Z.F.~Ezawa and K.~Hasebe, Phys. Rev. B {\bf 65}, 075311 (2002).

\bibitem{Gusynin1995PRD} V.P.~Gusynin, V.A.~Miransky, and I.A.~Shovkovy,
Phys. Rev. Lett. {\bf 73}, 3499 (1994);
Phys. Rev. D {\bf 52}, 4718 (1995);
Nucl.\ Phys.\ B {\bf 462}, 249 (1996).

\bibitem{Gorbar2007} E.V.~Gorbar, V.P.~Gusynin, and V.A.~Miransky,
arXiv:0710.3527 [cond-mat.mes-hall].

\bibitem{disorder0}
M.~Koshino and T.~Ando, Phys. Rev. B {\bf 75}, 033412 (2007).

\bibitem{disorder1}
P.~Goswami, X.~Jia, and S.~Chakravarty,
Phys. Rev. B {\bf 76}, 205408 (2007);
X.~Jia, P.~Goswami, and S.~Chakravarty,
Phys. Rev. Lett. ${\bf 101}$, 036805 (2008).

\bibitem{disorder2}
K.~Nomura, S.~Ryu, M.~Koshino, C.~Mudry, and A.~Furusaki,
arXiv:0801.3121 [cond-mat.mes-hall].

\bibitem{LL-width}
A.J.M.~Giesbers, U.~Zeitler, M.I.~Katsnelson, L.A.~Ponomarenko,
T.M.G.~Mohiuddin, and J.C.~Maan,
Phys. Rev. Lett. {\bf 99}, 206803 (2007).

\bibitem{Bolotin}
K.I.~Bolotin, K.J.~Sikes, Z.~Jiang, G.~Fudenberg,
J.~Hone, P.~Kim, and H. L. St\"ormer, Solid State Comm. {146} (2008) 351.

\bibitem{Andrei200803}
G.~Li, A.~Luican, and E.Y.~Andrei,
arXiv:0803.4016 [cond-mat.mes-hall].

\bibitem{Abanin2007PRL} D.A.~Abanin, K.S.~Novoselov, U.~Zeitler, P.A.~Lee, A.K.~Geim,
and L.S.~Levitov, Phys. Rev. Lett. {\bf 98}, 196806 (2007).

\bibitem{Ong2007} J.G.~Checkelsky, L.~Li, and N.P.~Ong,
Phys. Rev. Lett. {\bf 100}, 206801 (2008).

\bibitem{gamma} For Dirac matrices $\gamma^0$ and $\vec{\gamma}$, we use the same
representation as in Ref.~\onlinecite{Gusynin2006catalysis}.

\bibitem{Zeeman} The Zeeman coupling $\mu_{B}\vec{B}\vec{\sigma}$
can be always diagonalized in the spin space as $\mu_{B}B\sigma^3$.

\bibitem{book} V.A.~Miransky, {\it Dynamical Symmetry Breaking in
Quantum Field Theories} (World Scientific, Singapore, 1993).

\bibitem{footnote1a}
By definition, the values of the subcritical coupling constant are
those at which no dynamical gaps are generated without a magnetic field.

\bibitem{Katsnelson2006}
M.I.~Katsnelson, Phys. Rev. B {\bf 74}, 201401(R) (2006);
B.~Wunsch, T.~Stauber, F.~Sols, and  F.~Guinea, New J. Phys. {\bf 8}, 318 (2006).

\bibitem{Shitov2007}
A.V.~Shytov, M.I.~Katsnelson, and L.S.~Levitov,
Phys. Rev. Lett. {\bf 99}, 236801 (2007).

\bibitem{footnote1} The energy gap $\tilde{\Delta}_{\pm}$ is expressed through
the corresponding Dirac mass $\tilde{m}_{\pm}$ as $\tilde{\Delta}_{\pm} =
\tilde{m}_{\pm} v_{F}^2$. In what follows, we will ignore this difference between
them and use the term ``Dirac mass''.

\bibitem{Gusynin2007review} V.P.~Gusynin, S.G.~Sharapov, and
J.P.~Carbotte, Int. J. Mod. Phys. B {\bf 21}, 4611 (2007).


\bibitem{Aleiner2007} {I.L. Aleiner, D.E. Kharzeev, and A.M. Tsvelik,
Phys. Rev. B {\bf 76}, 195415 (2007).}

\bibitem{MW} N.D.~Mermin and H.~Wagner, Phys. Rev. Lett. {\bf 17}, 1133 (1966).

\bibitem{footnoteLLL}
For $\mbox{sign}(eB_{\perp}) > 0$, one has
$\Psi^{\dagger}P_{s}\Psi = \bar{\Psi}\gamma^3 \gamma^5 P_{s} \Psi =
\psi_{K  As}^\dagger\psi_{K As} +  \psi_{K^{\prime}Bs}^\dagger \psi_{K^{\prime} Bs}$
and $\Psi^{\dagger}\gamma^3\gamma^5 P_{s}\Psi = \bar{\Psi}P_{s} \Psi =
\psi_{K  As}^\dagger\psi_{K As} - \psi_{K^{\prime}Bs}^\dagger \psi_{K^{\prime} Bs}$.
For $\mbox{sign}(eB_{\perp}) < 0$, the relations are $\Psi^{\dagger}P_{s}\Psi =
-\bar{\Psi}\gamma^3 \gamma^5 P_{s} \Psi = \psi_{K^{\prime}  As}^\dagger\psi_{K^{\prime} As} +
\psi_{K Bs}^\dagger \psi_{K Bs}$ and
$\Psi^{\dagger}\gamma^3\gamma^5 P_{s}\Psi = -\bar{\Psi}P_{s} \Psi =
-\psi_{K^{\prime}  As}^\dagger\psi_{K^{\prime} As} + \psi_{K Bs}^\dagger \psi_{K Bs}$.

\bibitem{edge_states}
V.P.~Gusynin, V.A.~Miransky, S.G.~Sharapov, and I.A.~Shovkovy,
arXiv:0801.0708 [cond-mat.mes-hall].

\bibitem{edge_states_long}
V.P.~Gusynin, V.A.~Miransky, S.G.~Sharapov, and I.A.~Shovkovy,
Phys. Rev. B {\bf 77}, 205409 (2008).

\bibitem{footnote2} In dynamics in a magnetic field at zero temperature, there is
no one-to-one correspondence between electron density and chemical potential.
As a result, different values of the latter may correspond to the same physics,
as it takes place for this solution.

\bibitem{Herbut-scaling}
I.F.~Herbut and B.~Roy,
Phys. Rev. B {\bf 77}, 245438 (2008).

\bibitem{GR} I.S. Gradshtein, I.M. Ryzhik,  {\sl Tables of Integrals, Series,
and Products} (Academic Press, Orlando, 1980).

\bibitem{Zak} J. Zak, Phys. Rev. {\bf 134}, A1602 (1964).

\bibitem{potential}
J.M.~Luttinger and J.D.~Ward, Phys. Rev. {\bf 118}, 1417 (1960);
L.P.~Kadanoff and G.~Baym, {\it ibid}  {\bf 124}, 287 (1961);
G.~Baym, {\it ibid} {\bf 127}, 1391 (1962);
J.M.~Cornwall, R.~Jackiw, and E.~Tomboulis, Phys. Rev. D {\bf 10}, 2428 (1974).

\bibitem{Gorbar2003PLA} E.V.~Gorbar, V.P.~Gusynin, V.A.~Miransky,
and I.A.~Shovkovy, Phys. Lett. A {\bf 313}, 472 (2003).

\end{thebibliography}
\end{document}